\documentclass[aps,preprint,groupedaddress]{revtex4} 
\usepackage{amssymb}
\pdfoutput=1
\pdfoutput=1
\pdfoutput=1
\pdfoutput=1
\pdfoutput=1
\usepackage{amsmath}
\usepackage{amssymb}
\usepackage{graphicx}
\usepackage{subfigure}
\usepackage{color}
\usepackage{mathrsfs}
\usepackage[dvipsnames]{xcolor}

\usepackage[breaklinks=true,colorlinks=true]{hyperref}
\hypersetup{colorlinks=true,citecolor=blue,linkcolor=blue,urlcolor=blue}

\usepackage[utf8]{inputenc}

\begin{document}
\immediate\write16{<<WARNING: LINEDRAW macros work with emTeX-dvivers
                    and other drivers supporting emTeX \special's
                    (dviscr, dvihplj, dvidot, dvips, dviwin, etc.) >>}

\title{Oscillons from $Q$-balls in generalized models}

\author{E. da Hora$^{1,2}$}

%\email{carlos.hora@ufma.br}

\author{Fabiano C. Simas$^{1,3}$}

%\email{fc.simas@ufma.br}

\affiliation{$^{1}$Programa de Pós-graduação em Física, Universidade Federal do Maranhão, 65080-805, São Luís, Maranhão, Brazil.\\$^{2}$Coordenação do Curso de Bacharelado Interdisciplinar em Ciência e Tecnologia, Universidade Federal do Maranhão,  65080-805, São Luís, Maranhão, Brazil.\\$^{3}$Coordenação do Curso de Física-Bacharelado, Universidade Federal do Maranhão, 65080-805, São Luís, Maranhão, Brazil.}

%%%%%%%%%%%%%%%%%%%%%%%%%%%%%%%%%%%%%%%%%%%%%%%%%
 
\begin{abstract}

We study the %relation between oscillons and $Q$-balls
oscillon/$Q$-ball relation in an extended model with non-canonical kinematics. The model contains a single real scalar field whose kinetic term is enlarged to include a generalizing function. %of the field itself.
We approximate the real sector up to the third order in a book-keeping parameter. In this context, we implement the Renormalization Group Perturbation Expansion (RGPE), from which we conclude that the relation between oscillons and underlying $Q$-balls holds even in the presence of nontrivial kinematics. %To corroborate our results,
We apply our results to the study of three different effective cases. In all of them, %we observe that
our expressions mimic the numerical evolution of nonstandard oscillons with great accuracy, especially for small and moderate amplitudes. As the initial amplitude increases, the numerical profile develops a modulated behavior, and we use a two $Q$-balls solution to seed our analytical oscillon. %and then preserve the accuracy of the mapping.
We discuss how our non-canonical construction allows the existence of a well-behaved oscillon in connection to the simplest $\phi^2$-potential. This novel profile behaves in the same general way as the previous ones. %, including the emergence of modulated oscillations.
So, we argue that they belong to the same universality class. Finally, we extend our analysis to consider those contributions up to the fifth order in the approximation expansion. We explore an exotic $\phi^6$-scenario, and conclude that the relation between generalized oscillons and underlying $Q$-balls now belongs to a different universality class.

\end{abstract}

%%%%%%%%%%%%%%%%%%%%%%%%%%%%%%%%%%%%%%%%%%%%%%%%%

\maketitle

%%%%%%%%%%%%%%%%%%%%%%%%%%%%%%%%%%%%%%%%%%%%%%%%%
\section{Introduction} \label{secI}
%%%%%%%%%%%%%%%%%%%%%%%%%%%%%%%%%%%%%%%%%%%%%%%%%

%Some nonlinear field models give rise to long-lived, spatially localized configurations so-called \it{oscillons} stand for long-lived, spatially localized configurations that arise in some nonlinear models \cite{d1,d2,d3,d4,d5,d6,d7,d8,d9,d10,d11,d12,d13}. They exhibit a quasi-periodic profile that gain relevance in connection to areas such as condensed matter \cite{d21} and cosmology \cite{d14,d15,d16,d17,d18,d19,d20}. These solutions also appear in the electroweak model \cite{d22,d23}. However, in this case, their physical significance remains an open question.

Some nonlinear field models admit long-lived, spatially localized solutions so-called {\it{oscillons}} \cite{d1,d2,d3,d4,d5,d6,d7,d8,d10,d11,d12,d13}. These quasi-periodic configurations gain physical relevance in connection to condensed matter \cite{d21} and cosmology \cite{d14,d15,d16,d17,d18,d19,d20}. They also appear in the electroweak model \cite{d22,d23}. However, in this case, their physical meaning remains an open question.

Oscillons continue to be deeply enigmatic, although their notable similarity with topological solitons. In particular, their most distinctive aspects — extended lifetime and modulated amplitudes — have yet to be understood.

The stability of both topological solitons and $Q$-balls \cite{d24,d25,d26,d27} arises from their topological or non-topological charges. On the other hand, oscillons persist for remarkably long periods without a clear stabilizing mechanism. In general, their unusually slow decay is believed to be based on some hidden properties of the highly nonlinear field equations. However, no consensus on this has emerged. One hope is through $Q$-balls themselves. In this case, an approximately conserved $U(1)$ charge might provide temporary stability \cite{d28,d29,d30,d31}.

In the simplest case, oscillons continuously interpolate between two static maxima. This behavior indicates that there exists only one (fundamental) frequency. However, in complex scenarios, the maxima themselves also interpolate between two different values. The mechanism that gives rise to such additional frequency is still unknown. Nevertheless, it induces the oscillon to assume a modulated behavior whose description now requires two independent degrees of freedom. Here, it is worthwhile to note that the approximated treatment proposed by Fodor et al. applies to unmodulated oscillons (i.e. those with a single frequency only), see Ref. \cite{d32}. So, a detailed explanation of the modulated structure remains to be elaborated.

In this sense, Blaschke et al. recently argued that a bound state of two unmodulated oscillons may form a modulated one, see Ref. \cite{d33}. In such a scenario, each fundamental oscillon contributes with one degree of freedom. Then, the mutual interaction gives rise to the modulated behavior. However, in that work, the idea was applied in the context of a very specific model.

Even more recently, the same authors demonstrated the existence of a close relation between $(1+1)$-dimensional oscillons and $Q$-balls, see Ref. \cite{prl}. To find this conclusion, they applied the Renormalization Group Perturbation Expansion (RGPE) \cite{d34,d35}. Then, they showed that analytical oscillons are related to those $Q$-balls inherent to an underlying complex theory. To corroborate the construction, they demonstrated that these analytical solutions map the evolution of a numerical oscillon with reasonable accuracy, including the modulated behavior eventually exhibited by them. In other words, they clarified that a modulated oscillon can be seen indeed as a bound state of two unmodulated ones. For additional details, the reader is also referred to Ref. \cite{prd}.

The present manuscript serves as an extension of that work. Here, we go further and apply the same idea to a generalized model. That is, we investigate the oscillon/$Q$-ball relation in a theory with nonstandard kinematics. In such a highly nonlinear scenario, our aim is to study not only whether the relation still holds, but also how the generalization affects it. We also intend to check whether the novel analytical solution can be used to mimic the numerical evolution of a generalized oscillon accurately. The description of a modulated noncanonical profile in terms of two unmodulated oscillons is also investigated.

Oscillons with nonstandard kinematics were originally studied in \cite{amin}. In that work, the author added a quartic contribution to the standard (quadratic) kinetic term. However, the focus {\it{was not}} the oscillon/$Q$-ball relation. Here, on the other hand, we deal with a different generalization. We consider a kinetic term that includes an arbitrary function of the real scalar field. It is worthwhile to highlight that the original generalization finds important applications in connection to axions \cite{a37_1,a37_2}, Dirac-Born-Infeld inflation \cite{a1}, monodromy inflation \cite{a38,a39}, {\it{k}}-essence \cite{a4} and scalar-tensor theories \cite{a3}, while our idea has been used to study brane models with internal structures \cite{h54}, $(D+1)$-dimensional scenarios with stable global solutions \cite{h53}, domain-walls in magnetic materials submitted to geometric constraints \cite{h55}, fermions in a periodic bosonic background \cite{hnew3}, crystals as multi-kink arrangements \cite{hnew1,hnew2}, and double-kink configurations both numerically \cite{hnew4} and analytically \cite{hnew5}.

In order to present our results, this manuscript is organized as follows: In the Section II, we introduce the generalized one-field model, some definitions and conventions. We approximate the field up to the third order in a book-keeping parameter. Then, we implement the RGPE, from which we demonstrate that generalized analytical oscillons are closely related to those $Q$-balls inherent to an underlying theory. We apply our expressions to the study of three different effective scenarios. %They are defined by the $\phi^3$, inverse $\phi^4$, and double-well $\phi^4$ potentials.
In all cases, we compare the numerical evolution of a generalized oscillon %study the evolution of a generalized oscillon numerically. In the sequence, we compare it 
with our analytical result. We also use the two $Q$-ball profile to seed our renormalized oscillon, from which we allow it to describe those numerical solutions with modulated structure. Additionally, in Sec. III, we highlight that our expressions lead to a novel oscillon/$Q$-ball correspondence. We illustrate this point by focusing on the simplest $\phi^2$-potential. In the canonical scenario, it does not give rise to analytical oscillons. However, in our generalized case, we not only prove the existence of a well-behaved oscillon/$Q$-ball map, but we also use a two $Q$-ball profile to mimic the modulated numerical behavior. We extend our analysis to include those contributions up to the fifth order in the approximation expansion, from which we obtain a different relation between oscillons and $Q$-balls. We apply this extended result to the exotic $\phi^6$-potential. In Sec. IV, we present our final comments and perspectives.

%%%%%%%%%%%%%%%%%%%%%%%%%%%%%%%%%%%%%%%%%%%%%%%%%
\section{Main developments} \label{secII}
%%%%%%%%%%%%%%%%%%%%%%%%%%%%%%%%%%%%%%%%%%%%%%%%%

\subsection{Oscillons from $Q$-balls} \label{subsecII}

We consider a non-canonical model with a single real scalar field $\phi$. Its kinematics is enlarged to include an arbitrary function $f(\phi)$ that depends solely on $\phi$. The model is defined in a $(1+1)$-dimensional spacetime whose Minkowski metric is $\eta=$ diag$(+~-)$. Its Lagrange density reads%
\begin{equation}
\mathcal{L}=\frac{1}{2}f\left( \phi \right) \partial _{\mu }\phi \partial
^{\mu }\phi -V\left( \phi \right) \text{.}\label{gg1}
\end{equation}
Here, the generalizing function $f$ is assumed to be positive to ensure positiveness of the energy. As usual, $V(\phi)$ is the potential that controls the field self-interaction.

%It also defines the vacuum structure of the effective scenario.

The equation of motion (EoM) that comes from (\ref{gg1}) can be written in terms of $x$ and $t$ as%
%is%
%\begin{equation}
%f\partial _{\mu }\partial ^{\mu }\phi +\frac{1}{2}f_{\phi}\partial
%_{\mu }\phi \partial ^{\mu }\phi =-V_{\phi}\text{,} \label{gg2}
%\end{equation}
%where $f_{\phi}=df/d \phi$ and $V_{\phi} = dV/d \phi$.
%Alternatively, Eq. (\ref{gg2}) can also be written in terms of $x$ and $t$ as%
\begin{equation}
f\left( \frac{\partial ^{2}\phi }{\partial t^{2}}-\frac{\partial ^{2}\phi }{%
\partial x^{2}}\right) +\frac{f_{\phi}}{2}\left( \left( \frac{%
\partial \phi }{\partial t}\right) ^{2}-\left( \frac{\partial \phi }{%
\partial x}\right) ^{2}\right) =-V_{\phi}\text{,}  \label{gceomx1}
\end{equation}
where $f_{\phi}=df/d \phi$ and $V_{\phi} = dV/d \phi$.

We intend to apply the RGPE algorithm to the enlarged Eq. (\ref{gceomx1}). %model. Our aim is to study how a nontrivial kinematics affects the fundamental relation between oscillons and \textit{Q}-balls.
With this aim in mind, we follow the steps proposed in Ref. \cite{prd}. First, it proves useful to introduce the new variable%
\begin{equation}
\theta =Qx-\Omega t\text{.}
\end{equation}%
%from which we rewrite Eq. (\ref{gceomx1}) in the form%
%\begin{equation}
%f \frac{d^{2}\phi }{d\theta ^{2}}+\frac{f_{\phi}}{2}%
% \left( \frac{d\phi }{%
%d\theta }\right) ^{2}=-\frac{V_{\phi}}{\Omega ^{2}-Q^{2}}\text{.}\label{gg3}
%\end{equation}%

In the sequence, we choose $\Omega =\sqrt{Q^{2}+1}$, for the sake of simplicity. So, Eq. (\ref{gceomx1}) assumes the form
\begin{equation}
f\frac{d^{2}\phi }{d\theta ^{2}}+\frac{f_{\phi}}{2}\left( \frac{%
d\phi }{d\theta }\right) ^{2}=-V_{\phi}\text{,}  \label{gceom0x1}
\end{equation}%
which promptly recovers the canonical structure for $f=1$.

Now, it is necessary to specify the potential $V(\phi)$. We assume that it has a minimum at $\phi=0$ around which the oscillon oscillates. Also, the field mass is assumed to be $1$, for the sake of illustration. In view of these arguments, we write the potential as%
\begin{equation}
V\left( \phi \right) = \frac{\phi ^{2}}{2}%
-a_{3}\frac{\phi ^{3}}{3}-a_{4}\frac{\phi ^{4}}{4}+...\text{ ,}\label{gg4}
\end{equation}%
where the coefficients $a_{i}$'s are real numbers.

In our generalized scenario, we also need to fix the function $f(\phi)$. In the context of time-independent fields, $f$ can be used to modify the vacuum structure inherent to the potential. As a consequence, novel structures may appear. %It sounds reasonable to imagine that the same argument applies to non-BPS models.
In any case, it is important to ensure the possibility of recovering the canonical $f=1$ case. So, inspired by Eq. (\ref{gg4}), we propose%
\begin{equation}
f\left( \phi \right) = b_{0}+b_{1}\phi +b_{2}\frac{\phi ^{2}}{2}-b_{3}%
\frac{\phi ^{3}}{3}-b_{4}\frac{\phi ^{4}}{4}+...\text{ ,}\label{gg5}
\end{equation}%
where all $b_{i}$'s are real. Naturally, it reduces to the standard case for $b_{0}=1$, with the additional coefficients vanishing.

%In view of Eqs. (\ref{gg4}) and (\ref{gg5}), we rewrite Eq. (\ref{gceom0x1}) as%
%\begin{equation}
%\left( b_{0}+b_{1}\phi +b_{2}\frac{\phi ^{2}}{2}-b_{3}\frac{\phi ^{3}}{3}%
%-b_{4}\frac{\phi ^{4}}{4}+...\right) \frac{d^{2}\phi }{d\theta ^{2}}+\frac{1}{2}%
%\left( \frac{d\phi }{d\theta }\right) ^{2}\left( b_{1}+b_{2}\phi -b_{3}\phi
%^{2}-b_{4}\phi ^{3}+...\right) =-\phi +a_{3}\phi ^{2}+a_{4}\phi ^{3}+...\text{.}
%\label{gceom2x}
%\end{equation}
%\begin{eqnarray}
%&&\left( b_{0}+b_{1}\phi +b_{2}\frac{\phi ^{2}}{2}-b_{3}\frac{\phi ^{3}}{3}%
%-b_{4}\frac{\phi ^{4}}{4}+...\right) \frac{d^{2}\phi }{d\theta ^{2}}+\frac{1}{2}%
%\left( \frac{d\phi }{d\theta }\right) ^{2}\left( b_{1}+b_{2}\phi -b_{3}\phi
%^{2}-b_{4}\phi ^{3}+...\right)   \notag \\
%&&\text{ \ \ \ \ \ \ \ \ \ \ \ \ \ \ \ \ \ \ \ \ \ \ \ \ \ \ \ \ \ \ \ \ \ \
%\ \ \ \ \ \ \ \ \ \ \ \ \ \ \ \ \ \ \ \ \ \ \ \ \ \ \ \ \ \ \ \ \ \ }\left.
%=-\phi +a_{3}\phi ^{2}+a_{4}\phi ^{3}+...\right. \text{,}  \label{gceom2x}
%\end{eqnarray}

In the sequence, we approximate the field $\phi$ itself via a small-amplitude perturbation expansion, i.e.%
\begin{equation}
\phi = \varepsilon \phi _{1}+\varepsilon ^{2}\phi _{2}+\varepsilon
^{3}\phi _{3}+...\text{ ,}  \label{epx1}
\end{equation}%
where $\varepsilon$ stands for a book-keeping parameter.

In view of Eqs. (\ref{gg4}), (\ref{gg5}) and (\ref{epx1}), {\textit{up to the third order in $\varepsilon$}}, Eq. (\ref{gceom0x1}) leads to%
\begin{equation}
\frac{d^{2}\phi _{1}}{d\vartheta ^{2}}+\phi _{1}=0\text{,}  \label{ep1x0_y1}
\end{equation}%
\begin{equation}
\frac{d^{2}\phi _{2}}{d\vartheta ^{2}}+\phi _{2}+B_{1}\phi _{1}\frac{%
d^{2}\phi _{1}}{d\vartheta ^{2}}+\frac{B_{1}}{2}\left( \frac{d\phi _{1}}{%
d\vartheta }\right) ^{2}=a_{3}\phi _{1}^{2}\text{,}  \label{ep2x0_y1}
\end{equation}%
\begin{equation}
\frac{d^{2}\phi _{3}}{d\vartheta ^{2}}+\phi _{3}+B_{1}\phi _{1}\frac{%
d^{2}\phi _{2}}{d\vartheta ^{2}}+\left( \frac{B_{2}}{2}\phi
_{1}^{2}+B_{1}\phi _{2}\right) \frac{d^{2}\phi _{1}}{d\vartheta ^{2}}+B_{1}%
\frac{d\phi _{1}}{d\vartheta }\frac{d\phi _{2}}{d\vartheta }+\frac{B_{2}}{2}%
\phi _{1}\left( \frac{d\phi _{1}}{d\vartheta }\right) ^{2}=2a_{3}\phi
_{1}\phi _{2}+a_{4}\phi _{1}^{3}\text{.}  \label{ep3x0_y1}
\end{equation}%
Here, where we have introduced the rescaled coordinate $\vartheta =\theta / \sqrt{b_{0}}$. We have also defined the novel parameters
\begin{equation}
B_{1}=\frac{b_{1}}{b_{0}}\text{ \ \ and \ \ }B_{2}=\frac{b_{2}}{b_{0}}\text{,%
}
\end{equation}%
which carry the influence of the generalized kinematics from now on.

We study the solutions that emerge from Eqs. (\ref{ep1x0_y1}), (\ref{ep2x0_y1}) and (\ref{ep3x0_y1}). First, we note that, except for the presence of the rescaled coordinate, Eq. (\ref{ep1x0_y1}) is the same Eq. (21a) of Ref. \cite{prd}. So, its solution can be written automatically as%
\begin{equation}
\phi _{1}\left( \vartheta \right) =A_{0}e^{i\vartheta }+\text{c.c..}  \label{sp1x0_y1}
\end{equation}
%which describes a monochromatic wave in the very same way of Eq.~(22) of Ref. XXX.

On the other hand, Eq. (\ref{ep2x0_y1}) differs dramatically from its canonical $f=1$ version, see Eq. (21b) of Ref. \cite{prd}. The difference is due to those terms that are proportional to $B_1$ (in other words, to the influence of nonstandard kinematics). Surprisingly, the non-trivial scenario still admits an exact solution $\phi _{2}\left( \vartheta \right)$. The point is that, in view of Eq. (\ref{sp1x0_y1}), Eq. (\ref{ep2x0_y1}) can be written in the form%
\begin{equation}
\frac{d^{2}\phi _{2}}{d\vartheta ^{2}}+\phi _{2}=\left( a_{3}+\frac{3B_{1}}{2%
}\right) A_{0}^{2}e^{2i\vartheta } + \left\vert A_{0}\right\vert ^{2}\left(
a_{3}+\frac{B_{1}}{2}\right) +\text{c.c.,}
\end{equation}%
whose simplest solution reads%
\begin{equation}
\phi _{2}\left( \vartheta \right) =-\frac{1}{3}\left( a_{3}+\frac{3B_{1}}{2}%
\right) A_{0}^{2}e^{2i\vartheta }+\left\vert A_{0}\right\vert ^{2}\left(
a_{3}+\frac{B_{1}}{2}\right)+\text{c.c..}  \label{sp2x00_y1}
\end{equation}%
Here, the two integration constants were assumed to vanish, for the sake of simplicity. Note that Eq. (\ref{sp2x00_y1}) recovers the usual result in the limit $B_1 \rightarrow 0$, see Eq. (23a) of Ref. \cite{prd}.

Now, we focus on Eq. (\ref{ep3x0_y1}). It has a total of \textit{five} new terms. They carry not only $B_1$, but also $B_2$. In particular, the nonstandard kinematics leads to a novel crossed term of the form $d_{\vartheta} \phi _{1}d_{\vartheta} \phi _{2}$.

Despite its intricate structure, Eq. (\ref{ep3x0_y1}) can be verified to support an analytical solution. With this aim in mind, we use Eqs. (\ref{sp1x0_y1}) and (\ref{sp2x00_y1}). Then, after some algebraic work, we rewrite Eq. (\ref{ep3x0_y1}) as%
\begin{equation}
\frac{d^{2}\phi _{3}}{d\vartheta ^{2}}+\phi _{3}=\beta A_{0}\left\vert
A_{0}\right\vert ^{2}e^{i\vartheta }-8\alpha A_{0}^{3}e^{3i\vartheta }+\text{c.c.,}  \label{ep3z1_y1}
\end{equation}%
where we have defined%
\begin{equation}
\beta =\beta
_{u}+B_{2}+\left( 2a_{3}-\frac{B_{1}}{2}\right) B_{1}\text{ \ \ and \ \ }\beta _{u}=\frac{10a_{3}^{2}}{3}+3a_{4}\text{,}\label{bet}
\end{equation}%
\begin{equation}
\alpha =\alpha _{u}-\frac{B_{2}}{8}+%
\frac{B_{1}}{8}\left( \frac{10a_{3}}{3}+\frac{7B_{1}}{2}\right)\text{ \ \ and \ \ }\alpha _{u}=\frac{1}{24}\left( 2a_{3}^{2}-3a_{4}\right)\text{.}\label{alf}
\end{equation}
Here, $\beta _{u}$ and $\alpha _{u}$ stand for the \textit{usual} parameters, see the discussion right after Eq. (6) of Ref. \cite{prl}. Naturally, the generalized constants $\beta$ and $\alpha$ reduce to the canonical ones for $B_1=B_2=0$.

The noticeable conclusion is that, despite all non-standard terms in original Eq. (\ref{ep3x0_y1}), Eq. (\ref{ep3z1_y1}) gives rise to the exact solution%
\begin{equation}
\phi _{3}\left( \vartheta \right) =\alpha A_{0}^{3}e^{3i\vartheta }+\beta
A_{0}\left\vert A_{0}\right\vert ^{2}\mathcal{S}e^{i\vartheta }+\text{c.c.,}
\label{sp3x0_y1}
\end{equation}%
where we have again settled on the integration constants as $0$. This generalized solution presents the same structure as the standard result, see Eq. (6) of Ref. \cite{prl}. However, both $\beta$ and $\alpha$ now enclose the influence of the non-usual kinematics by means of $B_1$ and $B_2$.

In Eq. (\ref{sp3x0_y1}), $\mathcal{S}=\mathcal{S}(\vartheta,\overline{\vartheta })$ represents the so-called secular term. As in the usual scenario, this function satisfies%
\begin{equation}
\frac{\partial
^{2}\mathcal{S}}{\partial \vartheta ^{2}}-\frac{\partial ^{2}\mathcal{S}}{%
\partial \overline{\vartheta }^{2}}+2i\frac{\partial \mathcal{S}}{\partial
\vartheta }=1\text{,}  \label{fe_0x_y1}
\end{equation}
whose solution is not unique. Here, we have defined $\overline{\vartheta }=\overline{\theta}/\sqrt{b_{0}}$, with $\overline{\theta} =\Omega x-Qt$.

Then, up to the third order, the \textit{bare} solution Eq. (\ref{epx1}) can be written as%
\begin{eqnarray}
\phi _{B}\left( \vartheta \right) &=&\varepsilon A_{0}e^{i\vartheta
}-\frac{\varepsilon ^{2}}{3}\left( a_{3}+\frac{3B_{1}}{2}\right)
A_{0}^{2}e^{2i\vartheta }+\varepsilon ^{2}\left( a_{3}+\frac{B_{1}}{2}%
\right) \left\vert A_{0}\right\vert ^{2}  \notag \\
&&+\varepsilon ^{3}\alpha
A_{0}^{3}e^{3i\vartheta }+\varepsilon ^{3}\beta A_{0}\left\vert
A_{0}\right\vert ^{2}\mathcal{S}e^{i\vartheta }+\text{c.c.,}\label{fe_0x_y11}
\end{eqnarray}%
where we have used Eqs. (\ref{sp1x0_y1}), (\ref{sp2x00_y1}) and (\ref{sp3x0_y1}).

The secular term $\mathcal{S}$ is a mathematical artifact. So, physics is expected not to depend on it. Therefore, it is necessary to remove it from Eq. (\ref{fe_0x_y11}). To do so, we propose a connection between the \textit{bare} amplitude $A_0$ and its \textit{dressed} version $A$. The connection reads%
\begin{equation}
A_{0}=A\left( 1-\varepsilon ^{2}\beta \mathcal{S}_{0}\left\vert A\right\vert
^{2}+\mathcal{O}\left( \varepsilon ^{3}\right) \right) \text{.}\label{dA}
\end{equation}%
Here, both $A$ and $\mathcal{S}_{0}$ are functions of the renormalized scales $\vartheta_{0}$ and $\overline{\vartheta}_{0}$, simultaneously.

We then use Eq. (\ref{dA}) to rewrite Eq. (\ref{fe_0x_y11}) as%
\begin{eqnarray}
\phi _{D}\left( \vartheta \right) &=&\varepsilon Ae^{i\vartheta
}-\frac{\varepsilon ^{2}}{3}\left( a_{3}+\frac{3B_{1}}{2}\right)
A^{2}e^{2i\vartheta }+\varepsilon ^{2}\left( a_{3}+\frac{B_{1}}{2}\right)
\left\vert A\right\vert ^{2}  \notag \\
&&+\varepsilon ^{3}\alpha A^{3}e^{3i\vartheta
}+\varepsilon ^{3}\beta A\left\vert A\right\vert ^{2}\left( \mathcal{S}-%
\mathcal{S}_{0}\right) e^{i\vartheta }+\text{c.c..}\label{ds}
\end{eqnarray}

In addition, the dressed solution $\phi _{D}\left( \vartheta \right)$ is required to be independent of the renormalized scales. So, we impose
\begin{equation}
\frac{\partial \phi _{D}}{\partial \vartheta _{0}}=\frac{\partial \phi _{D}}{\partial \overline{\vartheta }_{0}}=0\text{,}
\end{equation}%
which leads to RG equations of the form%
\begin{equation}
\frac{\partial A}{\partial \vartheta _{0}}=\varepsilon ^{2}\beta A\left\vert
A\right\vert ^{2}\frac{\partial \mathcal{S}_{0}}{\partial \vartheta _{0}}+\mathcal{O}\left( \varepsilon ^{3}\right)\text{ \ \ and \ \ }\frac{\partial A}{\partial \overline{\vartheta }_{0}}=\varepsilon ^{2}\beta
A\left\vert A\right\vert ^{2}\frac{\partial \mathcal{S}_{0}}{\partial 
\overline{\vartheta }_{0}}+\mathcal{O}\left( \varepsilon ^{3}\right)
\text{.}\label{rge1}
\end{equation}%
%\begin{equation}
%\frac{\partial A}{\partial \overline{\vartheta }_{0}}=\varepsilon ^{2}\beta
%A\left\vert A\right\vert ^{2}\frac{\partial \mathcal{S}_{0}}{\partial 
%\overline{\vartheta }_{0}}+\mathcal{O}\left( \varepsilon ^{3}\right)\text{.}
%\end{equation}

The minimal subtraction scheme is defined by $\mathcal{S}=\mathcal{S}_{0}$. To attain this regime, it is necessary to choose $\vartheta=\vartheta_{0}$ and $\overline{\vartheta}=\overline{\vartheta}_{0}$. As a consequence, the secular term is entirely removed from Eq. (\ref{ds}).

In view of $\mathcal{S}=\mathcal{S}_{0}$, and to make the RG equation independent of the secular function, we adopt%
\begin{equation}
2i\frac{\partial A}{\partial \vartheta _{0}}+\frac{\partial ^{2}A}{\partial
\vartheta _{0}^{2}}-\frac{\partial ^{2}A}{\partial \overline{\vartheta }%
_{0}^{2}}=\varepsilon ^{2}\beta A\left\vert A\right\vert ^{2}\left( 2i\frac{%
\partial \mathcal{S}_{0}}{\partial \vartheta _{0}}+\frac{\partial ^{2}%
\mathcal{S}_{0}}{\partial \vartheta _{0}^{2}}-\frac{\partial ^{2}\mathcal{S}%
_{0}}{\partial \overline{\vartheta }_{0}^{2}}\right) \text{,}
\end{equation}%
which promptly reduces to%
\begin{equation}
2i\frac{\partial A}{\partial \vartheta _{0}}+\frac{\partial ^{2}A}{\partial
\vartheta _{0}^{2}}-\frac{\partial ^{2}A}{\partial \overline{\vartheta }%
_{0}^{2}}=\varepsilon ^{2}\beta A\left\vert A\right\vert ^{2}\text{,}
\label{rgex_y1}
\end{equation}%
where we have used Eq. (\ref{fe_0x_y1}).

It is interesting to note that, whether we define a new complex field $\Psi$ as%
\begin{equation}
\Psi \left( \vartheta \right) =\varepsilon \sqrt{\frac{\beta }{2}}%
Ae^{i\vartheta }\text{,}  \label{ncfx_y1}
\end{equation}%
Eq. (\ref{rgex_y1}) can be written in the form%
\begin{equation}
\frac{\partial ^{2}\Psi }{\partial \vartheta ^{2}}+\Psi =2\Psi \left\vert
\Psi \right\vert ^{2}\text{.}  \label{eom1}
\end{equation}

Equation (\ref{eom1}) can be seen as the EoM for $\Psi$ obtained from the Lagrange density%
\begin{equation}
\mathcal{L}=\left\vert \partial _{\mu }\Psi \right\vert ^{2}-\left\vert \Psi
\right\vert ^{2}+\left\vert \Psi \right\vert ^{4}\text{,}\label{lqb}
\end{equation}%
which is defined in a $(1+1)$-dimensional spacetime whose coordinates are $T=b_{0}^{-1/2}t$ and $X=b_{0}^{-1/2}x$. It follows from Eq. (\ref{lqb}) that the complex field self-interacts according to an upside-down wine-bottle potential. As a consequence, it supports $Q$-balls as its spatially localized periodic solutions.

In particular, whether we assume $\Omega =1$ and $Q=0$, %(i.e. $\vartheta=T$ and $\overline{\vartheta}=X$),
a stationary $Q$-ball emerges, i.e.%
\begin{equation}
\Psi \left( X,T \right)=\frac{\lambda e^{i\omega T}}{\cosh \left( \lambda X\right) }\text{.}
\label{cosh1}
\end{equation}%
Here, $\lambda$ represents a scale parameter, while $\omega =\sqrt{1-\lambda ^{2}}$ stands for the $Q$-ball's frequency.

At the same time, in the minimal choice regime, the dressed solution Eq. (\ref{ds}) assumes the \textit{renormalized} form%
%\begin{equation}
%\phi _{R}\left( \vartheta \right) =\varepsilon Ae^{i\vartheta
%}-\frac{\varepsilon ^{2}}{3}\left( a_{3}+\frac{3B_{1}}{2}\right)
%A^{2}e^{2i\vartheta }+\varepsilon ^{2}\left( a_{3}+\frac{B_{1}}{2}\right)
%\left\vert A\right\vert ^{2}+\varepsilon ^{3}\alpha A^{3}e^{3i\vartheta }+\text{c.c.,}
%\end{equation}%
%which can also be written as%
\begin{equation}
\phi _{R}\left( \vartheta \right) =\sqrt{\frac{2}{\beta }}\Psi -\frac{2}{%
3\beta }\left( a_{3}+\frac{3B_{1}}{2}\right) \Psi ^{2}+\frac{2}{\beta }%
\left( a_{3}+\frac{B_{1}}{2}\right) \left\vert \Psi \right\vert ^{2}+\alpha
\left( \frac{2}{\beta }\right) ^{\frac{3}{2}}\Psi ^{3}+\text{c.c.,}  \label{de1x}
\end{equation}
where we have used Eq. (\ref{ncfx_y1}).

Equation (\ref{de1x}) summarizes the main conclusion of our manuscript. That is, as in the canonical case, the renormalized oscillons that come from a $(1+1)$-dimensional theory \textit{with non-standard kinematics} are generated via those $Q$-ball solutions inherent to a subjacent model.% with a single complex scalar field.

The retention of such a property is indeed surprising once a non-trivial $f(\phi)$ has the power to affect drastically the original model, including the existence of novel solutions. Moreover, despite the generalized kinematics of the theory (\ref{gg1}), the subjacent model (\ref{lqb}) exhibits standard dynamics. In other words, except for the use of rescaled coordinates, canonical $Q$-balls serve as seeds to both the usual and generalized oscillons.

Remarkably, our renormalized result Eq. (\ref{de1x}) exhibits the very same mathematical structure as the usual one, see Eq. (40) of Ref. \cite{prd}. In other words, the generalized dynamics causes the rescaling of the original parameters. In this sense, the influence of the non-usual kinematics is enclosed by the novel constants, from which our results promptly recover the standard ones in the limit $f \rightarrow 1$.

The square-root in Eq. (\ref{ncfx_y1}) requires $\beta>0$. So, Eq. (\ref{de1x}) itself only holds when
\begin{equation}
\frac{10a_{3}^{2}}{3}+3a_{4}+B_{2}+\left(
2a_{3}-\frac{B_{1}}{2}\right) B_{1}>0 \text{}  \label{b1}
\end{equation}%
is satisfied, see Eq. (\ref{bet}). It is interesting to note that Eq. (\ref{b1}) offers possibilities that are absent in the usual case. For instance, in our enlarged scenario, oscillons may exist even for $a_3=a_4=0$. This is a clear new effect due to the generalized kinematics. We explore such a novel situation later below, when we also extend our calculations to cover new issues.

\subsection{Effective examples} \label{subsecIIa}

We now apply the expressions above to investigate effective non-trivial scenarios. To do so, we first need to choose a particular expression for $f(\phi)$. In this sense, we focus on
\begin{equation}
f\left( \phi \right) =\frac{1+\eta }{1+\eta \phi ^{2}}\text{.} \label{fphi}
\end{equation}
Here, $\eta \ge0$ is a real parameter that controls the strength of the effects due to the non-usual kinematics. As $\eta$ vanishes, one gets $f \rightarrow 1$. Then, kinematics reduces to the usual case, and the system behaves standardly. On the other hand, the limit $\eta \rightarrow \infty$ leads to $f\left( \phi \right) \rightarrow 1/\phi ^{2}$. In this regime, the novel effects possess the highest intensity possible.

Equation (\ref{fphi}) has been applied to describe geometrically constrained fields and some of their aspects, see Refs. \cite{balbazmar,jhepb,epjch}. Here, however, we are focused on the small-amplitude regime. So, we approximate Eq. (\ref{fphi}) as%
\begin{equation}
f\left( \phi \right) \approx \left( 1+\eta \right) -\eta \left( 1+\eta
\right) \phi ^{2}+\eta ^{2}\left( 1+\eta \right) \phi ^{4}\text{,}
\end{equation}%
from which we identify $b_{0}=1+\eta $, $b_{1}=b_{3}=0$, $b_{2}=-2\eta \left(
1+\eta \right) $, and $b_{4}=-4\eta ^{2}\left( 1+\eta \right) $, see Eq. (\ref{gg5}). These values lead to $B_{1}=0$ and $B_{2}=-2\eta$. As a consequence, we get%
\begin{equation}
\beta =\frac{10a_{3}^{2}}{3}+3a_{4}-2\eta\text{ \ \ and \ \ }\alpha =\frac{a_{3}^{2}}{12}-\frac{a_{4}}{8}+\frac{2\eta }{8} \text{,}
\label{bna}
\end{equation}%
see Eqs. (\ref{bet}) and (\ref{alf}).

In addition, Eq. (\ref{b1}) reduces to
\begin{equation}
\eta<\frac{5a_{3}^{2}%
}{3}+\frac{3a_{4}}{2} \text{.}  \label{bnn1}
\end{equation}
It represents an upper bound to be imposed on the values of $\eta$. Once this limit is defined in terms of $a_{3}$ and $a_{4}$ (i.e. the coefficients in Eq. (\ref{gg4})), we conclude that the structure of the potential restricts the effects due to the generalized kinematics. This conclusion emerges in a rather natural way from Eq. (\ref{b1}), and its implementation is explored in what follows.

%i.e. the coefficients that appear in Eq. (\ref{gg4}). Therefore, it indicates that the structure of the potential restricts the effects due to the generalized kinematics. Note that this conclusion emerges in a rather natural way from Eq. (\ref{b1}), and its implementation is explored in what follows.

\subsubsection{The $\phi ^{3}$-potential.} \label{subsecIIa1}

As an initial example, we consider the effective potential%
\begin{equation}
V\left( \phi \right) =\frac{\phi ^{2}}{2}-\frac{\phi ^{3}}{3}\label{p3}\text{,}
\end{equation}%
from which we get $a_{3}=1$, while all other coefficients in Eq. (\ref{gg4}) vanish. This particular potential was used previously in Ref. \cite{prd36} to study the connection between oscillons and sphalerons.

In view of $a_{3}=1$ and $a_{4}=0$, Eqs. (\ref{bna}) lead to%
\begin{equation}
\beta =\frac{2}{3}\left( 5-3\eta \right) \text{ \ \ and \ \ }\alpha =\frac{%
1+3\eta }{12}\label{bap3}\text{,}
\end{equation}%
which we use to write the renormalized profile Eq. (\ref{de1x}) as%
\begin{equation}
\phi _{R}\left( x,t\right) =\sqrt{\frac{3}{5-3\eta }}\frac{2\lambda \cos
\left( \frac{\omega }{\sqrt{1+\eta }}t\right) }{\cosh \left( \frac{\lambda }{%
\sqrt{1+\eta }}x\right) }+\frac{2\lambda ^{2}}{5-3\eta }\frac{3-\cos \left( 
\frac{2\omega }{\sqrt{1+\eta }}t\right) }{\cosh ^{2}\left( \frac{\lambda }{%
\sqrt{1+\eta }}x\right) }+\frac{1+3\eta }{2\left( 5-3\eta \right) }\sqrt{%
\frac{3}{5-3\eta }}\frac{\lambda ^{3}\cos \left( \frac{3\omega }{\sqrt{%
1+\eta }}t\right) }{\cosh ^{3}\left( \frac{\lambda }{\sqrt{1+\eta }}x\right) 
}\label{sp3}\text{,}
\end{equation}%
where we have also implemented the $Q$-ball solution Eq. (\ref{cosh1}).

%%%%%%%%%%%%%%%%%%%%%%%%%%%%%%%%%%%%%%%%%%%%%%%%%%%%%%%%%%%%%%%%%%%%%%%%%%% 
\begin{figure*}[!ht]
\begin{center}
  \centering
    \includegraphics[{angle=0,width=8cm,height=4cm}]{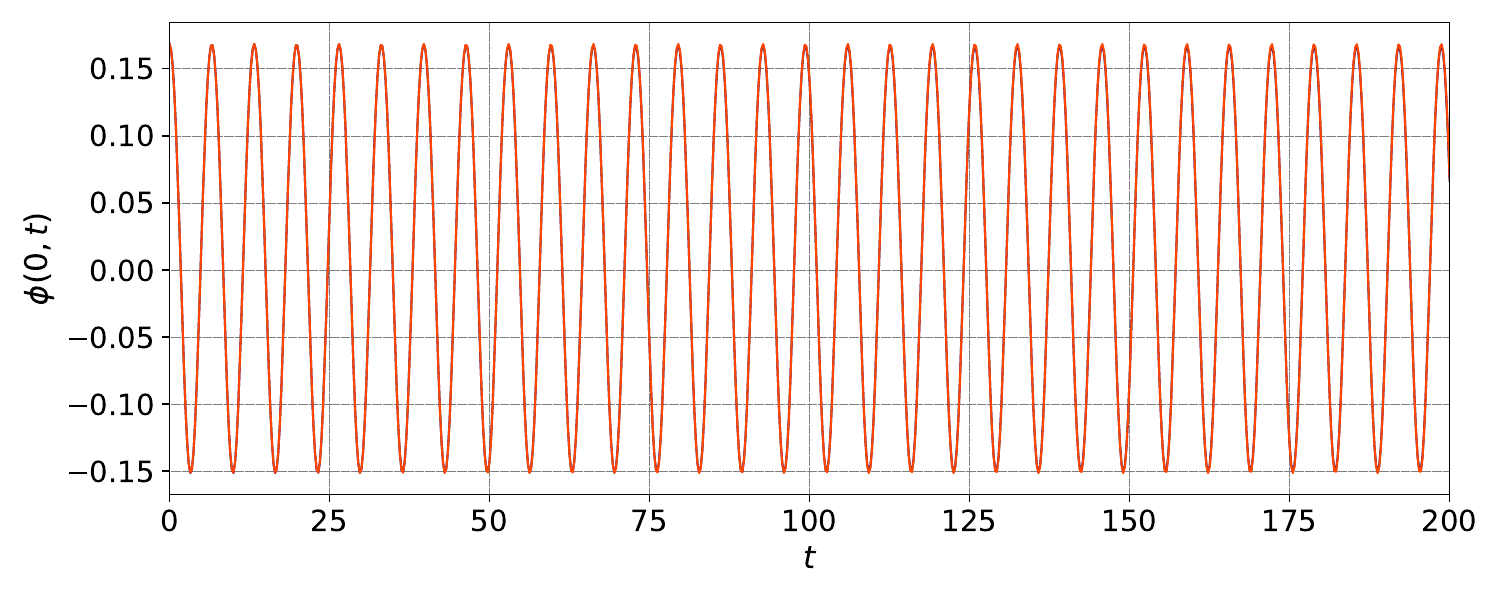}\label{phi3_01_01}
    \includegraphics[{angle=0,width=8cm,height=4cm}]{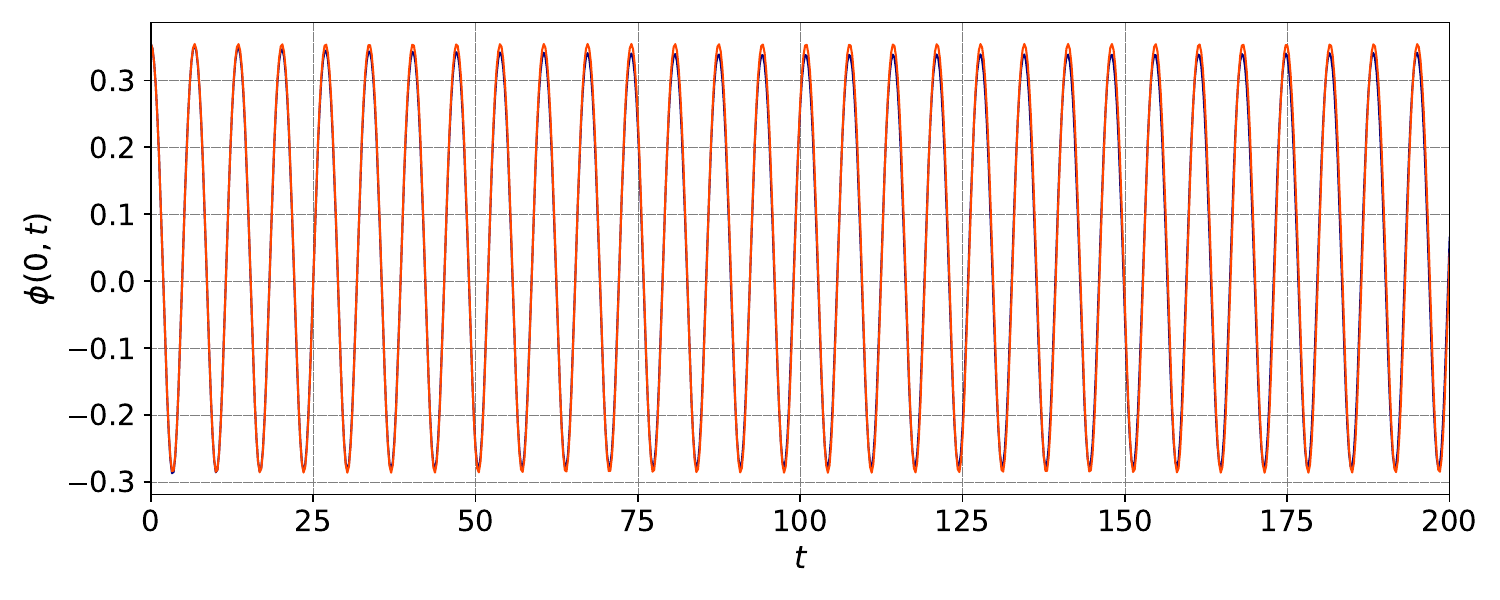}\label{phi3_01_02}
    \includegraphics[{angle=0,width=8cm,height=4cm}]{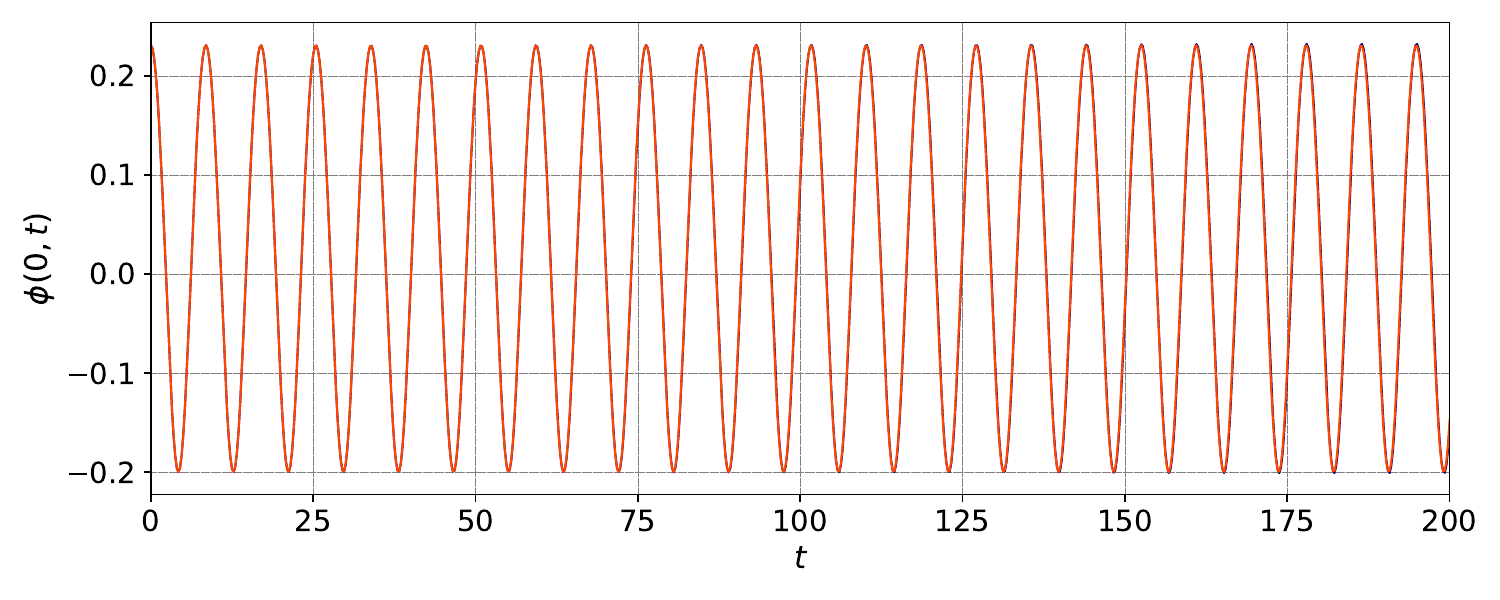}\label{phi3_08_01}
    \includegraphics[{angle=0,width=8cm,height=4cm}]{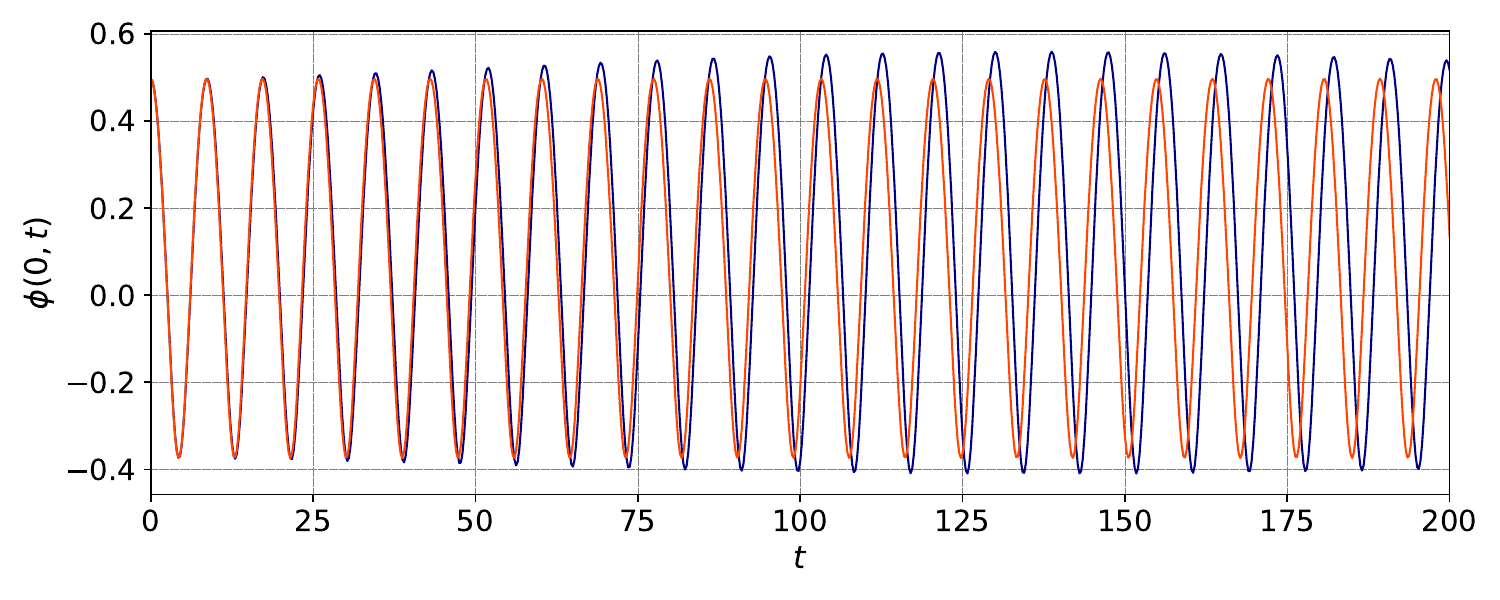}\label{phi3_08_02}
    \vspace{-0.5cm}
  \caption{Results for the $\phi^3$-potential Eq. (\ref{p3}). The numerical oscillon (black line) is compared to the renormalized analytical one (red line). Here, we have chosen $\eta=0.10$ (upper line) and $\eta=0.80$ (lower line), with $\lambda=0.10$ (left column) and $\lambda=0.20$ (right column). %Upper: $\eta=0.10$, with $\lambda=0.10$ (left) and $\lambda=0.20$ (right). Lower: $\eta=0.80$, with $\lambda=0.10$ (left) and $\lambda=0.20$ (right).}
  }
  \label{fig_1A}
\end{center}
\end{figure*}
%%%%%%%%%%%%%%%%%%%%%%%%%%%%%%%%%%%%%%%%%%%%%%%%%%%%%%%%%%%%%%%%%%%%%%%%%%

Equation (\ref{sp3}) emphasizes a central insight of our development, i.e. renormalized oscillons within the generalized framework are fundamentally generated from the $Q$-ball solutions of an underlying complex theory. Notably, the non-standard kinematics does not break this correspondence. Instead, it rescales the parameters while preserving the core mechanism of the mapping. So, %demonstrating that
the oscillon/$Q$-ball relation stands for a robust feature even in highly nonlinear scenarios.

%Here, we have used $T=b_{0}^{-1/2}t$ and $X=b_{0}^{-1/2}x$, with $b_{0}=1+\eta$.

As we have discussed, the potential now restricts the effects of the non-standard kinematics. In the present case, Eq. (\ref{p3}) leads to
\begin{equation}
\frac{5}{3}>\eta \label{up001}\text{,}
\end{equation}
see Eq. (\ref{bnn1}). %with $a_{3}=1$ and $a_{4}=0$.
This condition ensures the existence of %the $Q$-ball solution Eq. (\ref{cosh1}) and, as a consequence, also that of the
the renormalized oscillon Eq. (\ref{sp3}).

To corroborate our construction, we now verify whether our renormalized oscillon mimics the evolution of a generalized numerical one. With this aim in mind, we consider Eq. (\ref{gceomx1}) in the presence of Eqs. (\ref{fphi}) and (\ref{p3}). We adopt Eq. (\ref{sp3}) at $t=0$ as the initial configuration, and solve the numerical problem for different values of $\lambda$ and $\eta$.

%%%%%%%%%%%%%%%%%%%%%%%%%%%%%%%%%%%%%%%%%%%%%%%%%%%%%%%%%%%%%%%%%%%%%%%%%%% 
\begin{figure*}[!ht]
\begin{center}
  \centering
    \includegraphics[{angle=0,width=8cm,height=4cm}]{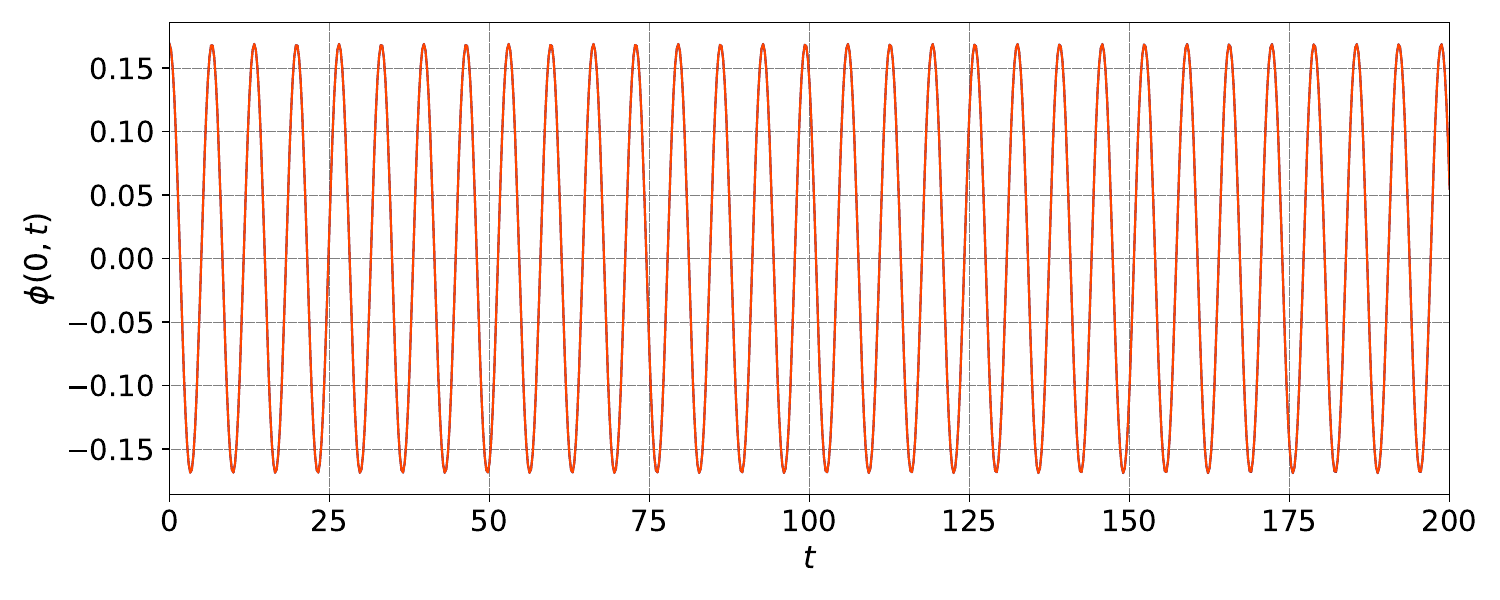}\label{phi4inv_01_01}
    \includegraphics[{angle=0,width=8cm,height=4cm}]{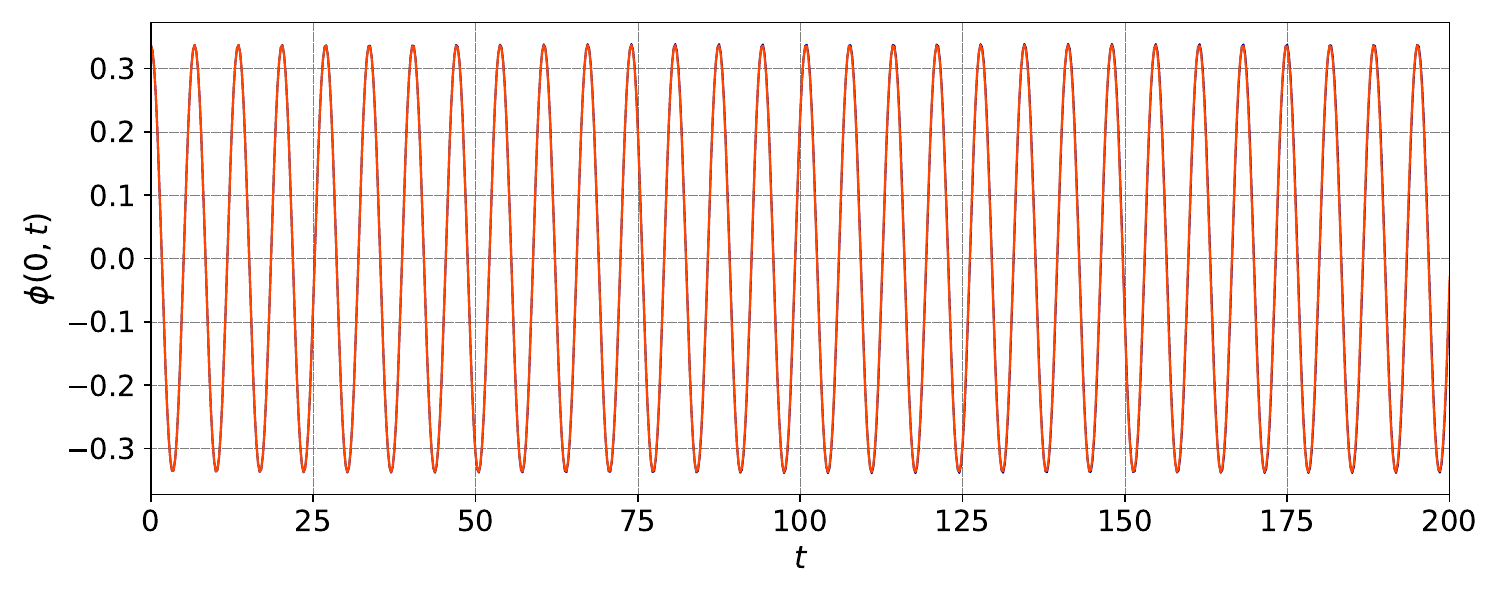}\label{phi4inv_01_02}
    \includegraphics[{angle=0,width=8cm,height=4cm}]{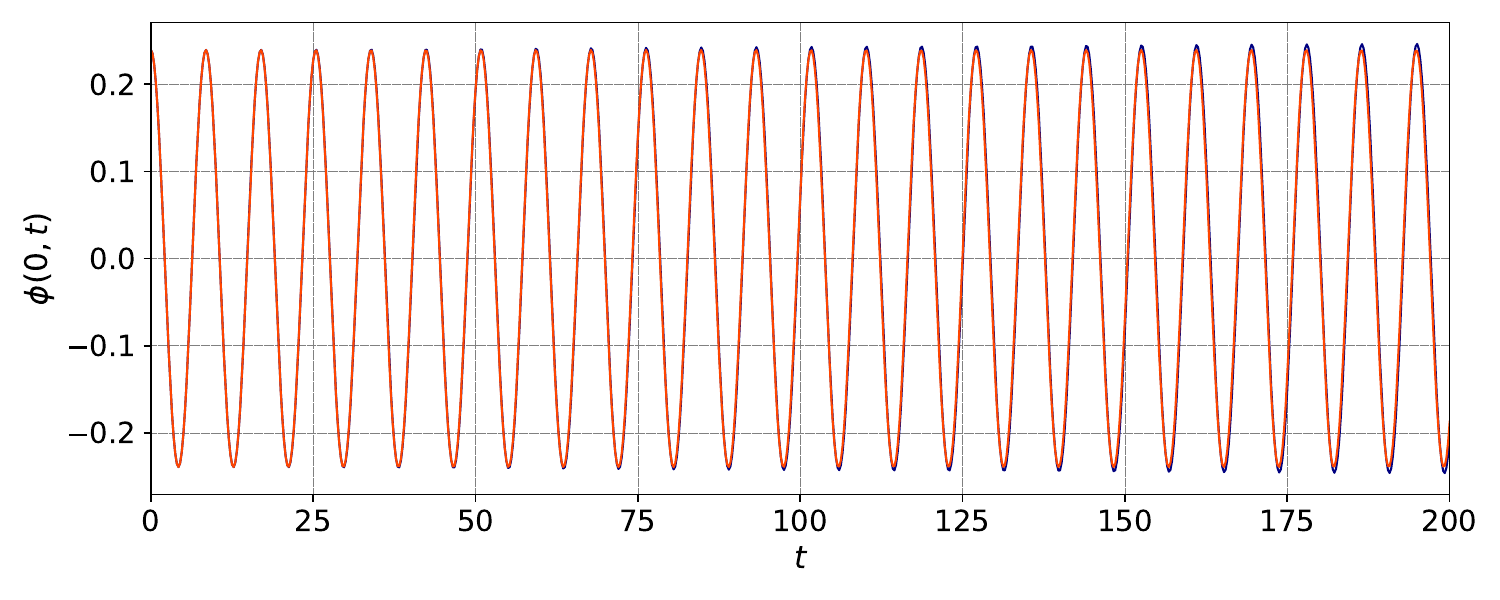}\label{phi4inv_08_01}
    \includegraphics[{angle=0,width=8cm,height=4cm}]{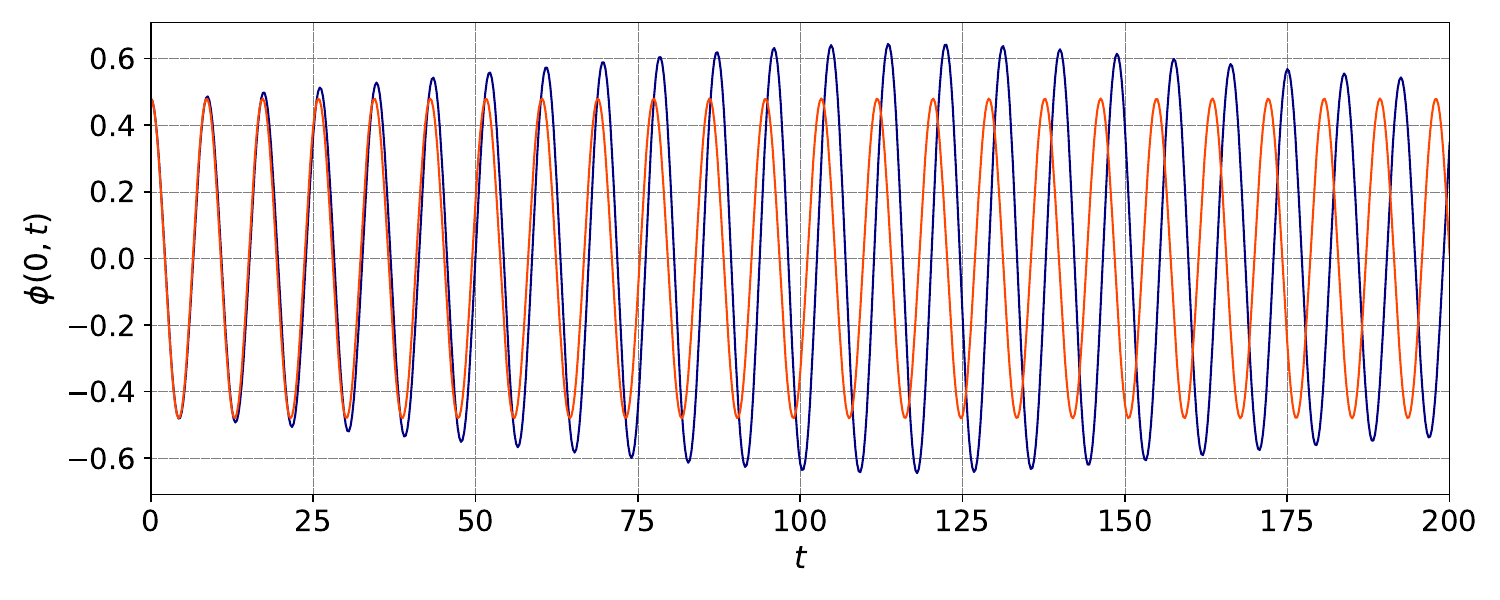}\label{phi4inv_08_02}
    \vspace{-0.5cm}
  \caption{Results for the inverse $\phi^4$-potential Eq. (\ref{pi4}). Conventions as in Fig. \ref{fig_1A}.}
  \label{fig_2A}
\end{center}
\end{figure*}
%%%%%%%%%%%%%%%%%%%%%%%%%%%%%%%%%%%%%%%%%%%%%%%%%%%%%%%%%%%%%%%%%%%%%%%%%%

The numerical results appear in Fig. \ref{fig_1A} as the black line. Here, we depict the field at the center of mass, i.e. $\phi(x=0,t)$. The renormalized oscillon Eq. (\ref{sp3}) is also shown, see the red line. The small-amplitude regime is defined in terms of both $\lambda$ and $\eta$.

As expected, there is a very good agreement between the numerical and analytical results in the small-amplitude limit. In this case, the numerical oscillon presents very small modulations. So, it can be very well approximated by the renormalized solution Eq. (\ref{sp3}) whose pillar is the single $Q$-ball Eq. (\ref{cosh1}).

In contrast, as the small-amplitude regime is annihilated, a large (excited) numerical oscillon emerges. The approximation based on the renormalized solution loses accuracy. The differences become more and more visible as the system evolves. They may appear as phase-gaps, amplitude fluctuations, or both.

The large-amplitude numerical oscillon develops an intricate modulated structure whose description requires two degrees of freedom. So, a renormalized oscillon based on a two $Q$-ball solution may be applied. We return to this point later below.

%%%%%%%%%%%%%%%%%%%%%%%%%%%%%%%%%%%%%%%%%%%%%%%%%%%%%%%%%%%%%%%%%%%%%%%%%%% 
\begin{figure*}[!ht]
\begin{center}
  \centering
    \includegraphics[{angle=0,width=8cm,height=4cm}]{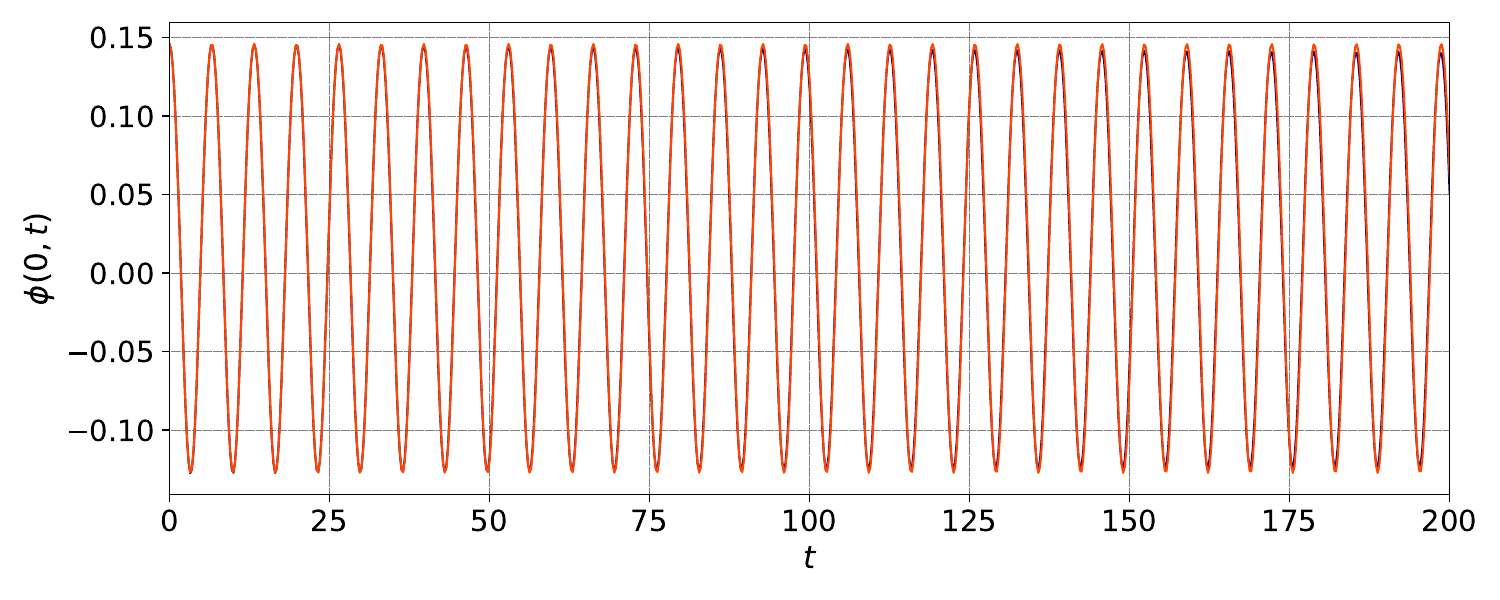}\label{phi4_01_01}
    \includegraphics[{angle=0,width=8cm,height=4cm}]{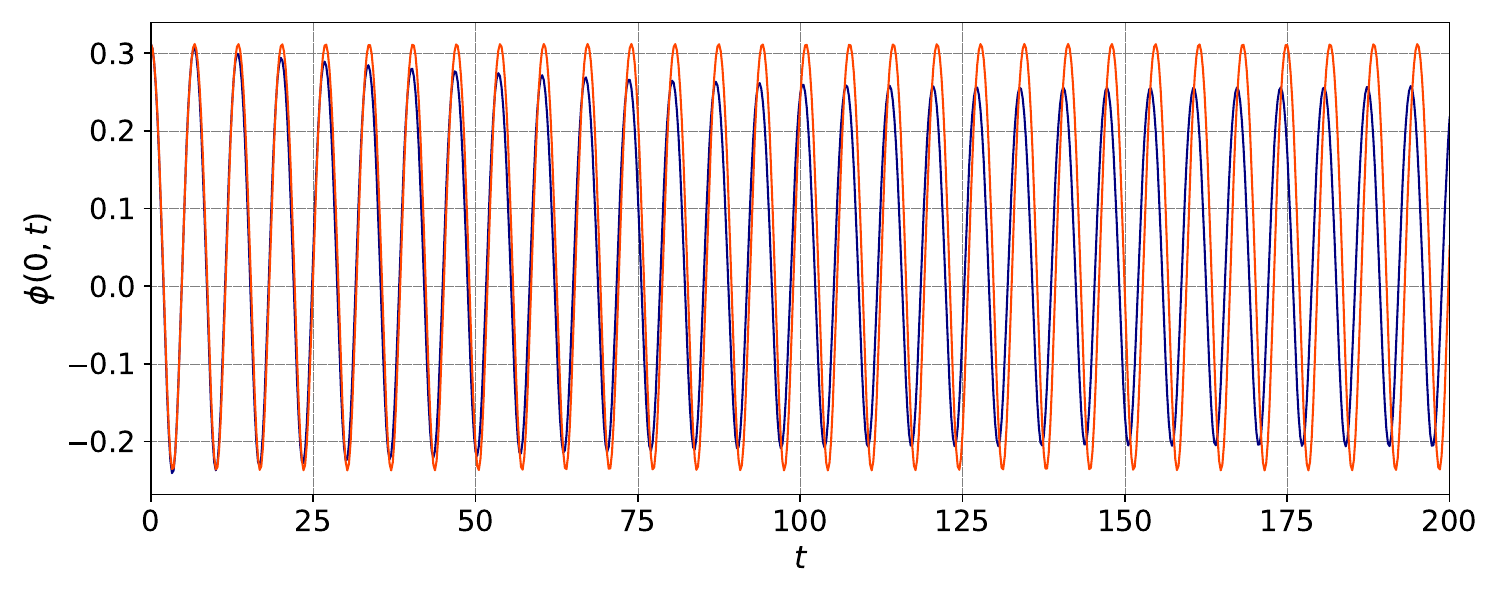}\label{phi4_01_02}
    \includegraphics[{angle=0,width=8cm,height=4cm}]{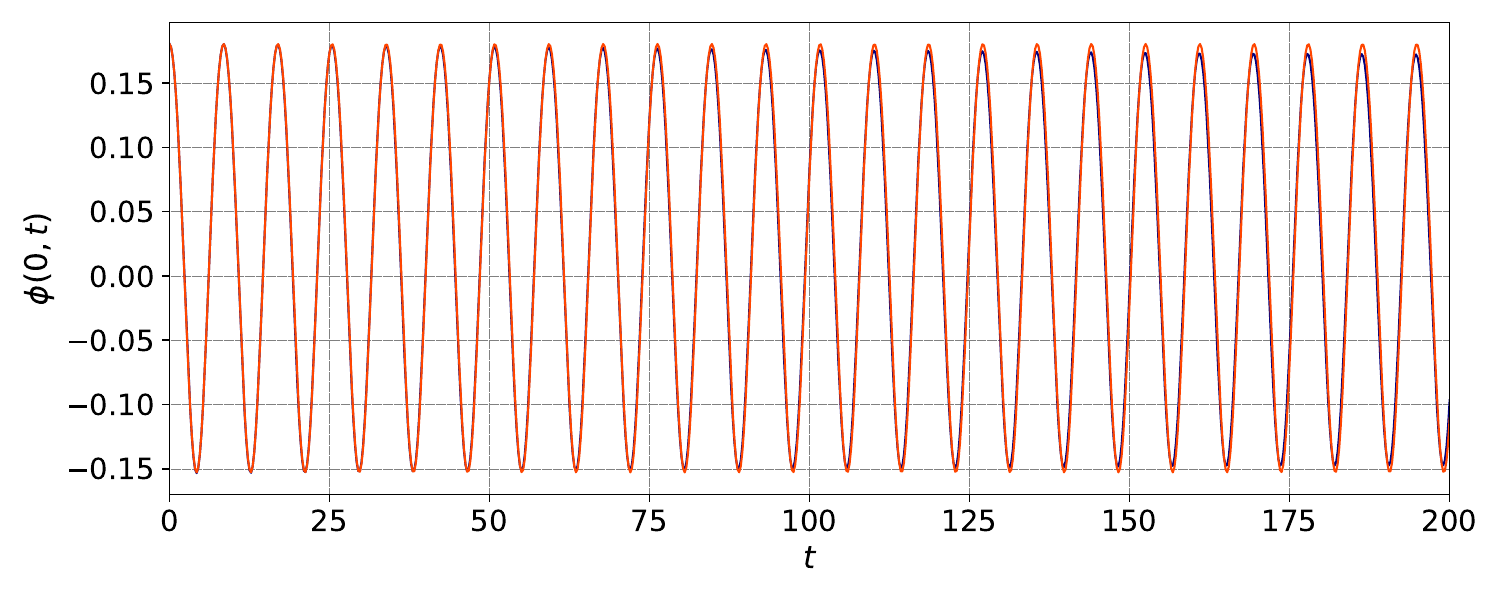}\label{phi4_08_01}
    \includegraphics[{angle=0,width=8cm,height=4cm}]{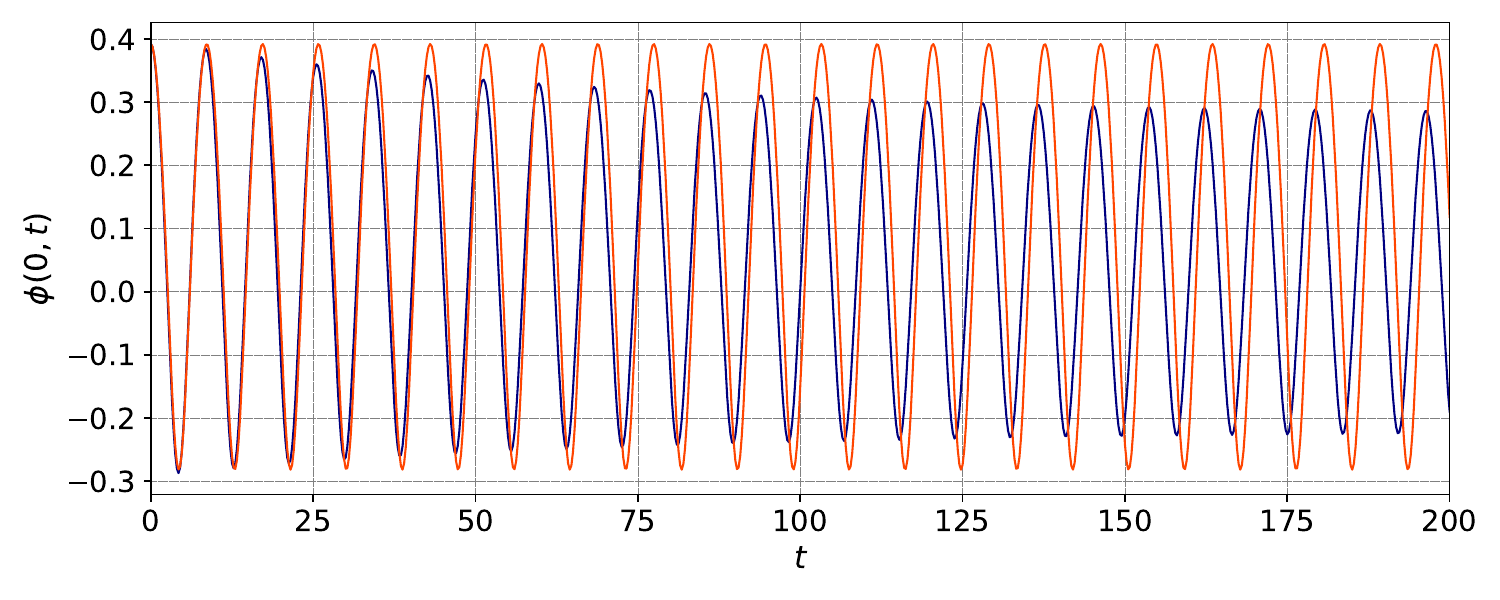}\label{phi4_08_02}
    \vspace{-0.5cm}
  \caption{Results for the prototypical double-well $\phi^4$-potential Eq. (\ref{pdwp}). Conventions as in Fig. \ref{fig_1A}.}
  \label{fig_3A}
\end{center}
\end{figure*}
%%%%%%%%%%%%%%%%%%%%%%%%%%%%%%%%%%%%%%%%%%%%%%%%%%%%%%%%%%%%%%%%%%%%%%%%%%

\subsubsection{The inverse $\phi ^{4}$-potential.} \label{subsecIIa2}

As a second example, we consider%
\begin{equation}
V\left( \phi \right) =\frac{\phi ^{2}}{2}-\frac{\phi ^{4}}{4}\label{pi4}\text{,}
\end{equation}%
i.e. the inverse version of the $\phi^4$-potential, see Ref. \cite{prd37}. It leads to $a_{3}=0$ and $a_{4}=1$. All additional terms in Eq. (\ref{gg4}) vanish.

In this case, %Eqs. (\ref{bna}) lead to%
%\begin{equation}
%\beta =3-2\eta \text{ \ \ and \ \ }\alpha =\frac{2\eta -1}{8}\label{baip4}\text{,}
%\end{equation}%
%via which
we write Eq. (\ref{de1x}) as%
\begin{equation}
\phi _{R}\left( x,t\right) =2\sqrt{\frac{2}{3-2\eta }}\frac{\lambda \cos
\left( \frac{\omega }{\sqrt{1+\eta }}t\right) }{\cosh \left( \frac{\lambda }{%
\sqrt{1+\eta }}x\right) }+\frac{2\eta -1}{2\left( 3-2\eta \right) }\sqrt{%
\frac{2}{3-2\eta }}\frac{\lambda ^{3}\cos \left( \frac{3\omega }{\sqrt{%
1+\eta }}t\right) }{\cosh ^{3}\left( \frac{\lambda }{\sqrt{1+\eta }}x\right) 
}\label{spi4}\text{,}
\end{equation}%
where we have used $\beta =3-2\eta$ and $\alpha =(2\eta -1)/8$, see Eqs. (\ref{bna}). Here, we have again implemented Eq. (\ref{cosh1}).
%lead to%
%\begin{equation}
%\beta =3-2\eta \text{ \ \ and \ \ }\alpha =\frac{2\eta -1}{8}\label{baip4}\text{,}
%\end{equation}%
%As before, we have implemented $T=t/\sqrt{1+\eta}$ and $X=x/\sqrt{1+\eta}$.

In view of Eq. (\ref{pi4}), Eq. (\ref{bnn1}) gives rise to the bound%
\begin{equation}
\frac{3}{2}>\eta \label{bep4i}\text{,}
\end{equation}
which sustains the existence of the renormalized profile Eq. (\ref{spi4}).

We again validate our approach by verifying whether %check the conditions under which
our analytical result approximates the behavior of a generalized numerical oscillon. We consider Eq. (\ref{gceomx1}) in the presence of Eqs. (\ref{fphi}) and (\ref{pi4}), and use Eq. (\ref{spi4}) at $t=0$ as the initial state. Then, we solve the problem numerically for different $\lambda$ and $\eta$.

The results appear in Fig. \ref{fig_2A}. Again, for %$\lambda$ and $\eta$ sufficiently small, the
small initial amplitudes, %limit is attained, from which
we find a very good agreement between the numerical and analytical solutions. On the other hand, when the small-amplitude scenario is extinguished, %when $\lambda$ and/or $\eta$ increase.
the initial state produces a large-amplitude oscillon with modulated structure. As a consequence, the approximation via the renormalized profile fails.

\subsubsection{The double-well $\phi^{4}$-potential.} \label{subsecIIa3}

We %now end the discussion of our first results by presenting
also explore a third example. It is defined by a prototype of the double-well $\phi ^{4}$-model. Here, the potential reads%
\begin{equation}
V\left( \phi \right) =\frac{\phi ^{2}}{2}-\frac{\phi ^{3}}{2}+\frac{\phi ^{4}%
}{4}\label{pdwp}\text{,}
\end{equation}%
from which we get $a_{3}=3/2$ and $a_{4}=-1$. Equation (\ref{pdwp}) can be obtained from the standard form $V ( \tilde{\phi} ) =(1/8)(1-\tilde{\phi}^2)^2$ by expanding $\tilde{\phi}$ around its vacuum value $\tilde{\phi} \equiv1-\phi$.

%Equations %In view of $a_{3}=3/2$ and $a_{4}=-1$,
%(\ref{bna}) provide%
%\begin{equation}
%\beta =\frac{9-4\eta }{2}\text{ \ \ and \ \ }\alpha =\frac{5+4\eta }{16}%
%\label{badwp}\text{,}
%\end{equation}%
%while
The renormalized profile (\ref{de1x}) assumes the form
\begin{equation}
\phi _{R}\left( x,t\right) =4\sqrt{\frac{1}{9-4\eta }}\frac{\lambda \cos
\left( \frac{\omega }{\sqrt{1+\eta }}t\right) }{\cosh \left( \frac{\lambda }{%
\sqrt{1+\eta }}x\right) }+\frac{4\lambda ^{2}}{9-4\eta }\frac{3-\cos \left( 
\frac{2\omega }{\sqrt{1+\eta }}t\right) }{\cosh ^{2}\left( \frac{\lambda }{%
\sqrt{1+\eta }}x\right) }+\frac{5+4\eta }{9-4\eta }\sqrt{%
\frac{1}{9-4\eta }}\frac{\lambda ^{3}\cos \left( \frac{3\omega }{\sqrt{%
1+\eta }}t\right) }{\cosh ^{3}\left( \frac{\lambda }{\sqrt{1+\eta }}x\right) 
}\label{rsdwp}\text{,}
\end{equation}%
where we have implemented $\beta =(9-4\eta)/2$ and $\alpha =(5+4\eta)/16$.
% Eq. (\ref{cosh1}).
It
%The solution above %renormalized oscillon Eq. (\ref{rsdwp})
only exists when $9/4>\eta$, %
%\begin{equation}
%\frac{9}{4}>\eta \label{edwp}\text{,}
%\end{equation}
see Eq. (\ref{bnn1}).

The results are shown in Fig. \ref{fig_3A}. As in the previous examples, numerical oscillons with small amplitudes are very well mapped by the analytical solution based on the single $Q$-ball. As the initial amplitude increases, modulations get more and more pronounced, and the description via the renormalized profile does not apply anymore.

In all three scenarios discussed above, the renormalized oscillons are related to the same RG Equation (\ref{eom1}). In other words, they are based on the same $Q$-ball solution Eq. (\ref{cosh1}). So, we argue that they belong to the same universality class. As in the usual case, a different universality class can also be identified in our generalized model. The corresponding renormalized oscillon will be derived later below. First, we discuss the mapping of the modulated structure by means of a two $Q$-ball profile.

%%%%%%%%%%%%%%%%%%%%%%%%%%%%%%%%%%%%%%%%%%%%%%%%%%%%%%%%%%%%%%%%%%%%%%%%%%% 
\begin{figure*}[!ht]
\begin{center}
  \centering
    \includegraphics[{angle=0,width=8cm,height=4cm}]{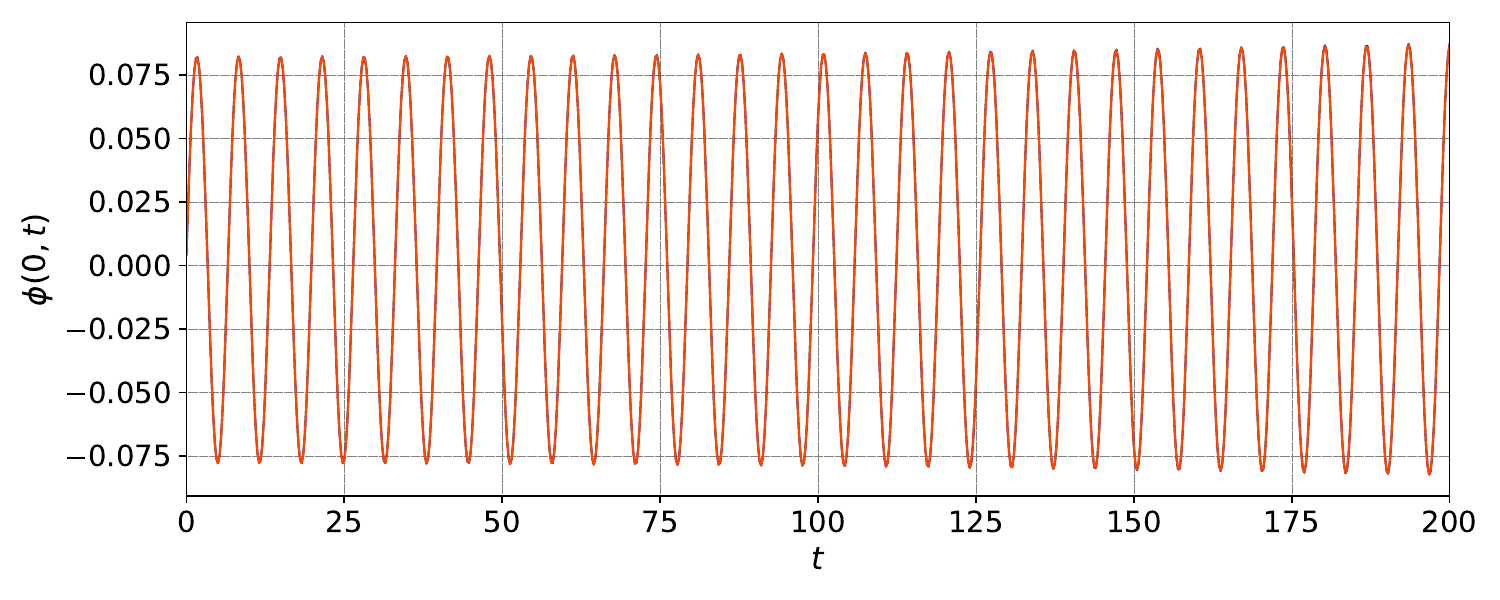}\label{phi3_01_01_005}
    \includegraphics[{angle=0,width=8cm,height=4cm}]{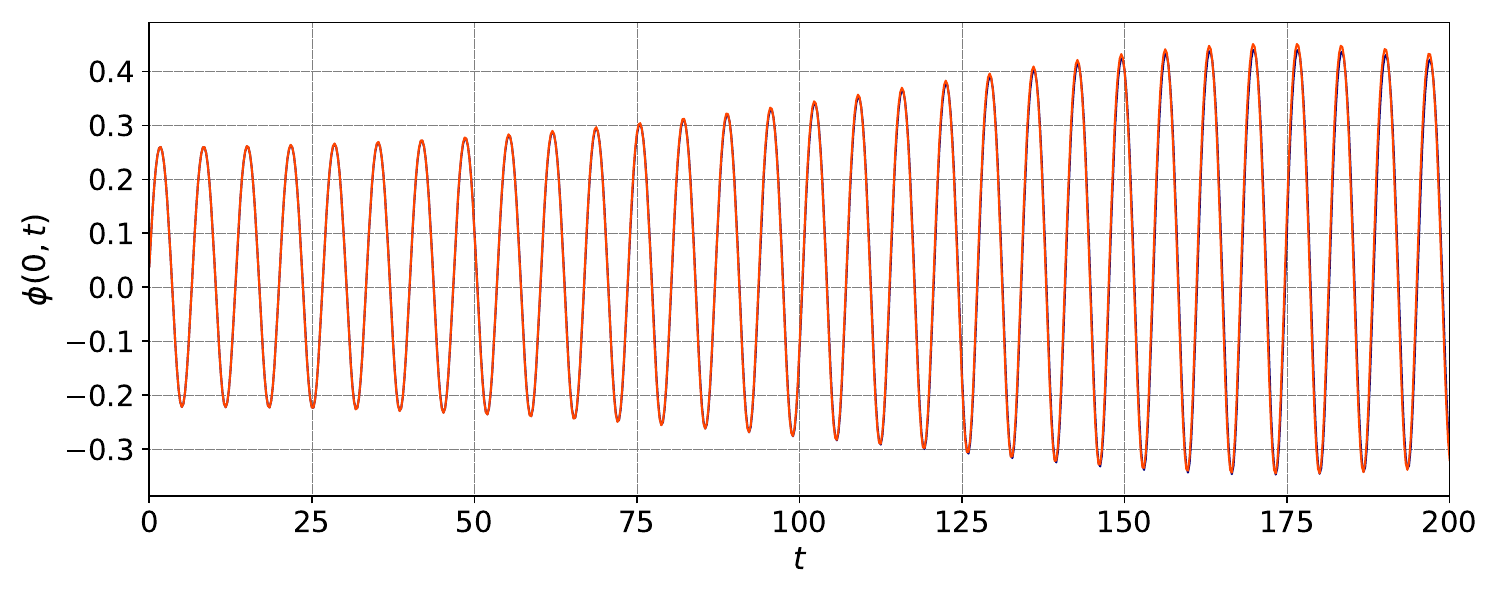}\label{phi3_01_02_005}
    \includegraphics[{angle=0,width=8cm,height=4cm}]{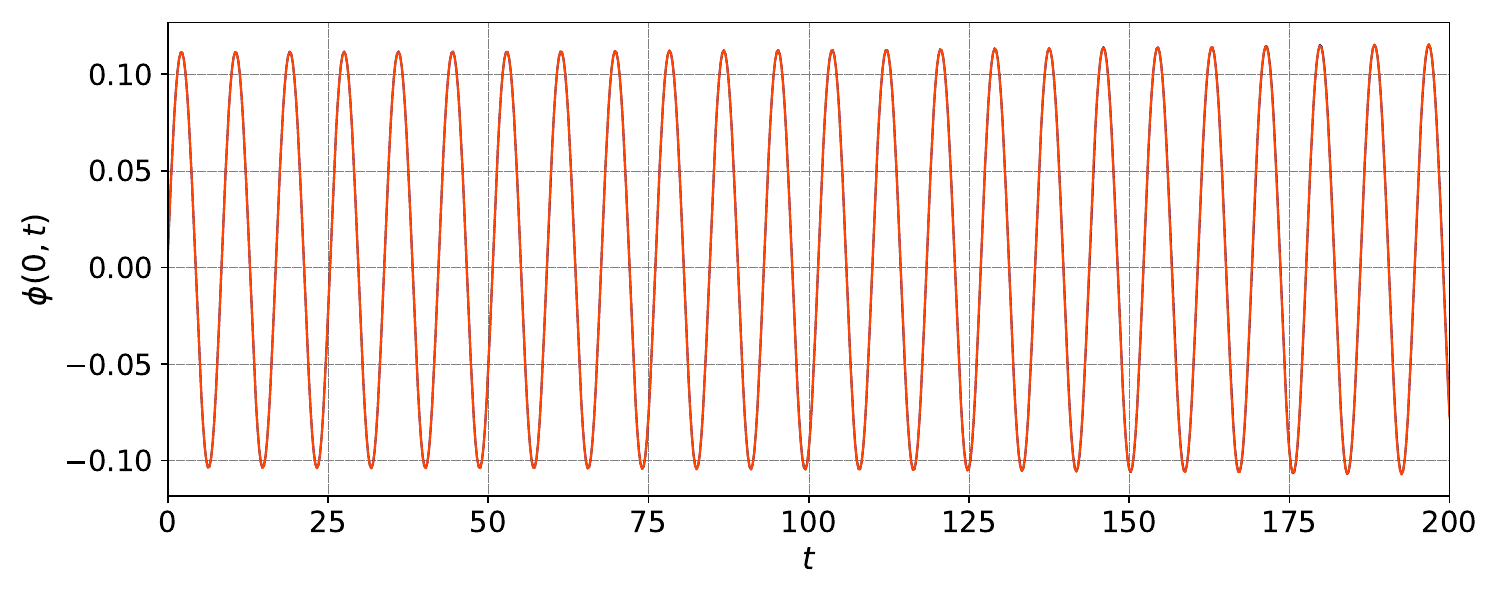}\label{phi3_08_01_005}
    \includegraphics[{angle=0,width=8cm,height=4cm}]{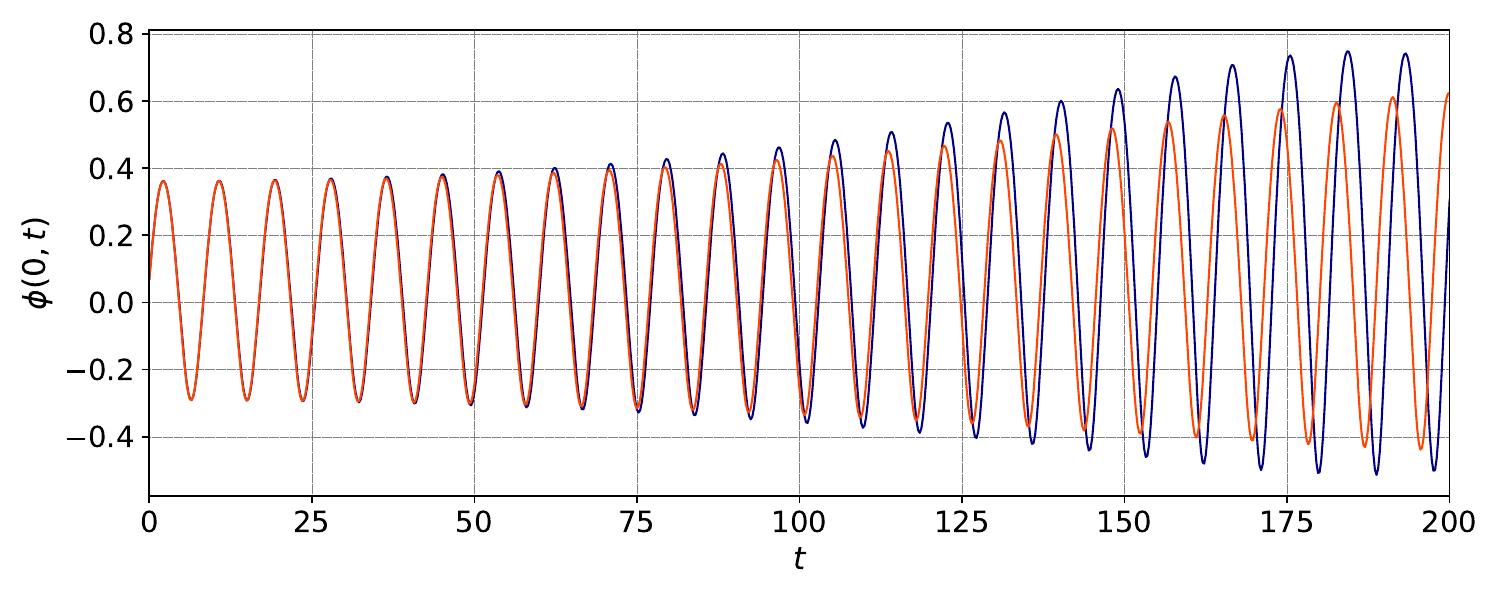}\label{phi3_08_02_005}
    \vspace{-0.5cm}
  \caption{Results for the $\phi^3$-potential Eq. (\ref{p3}), with $\lambda_2=-0.05$ fixed. The numerical oscillon (black line) is compared to the renormalized analytical one (red line). Here, we have chosen $\eta=0.10$ (upper line) and $\eta=0.80$ (lower line), with $\lambda_1=0.10$ (left column) and $\lambda_1=0.20$ (right column).%The numerical oscillon (black line) is compared to the renormalized analytical one (red line). Upper: $\eta=0.10$ with $\lambda_1=0.10$ (left) and $\lambda_1=0.20$ (right). Lower: $\eta=0.80$ with $\lambda_1=0.10$ (left) and $\lambda_1=0.20$ (right). For all figures, we fixed $\lambda_2=-0.05$.}
  }
  \label{fig_1B}
\end{center}
\end{figure*}
%%%%%%%%%%%%%%%%%%%%%%%%%%%%%%%%%%%%%%%%%%%%%%%%%%%%%%%%%%%%%%%%%%%%%%%%%%

\subsection{Modulated oscillons and the two $Q$-balls solution} \label{subsecIII}

Our simulations have revealed that, as the amplitude of the initial state increases, the generalized system evolves to a large-amplitude oscillon whose modulated structure can not be described accurately by the renormalized solution based on a single $Q$-ball.

We circumvent this issue by working with a two $Q$-balls solution. To introduce it, we observe that the original RG Equation (\ref{eom1}) can be approximated by%
\begin{equation}
\partial _{\mu }\partial ^{\mu }\Psi+\Psi =\Psi \left\vert
\Psi \right\vert ^{2}-\overline{\Psi}\frac{\partial_{\mu}\Psi\partial^{\mu}\Psi}{1-\left\vert
\Psi \right\vert ^{2}}\text{,}  \label{csg}
\end{equation}
i.e. the complex sine-Gordon equation.

%The point is that Eq. (\ref{csg}) admits exactly the same single $Q$-ball solution Eq. (\ref{cosh1}).
Equation (\ref{csg}) can be seen as the
%In particular, it stands for the
EoM that comes from the Lagrange density
\begin{equation}
\mathcal{L}_{\mathbb{C}\text{sG}}=\frac{1}{1-\left\vert
\Psi \right\vert ^{2}}\left( \left\vert \partial _{\mu }\Psi \right\vert ^{2}-\left\vert
\Psi \right\vert ^{2}+\left\vert
\Psi \right\vert ^{4}\right) \text{,}\label{lcsg}
\end{equation}
which differs from Eq. (\ref{lqb}) by a factor of the form $1+\mathcal{O}\left( \left\vert
\Psi \right\vert ^{2} \right)$. Here, the coordinates of the $(1+1)$-dimensional spacetime are $T$ and $X$.

%%%%%%%%%%%%%%%%%%%%%%%%%%%%%%%%%%%%%%%%%%%%%%%%%%%%%%%%%%%%%%%%%%%%%%%%%%% 
\begin{figure*}[!ht]
\begin{center}
 \centering
   \includegraphics[{angle=0,width=8cm,height=4cm}]{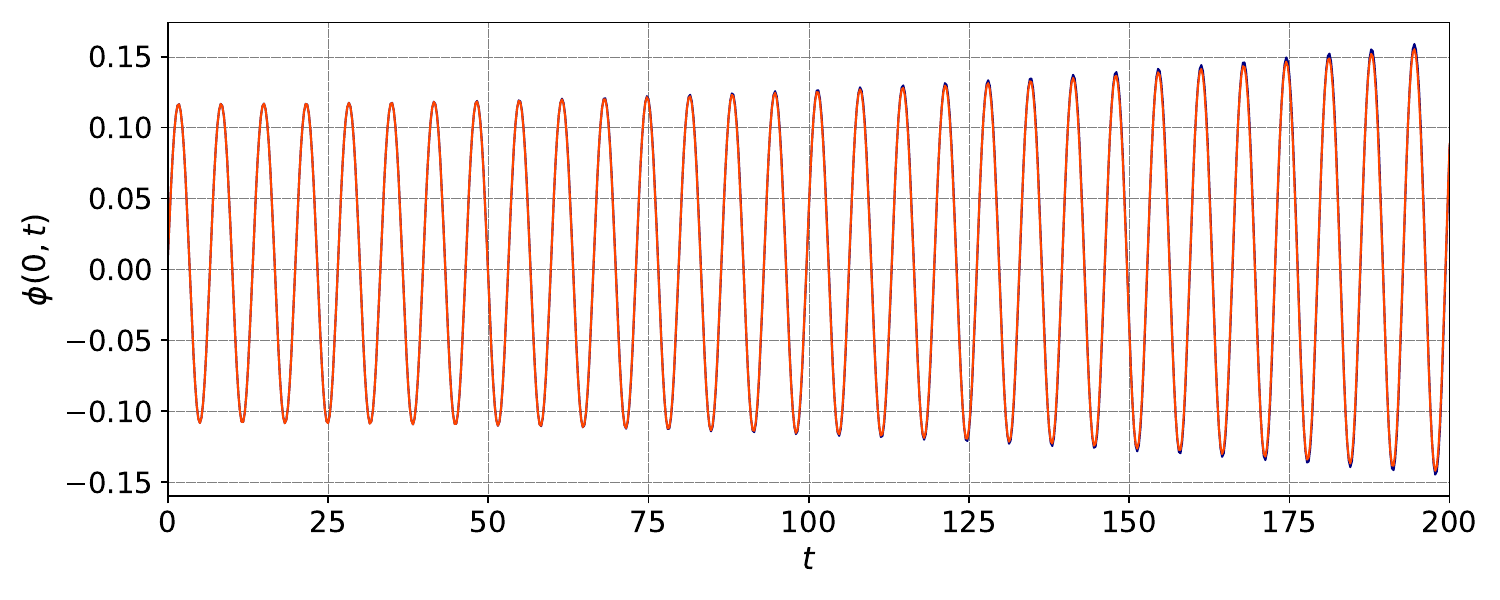}\label{phi3_01_015_008}
    \includegraphics[{angle=0,width=8cm,height=4cm}]{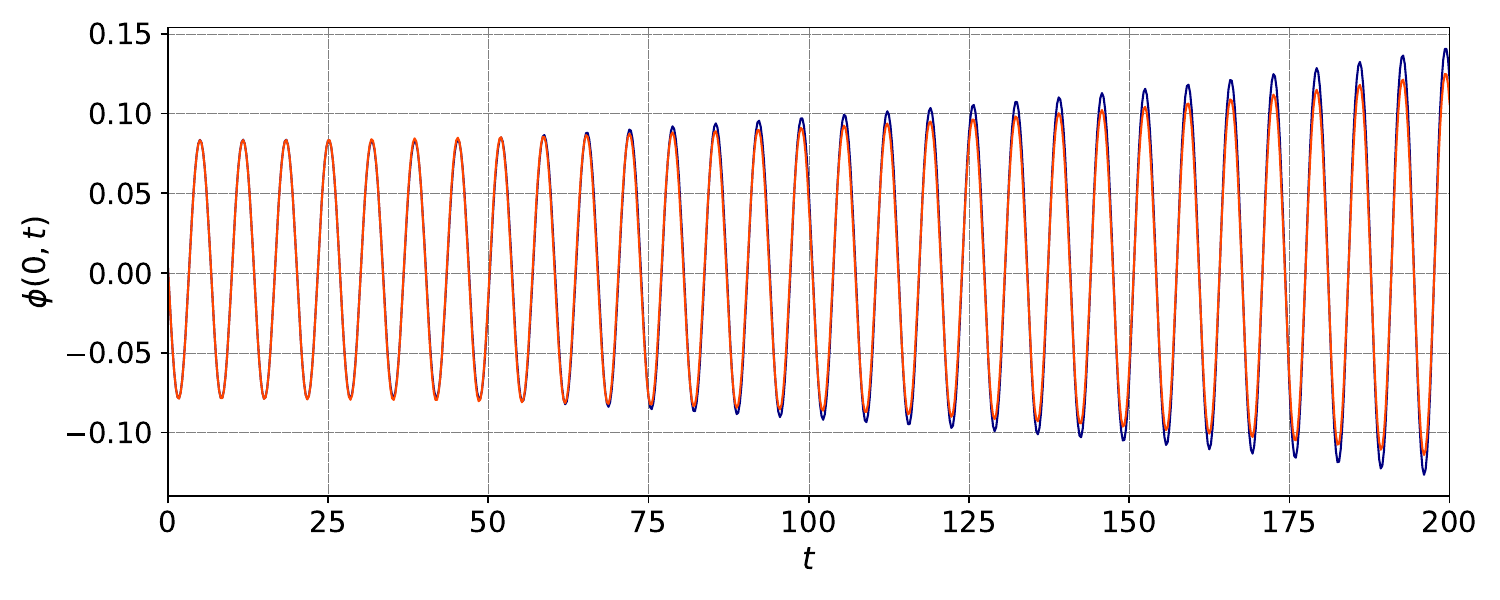}\label{phi3_01_015_02}
    \includegraphics[{angle=0,width=8cm,height=4cm}]{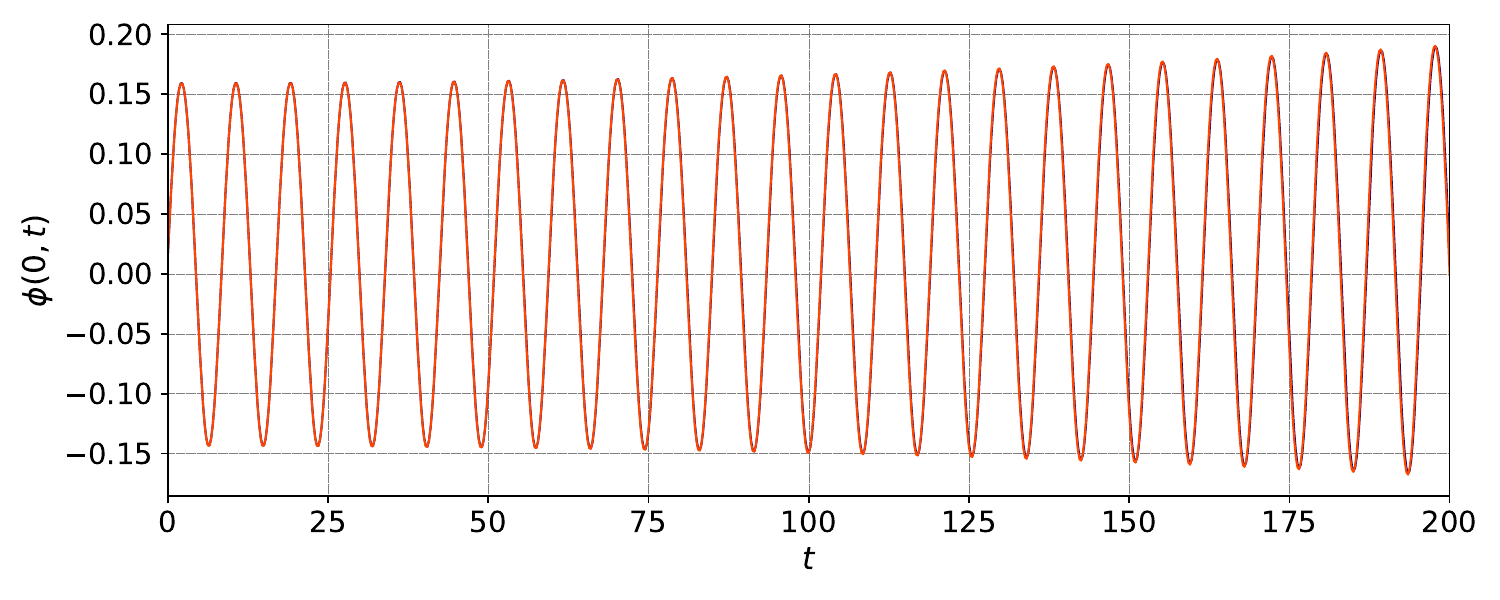}\label{phi3_08_015_008}
    \includegraphics[{angle=0,width=8cm,height=4cm}]{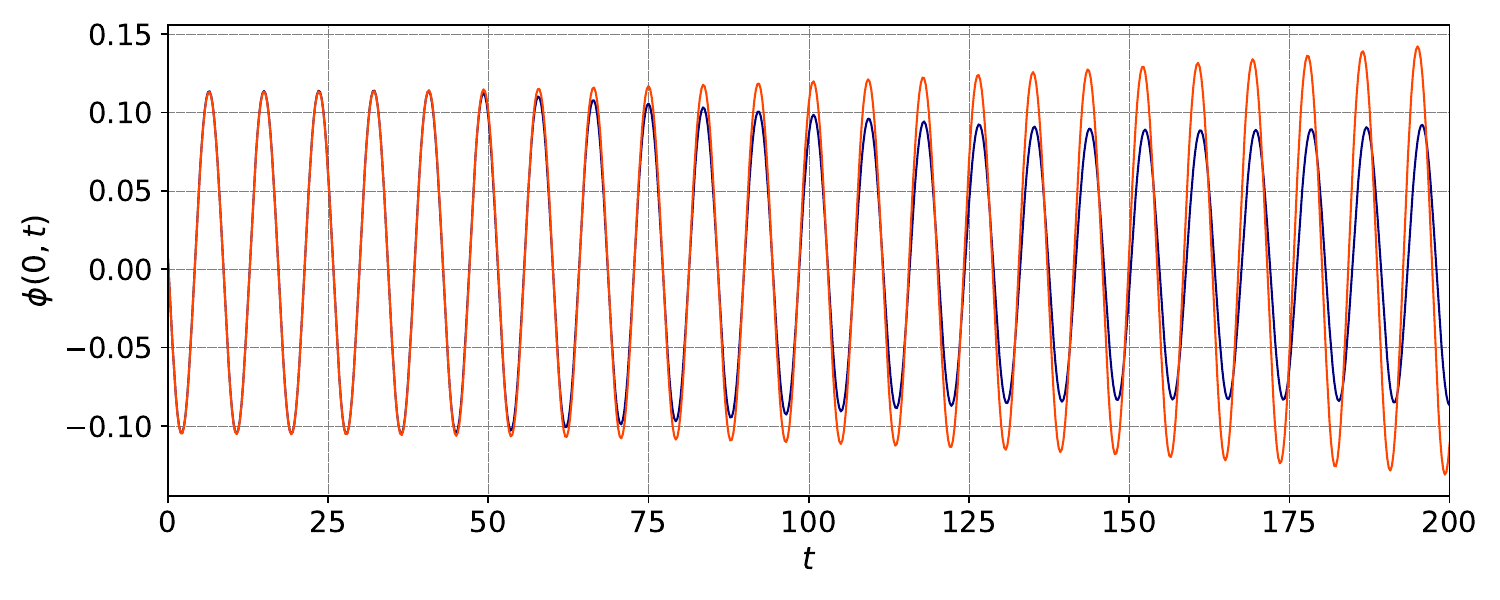}\label{phi3_08_015_02}
    \vspace{-0.5cm}
  \caption{Results for the $\phi^3$-potential Eq. (\ref{p3}), with $\lambda_1=0.15$ fixed. The numerical oscillon (black line) is compared to the renormalized analytical one (red line). Here, we have chosen $\eta=0.10$ (upper line) and $\eta=0.80$ (lower line), with $\lambda_2=-0.08$ (left column) and $\lambda_2=-0.20$ (right column). %Results obtained for the $\phi^3$-potential. The numerical oscillon (black line) is compared to the renormalized analytical one (red line). Upper: $\eta=0.10$ with $\lambda_2=-0.08$ (left) and $\lambda_2=-0.20$ (right). Lower: $\eta=0.80$ with $\lambda_2=-0.08$ (left) and $\lambda_2=-0.20$ (right). For all figures, we fixed $\lambda_1=0.15$.}
  }
  \label{fig_2Ba}
\end{center}
\end{figure*}
%%%%%%%%%%%%%%%%%%%%%%%%%%%%%%%%%%%%%%%%%%%%%%%%%%%%%%%%%%%%%%%%%%%%%%%%%%

The complex sine-Gordon model (\ref{lcsg}) supports exact multisoliton solutions. The simplest one is the two $Q$-ball profile \cite{prd38}. In this context, we focus on a bound state of two coincident $Q$-balls centered at the origin. The corresponding solution can be written explicitly as
\begin{equation}
\Psi \left( x,t\right) =\frac{i\left( \omega _{1}-\omega _{2}\right)
\left( \frac{\lambda _{1}e^{i\frac{\omega _{1}}{\sqrt{1+\eta }}t}}{\cosh
\left( \frac{\lambda _{1}}{\sqrt{1+\eta }}x\right) }-\frac{\lambda _{2}e^{i%
\frac{\omega _{2}}{\sqrt{1+\eta }}t}}{\cosh \left( \frac{\lambda _{2}}{\sqrt{%
1+\eta }}x\right) }\right) }{1-\omega _{1}\omega _{2}-\lambda _{1}\lambda
_{2}\left( \tanh \left( \frac{\lambda _{1}}{\sqrt{1+\eta }}x\right) \tanh
\left( \frac{\lambda _{2}}{\sqrt{1+\eta }}x\right) +\frac{\cos \left( \frac{%
\omega _{1}-\omega _{2}}{\sqrt{1+\eta }}t\right) }{\cosh \left( \frac{%
\lambda _{1}}{\sqrt{1+\eta }}x\right) \cosh \left( \frac{\lambda _{2}}{\sqrt{%
1+\eta }}x\right) }\right) }\label{tqbs}\text{,}
\end{equation}
where we have assumed that the total momentum of the composite configuration vanishes, for the sake of simplicity. As before, we have implemented $T=b_{0}^{-1/2}t$ and $X=b_{0}^{-1/2}x$, with $b_{0}=1+\eta$ (i.e. we are still working with the generalizing function (\ref{fphi})). In Eq. (\ref{tqbs}), $\omega _{1}=\sqrt{1-\lambda _{1}^{2}}$ and $\omega_{2}=\sqrt{1-\lambda _{2}^{2}}$ represent the frequencies of the two $Q$-balls, while $\lambda _{1}$ and $\lambda _{2}$ are the respective scalar parameters.

In the sequence, we return to the cases discussed previously. However, instead of the single $Q$-ball solution Eq. (\ref{cosh1}), we now use the two $Q$-ball profile Eq. (\ref{tqbs}) to generate the renormalized oscillon. Our aim is to demonstrate that the analytical result based on a two $Q$-ball solution may be applied to describe numerical oscillons with modulated structures.

%%%%%%%%%%%%%%%%%%%%%%%%%%%%%%%%%%%%%%%%%%%%%%%%%%%%%%%%%%%%%%%%%%%%%%%%%%% 
\begin{figure*}[!ht]
\begin{center}
 \centering
   \includegraphics[{angle=0,width=8cm,height=4.5cm}]{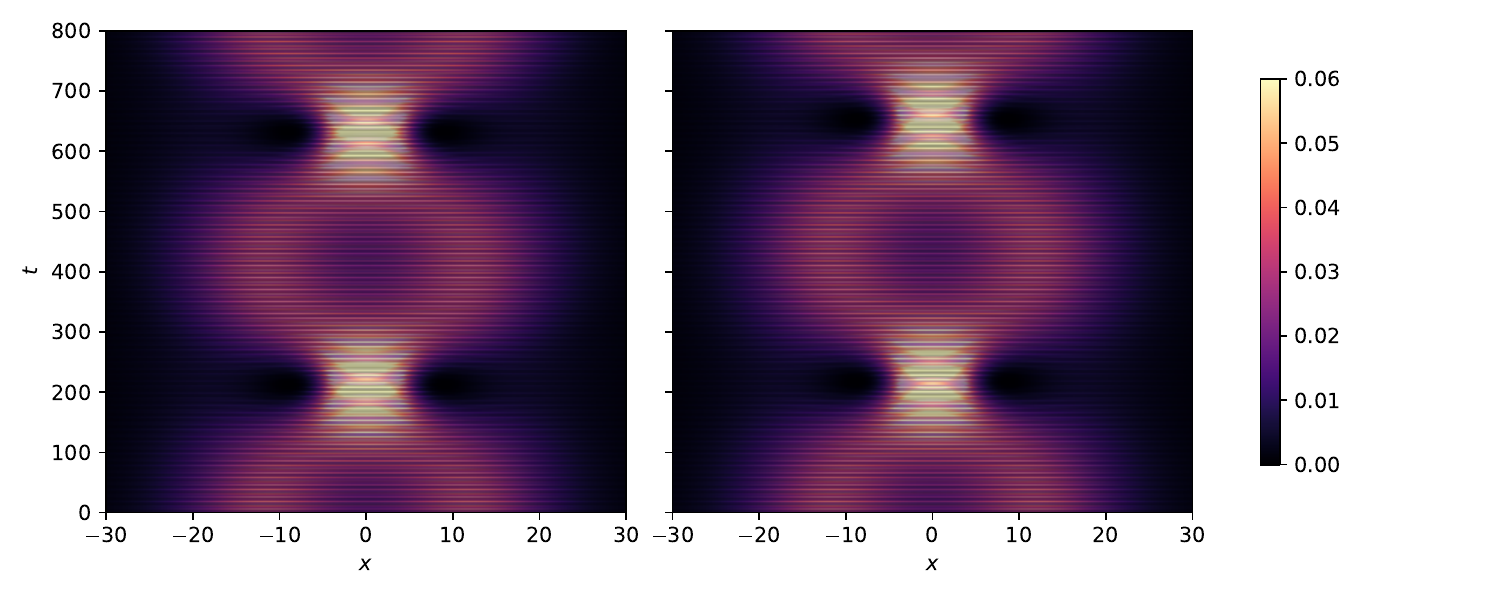}\label{3d_phi3_01_01_02}
   \includegraphics[{angle=0,width=8cm,height=4.5cm}]{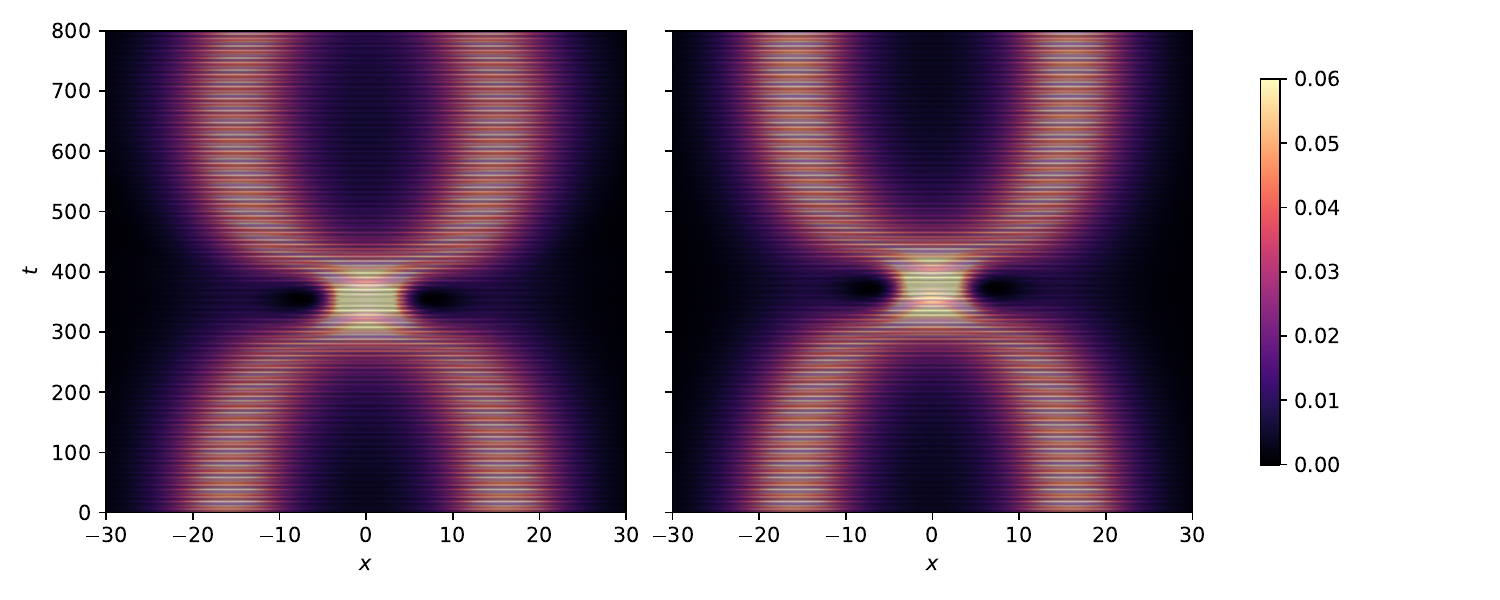}\label{3d_phi3_01_015_02}
   \includegraphics[{angle=0,width=8cm,height=4.5cm}]{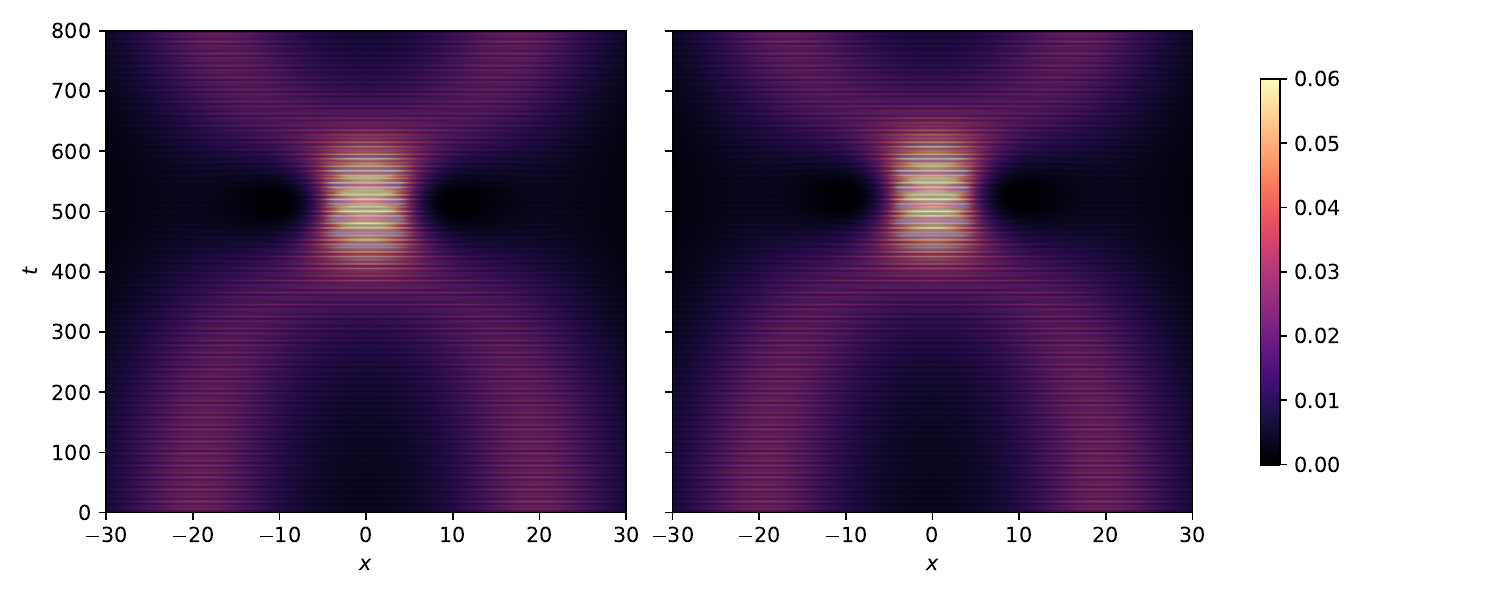}\label{3d_phi3_01_01_015}
   \includegraphics[{angle=0,width=8cm,height=4.5cm}]{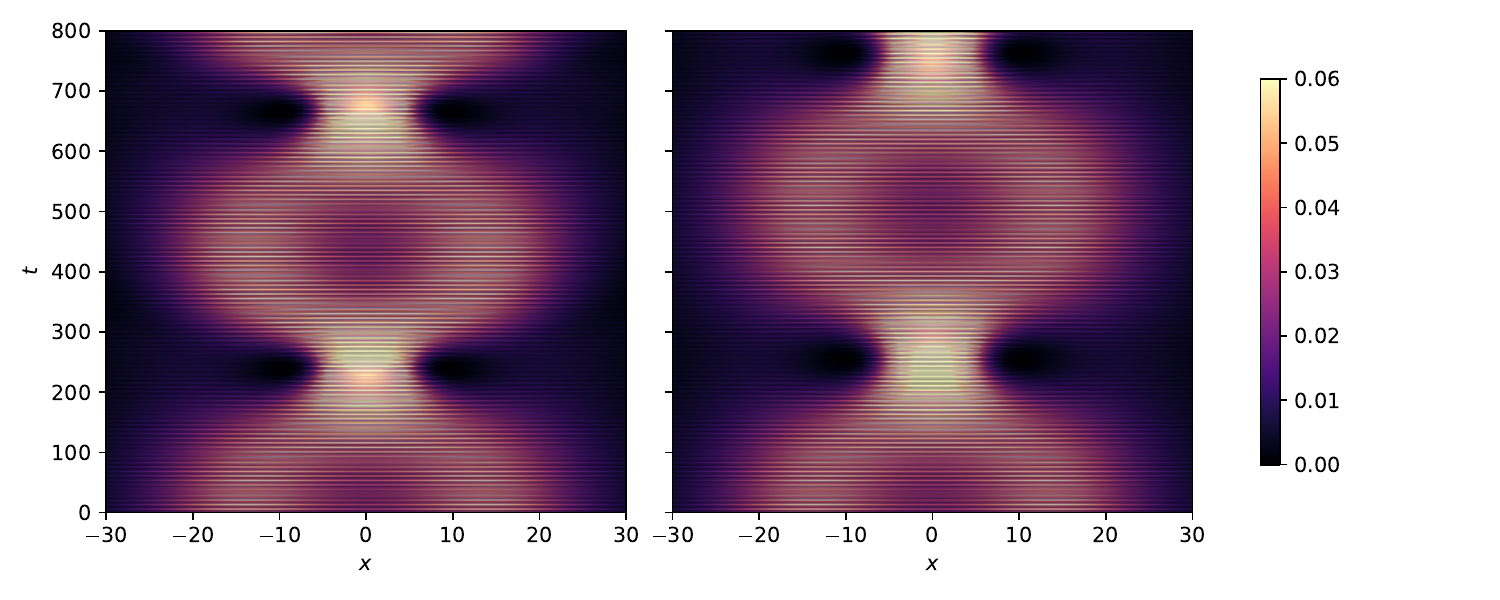}\label{3d_phi3_05_01_02}
    \vspace{-0.5cm}
  \caption{Results for the $\phi^3$-potential Eq. (\ref{p3}). Comparison between the numerical modulated oscillon (left) and the renormalized analytical profile based on the two $Q$-ball solution (right). Upper left: $\eta=0.10$, $\lambda_1=0.10$, and $\lambda_2=-0.20$. Upper right: $\eta=0.10$, $\lambda_1=0.15$, and $\lambda_2=-0.20$. Lower left: $\eta=0.10$, $\lambda_1=0.10$, and $\lambda_2=-0.15$. Lower right: $\eta=0.50$, $\lambda_1=0.10$, and $\lambda_2=-0.20$. We depict $|\partial^2 \phi + \phi|$, for the sake of visualization.} %versus $x$ and $t$ for the $\phi^3$ model.}
  \label{fig_1Ca}
\end{center}
\end{figure*}
%%%%%%%%%%%%%%%%%%%%%%%%%%%%%%%%%%%%%%%%%%%%%%%%%%%%%%%%%%%%%%%%%%%%%%%%%%

\subsubsection{The $\phi ^{3}$-potential.}

This case is defined by Eq. (\ref{p3}), which leads to $a_{3}=1$ and $a_{4}=0$. As a consequence, $\beta$ and $\alpha$ are given by Eqs. (\ref{bap3}). The renormalized solution (\ref{de1x}) can be written as
\begin{equation}
\phi _{R}\left( x,t\right) =\sqrt{\frac{3}{5-3\eta}}\Psi-\frac{1}{5-3\eta}\Psi^{2}+\frac{3}{5-3\eta}\left\vert \Psi \right\vert
^{2}+\frac{1+3\eta}{4(5-3\eta)} \sqrt{\frac{3}{5-3\eta} }\Psi ^{3}+\text{c.c.,}\label{mop3} 
\end{equation}
where $\Psi$ now represents the two $Q$-ball solution (\ref{tqbs}). As before, $\eta$ is constrained to satisfy Eq. (\ref{up001}), i.e. $ \eta<5/3 $.

We now verify whether the renormalized profile based on the two $Q$-ball solution maps a modulated oscillon accurately. To do so, we assume Eq. (\ref{mop3}) itself at $t=0$ as the initial state. We use it to solve again Eq. (\ref{gceomx1}) in the presence of Eqs. (\ref{fphi}) and (\ref{p3}). We depict the numerical results for different $\lambda_1$, $\lambda_2$ and $\eta$.

%%%%%%%%%%%%%%%%%%%%%%%%%%%%%%%%%%%%%%%%%%%%%%%%%%%%%%%%%%%%%%%%%%%%%%%%%%% 
\begin{figure*}[!ht]
\begin{center}
  \centering
    \includegraphics[{angle=0,width=8cm,height=4cm}]{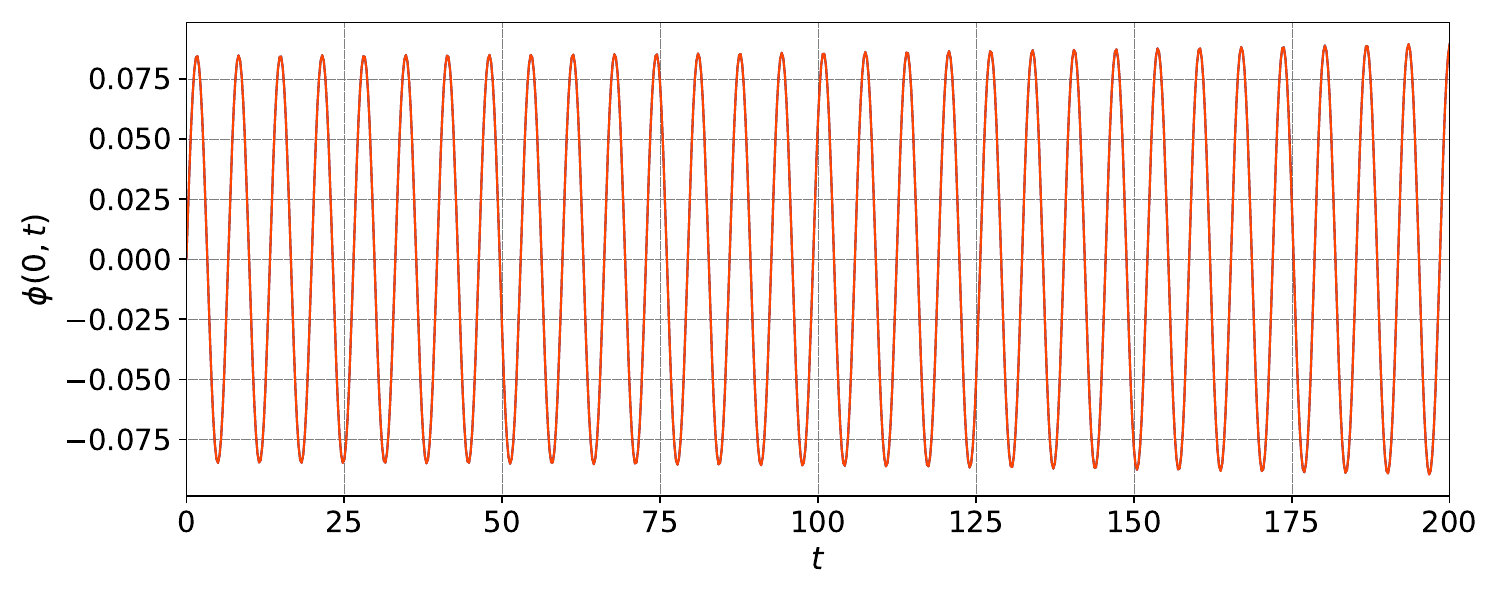}\label{phi4inv_01_01_005}
    \includegraphics[{angle=0,width=8cm,height=4cm}]{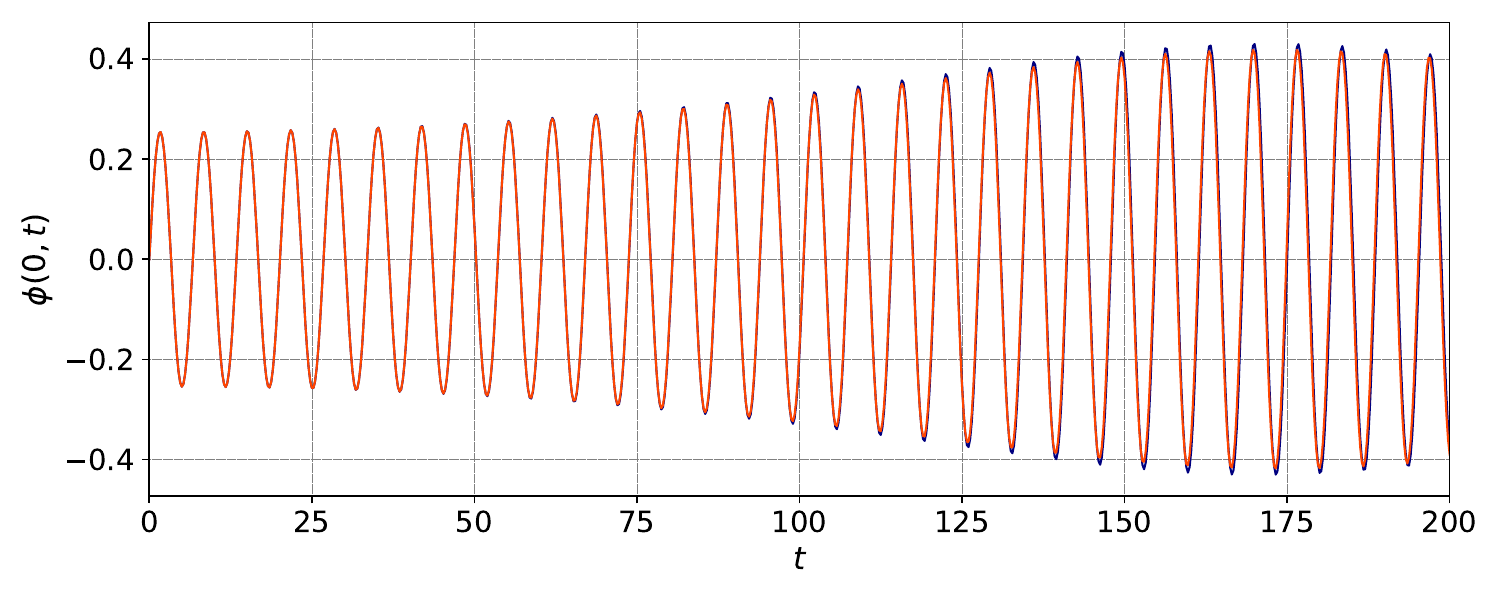}\label{phi4inv_01_02_005}
    \includegraphics[{angle=0,width=8cm,height=4cm}]{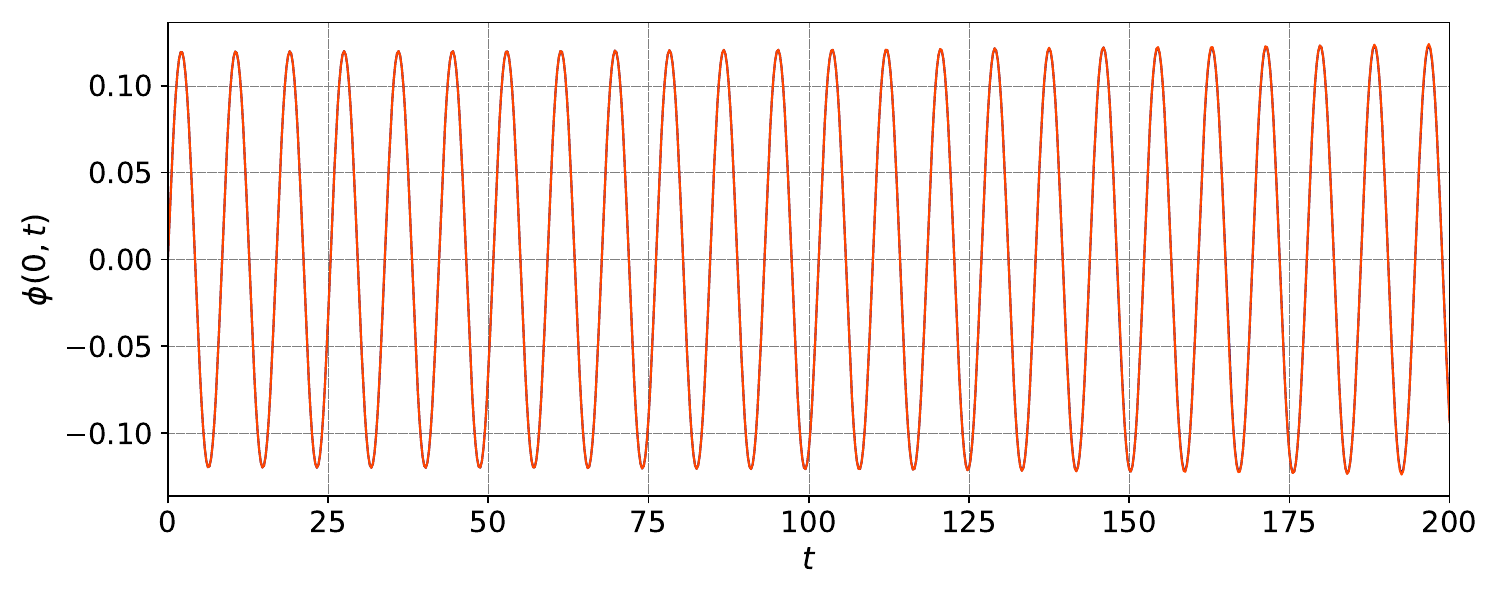}\label{phi4inv_08_01_005}
    \includegraphics[{angle=0,width=8cm,height=4cm}]{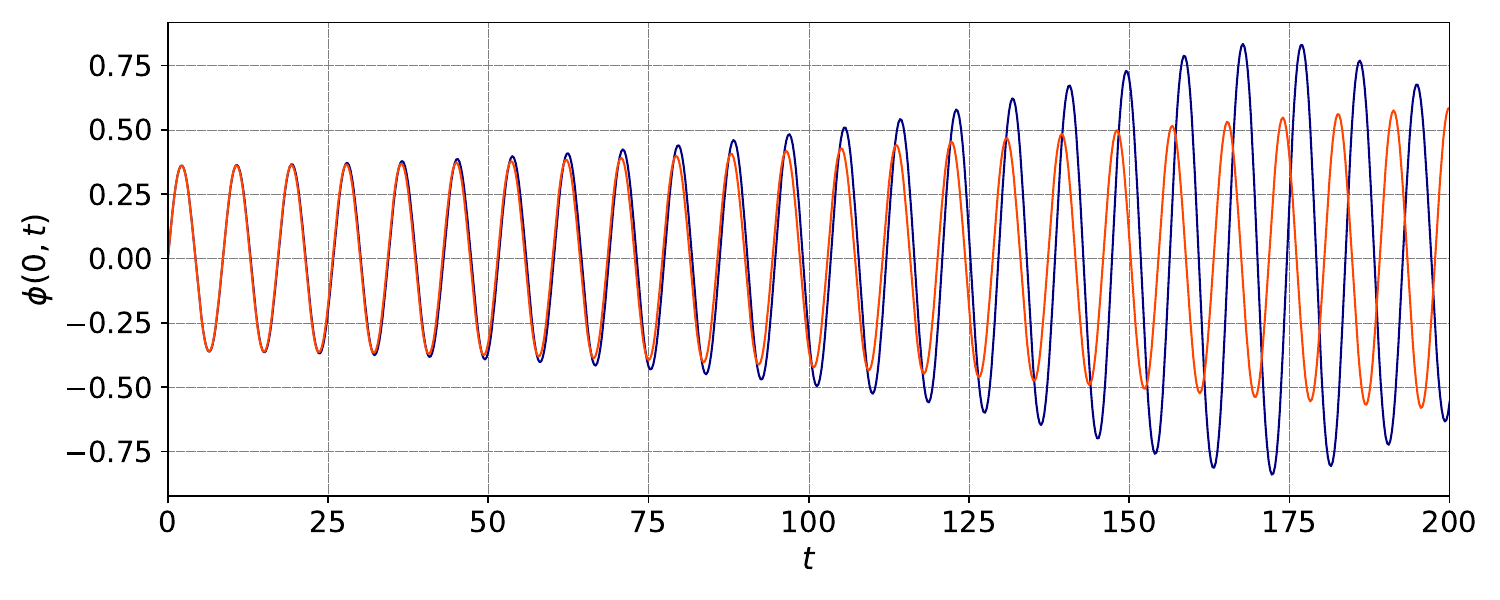}\label{phi4inv_08_02_005}
    \vspace{-0.5cm}
  \caption{Results for the inverse $\phi^4$-potential Eq. (\ref{pi4}). Conventions as in Fig. \ref{fig_1B}. %Upper: $\eta=0.10$ with $\lambda_1=0.10$ (left) and $\lambda_1=0.2$ (right). Lower: $\eta=0.80$ with $\lambda_1=0.10$ (left) and $\lambda_1=0.20$ (right). For all figures, we fixed $\lambda_2=-0.05$.
  }
  \label{fig_3B}
\end{center}
\end{figure*}
%%%%%%%%%%%%%%%%%%%%%%%%%%%%%%%%%%%%%%%%%%%%%%%%%%%%%%%%%%%%%%%%%%%%%%%%%%

Figure \ref{fig_1B} shows %the field at the center of mass
$\phi(x=0,t)$ for different $\lambda_1$ and $\eta$, with $\lambda_2$ fixed. The analytical profile (\ref{mop3}) is also depicted. In general, the renormalized oscillon mimics the numerical one with great accuracy. The coincidence gets better in the limit of %sufficiently small $\lambda_1$ and $\eta$ (i.e.
small and moderate modulations, but it also applies relatively well for large ones. The conclusion is that amplitude modulations are an effect related to the interaction between two generalized unmodulated oscillons.

The same conclusion emerges for $\lambda_1$ fixed, see Fig. \ref{fig_2Ba}. Now, we depict $\phi(x=0,t)$ for different $\lambda_2$ and $\eta$. Again, the analytical result maps the modulated numerical oscillon very precisely. This precision increases as the small modulation limit is attained. However, some accuracy still remains in the large modulation regime.

Naturally, the correspondence applies not only to the field at the center of mass, but also to the entire solution. To clarify it, Fig. \ref{fig_1Ca} presents the field evolution for various $\lambda_1$, $\lambda_2$ and $\eta$. %The reader can visualize how well the modulated behavior of a generalized numerical oscillon is mimicked by the renormalized analytical profile based on the two $Q$-ball solution.
Our analysis reveals that amplitude modulations are intrinsically captured by the interaction %and motion
of two $Q$-balls. This confirms that the intricate modulated profiles observed in generalized models can be accurately modeled as a nonlinear superposition of two %fundamental
unmodulated states.
%the It is possible to visualize that the modulated behavior of a generalized numerical oscillon is very well approximated by the renormalized analytical profile based on the two $Q$-ball solution.

%%%%%%%%%%%%%%%%%%%%%%%%%%%%%%%%%%%%%%%%%%%%%%%%%%%%%%%%%%%%%%%%%%%%%%%%%%% 
\begin{figure*}[!ht]
\begin{center}
 \centering
   \includegraphics[{angle=0,width=8cm,height=4cm}]{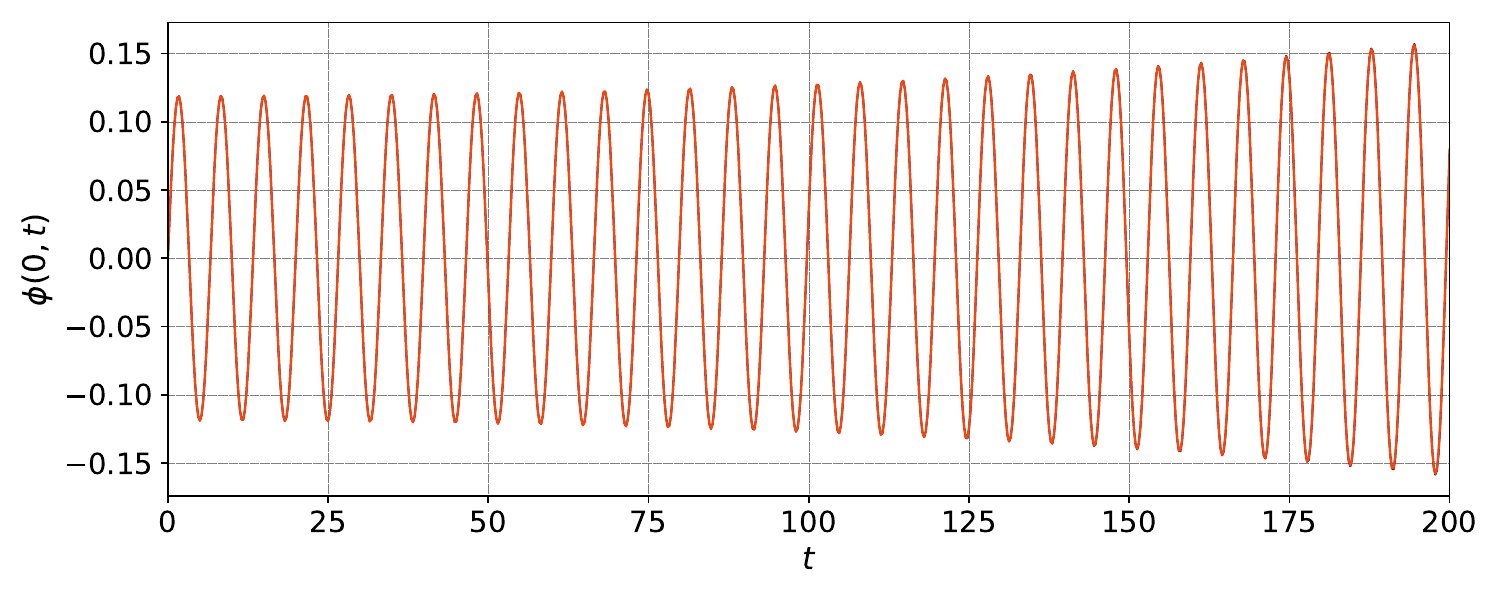}\label{phi4inv_01_015_008}
    \includegraphics[{angle=0,width=8cm,height=4cm}]{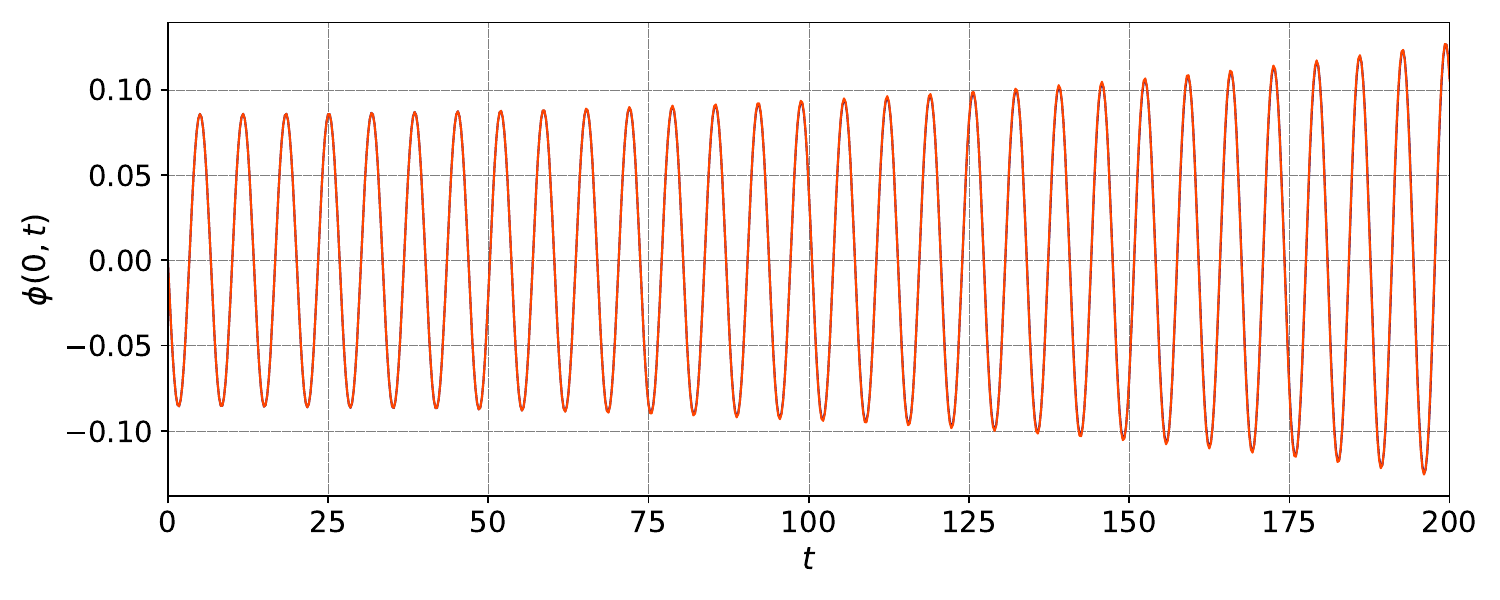}\label{phi4inv_01_015_02}
    \includegraphics[{angle=0,width=8cm,height=4cm}]{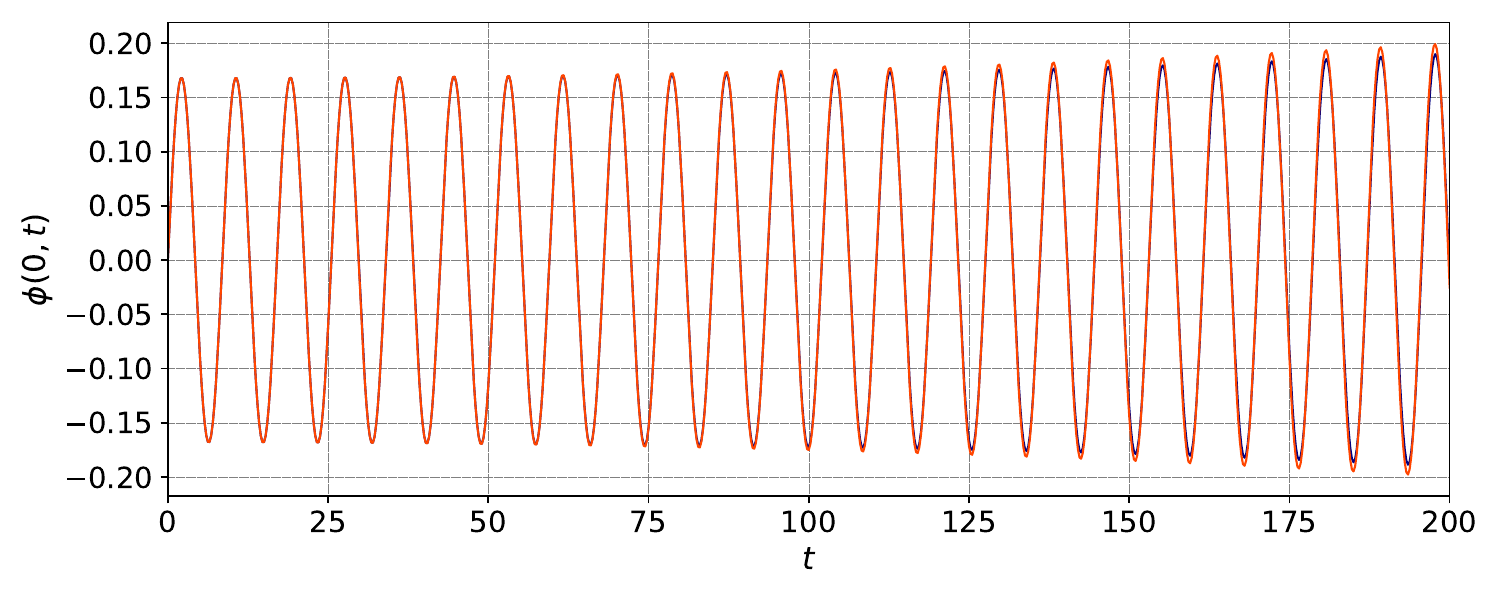}\label{phi4inv_08_015_008}
    \includegraphics[{angle=0,width=8cm,height=4cm}]{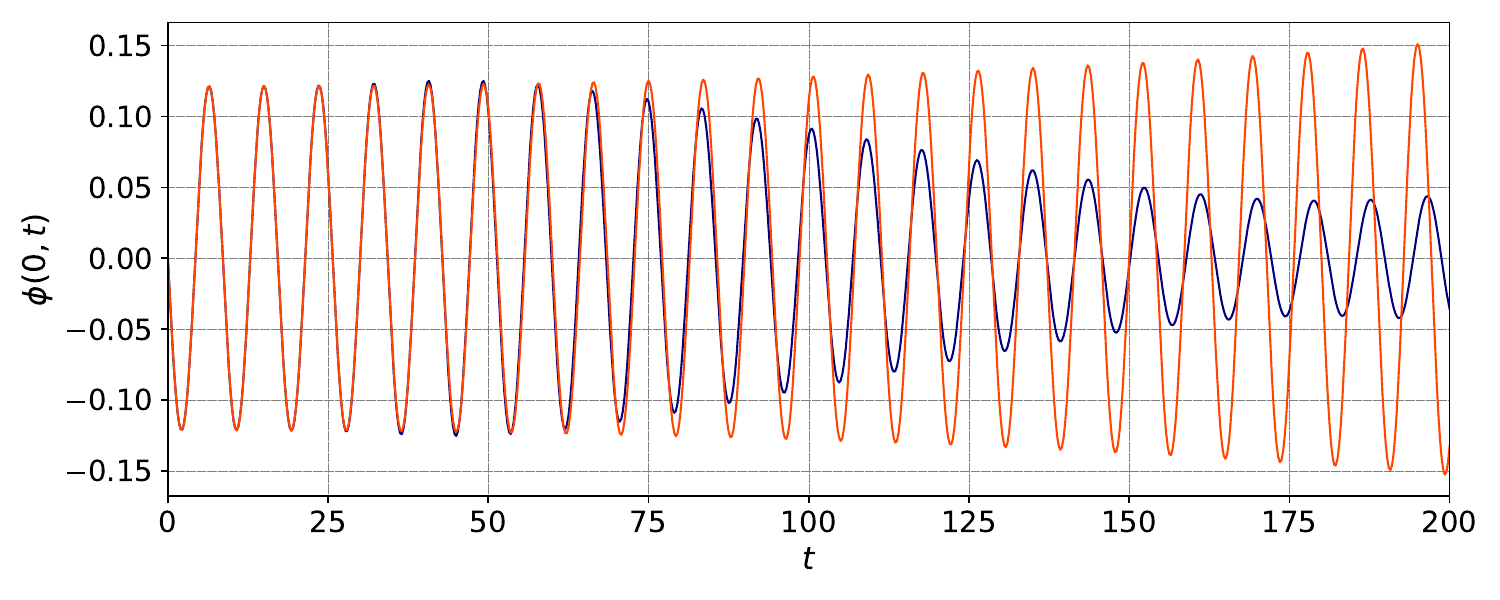}\label{phi4inv_08_015_02}
    \vspace{-0.5cm}
  \caption{Results for the inverse $\phi^4$-potential Eq. (\ref{pi4}). Conventions as in Fig. \ref{fig_2Ba}. %Upper: $\eta=0.10$ with $\lambda_2=-0.08$ (left) and $\lambda_2=-0.20$ (right). Lower: $\eta=0.80$ with $\lambda_2=-0.08$ (left) and $\lambda_2=-0.20$ (right). For all figures, we fixed $\lambda_1=0.15$.
  }
  \label{fig_2B}
\end{center}
\end{figure*}
%%%%%%%%%%%%%%%%%%%%%%%%%%%%%%%%%%%%%%%%%%%%%%%%%%%%%%%%%%%%%%%%%%%%%%%%%%

%{\color{blue}Naturally, the agreement also applies to the entire solution, see Fig. XXX. That is, the modulated oscillon is very well described by the renormalized profile Eq. (\ref{mop3}) for various $\lambda_2$ and $\eta$, with $\lambda_1$ fixed.}

\subsubsection{The inverse $\phi ^{4}$-potential vs. the double-well one.}

We also consider those oscillons with modulated behavior in the inverse $\phi ^{4}$ and double-well models, separately. Again, we show that the numerical results can be very well approximated by the analytical profile based on the two $Q$-ball solution.

The inverse $\phi ^{4}$-potential (\ref{pi4}) leads to $a_{3}=0$ and $a_{4}=1$, from which one gets $\beta=3-2\eta$ and $\alpha=(2\eta -1)/8$. %see Eqs. (\ref{baip4}).
Equation (\ref{de1x}) then assumes the form%
\begin{equation}
\phi _{R}\left( x,t\right) =\sqrt{\frac{2}{3-2\eta}}\Psi +\frac{2\eta -1}{4(3-2\eta)}  \sqrt{\frac{2}{3-2\eta}} 
\Psi ^{3}+\text{c.c.,}\label{roip4}
\end{equation}%
which only exists for $ \eta<3/2$, see Eq. (\ref{bep4i}). Here, $\Psi$ is again given by Eq. (\ref{tqbs}).

We now use the solution above to mimic the numerical oscillon. %We implement Eq. (\ref{roip4}) as the initial state, and solve Eq. (\ref{gceomx1}) numerically in the presence of Eqs. (\ref{fphi}) and (\ref{pi4}).
We consider Eqs. (\ref{gceomx1}), (\ref{fphi}) and (\ref{pi4}), and implement Eq. (\ref{roip4}) as the initial configuration. We study the evolution of the system for different $\lambda_1$, $\lambda_2$ and $\eta$.

The results for $\phi(x=0,t)$ appear in Fig. \ref{fig_3B}. Here, we have used various $\lambda_1$ and $\eta$, with $\lambda_2$ fixed. As in the previous example, the evolution of the numerical oscillon can be mapped via the analytical profile with great accuracy indeed. The correspondence is particularly remarkable for small and moderate modulations, but it can still be noted in the large modulations regime.
% {\color{blue}Naturally, such a correspondence also applies to the entire oscillon, see Fig XXX.}

\begin{figure*}[!ht]
\begin{center}
 \centering
   \includegraphics[{angle=0,width=8cm,height=4.5cm}]{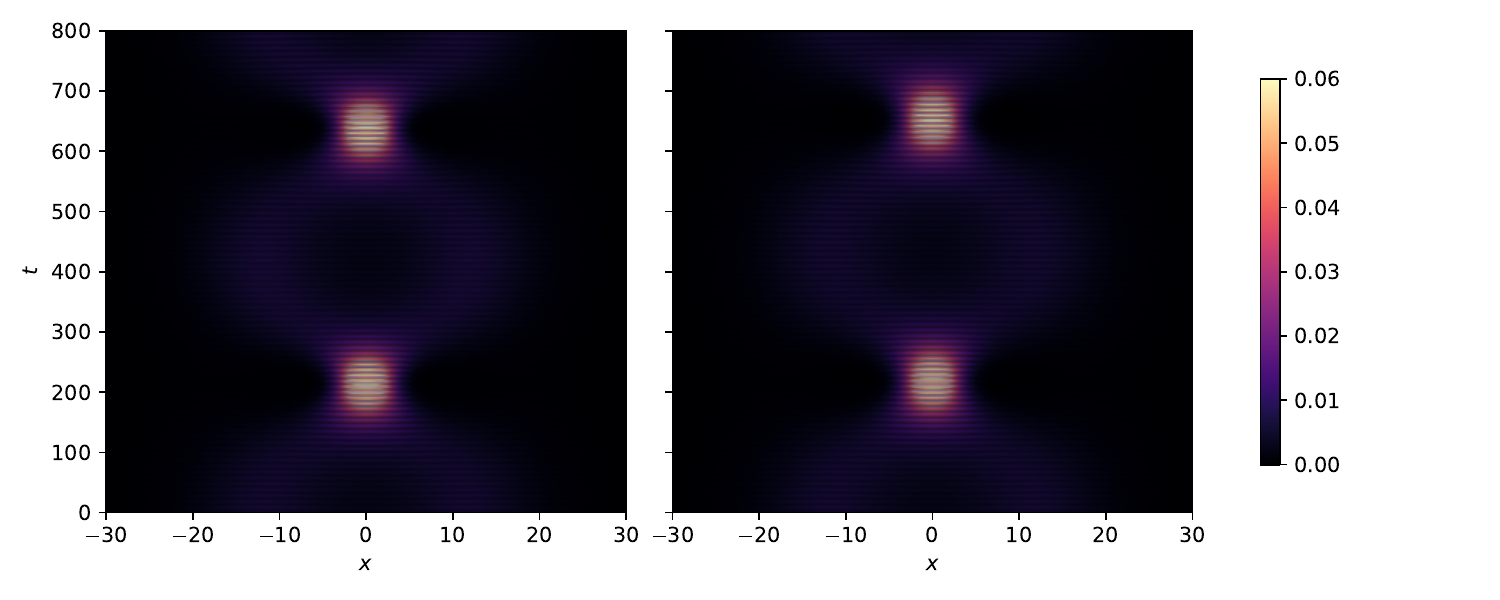}\label{3d_phi4inv_01_01_02}
   \includegraphics[{angle=0,width=8cm,height=4.5cm}]{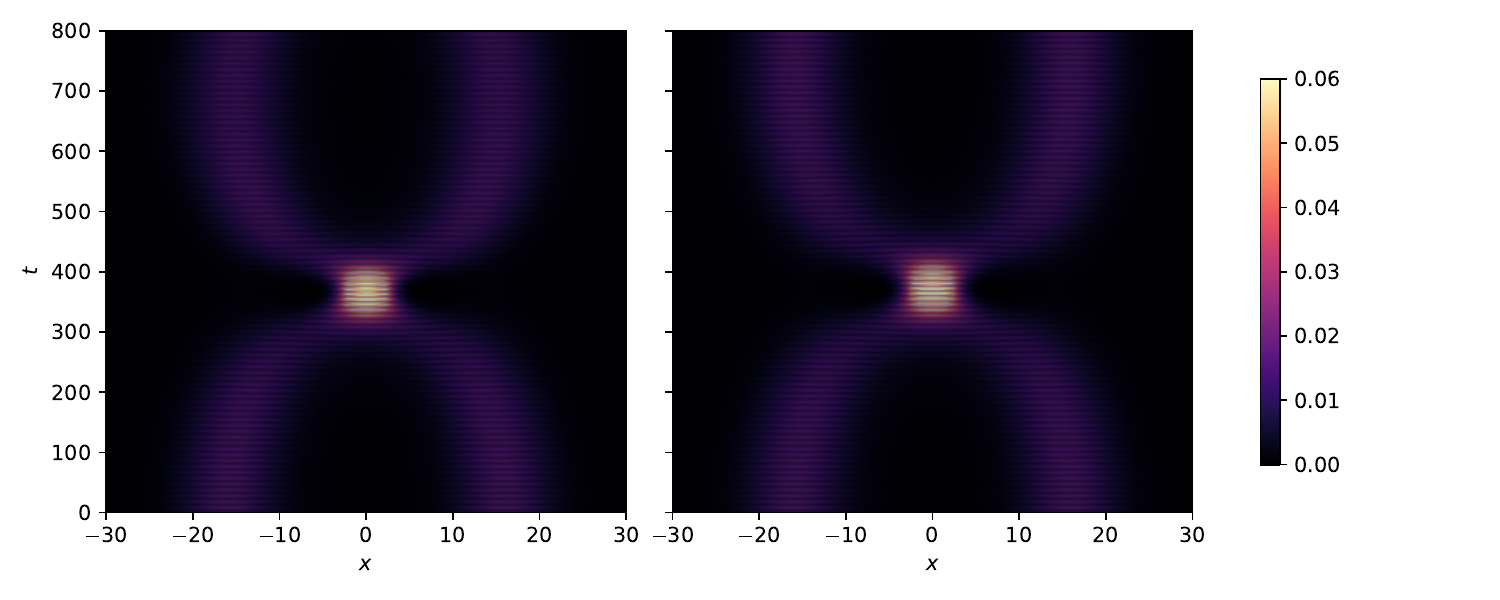}\label{3d_phi4inv_01_015_02}
   \includegraphics[{angle=0,width=8cm,height=4.5cm}]{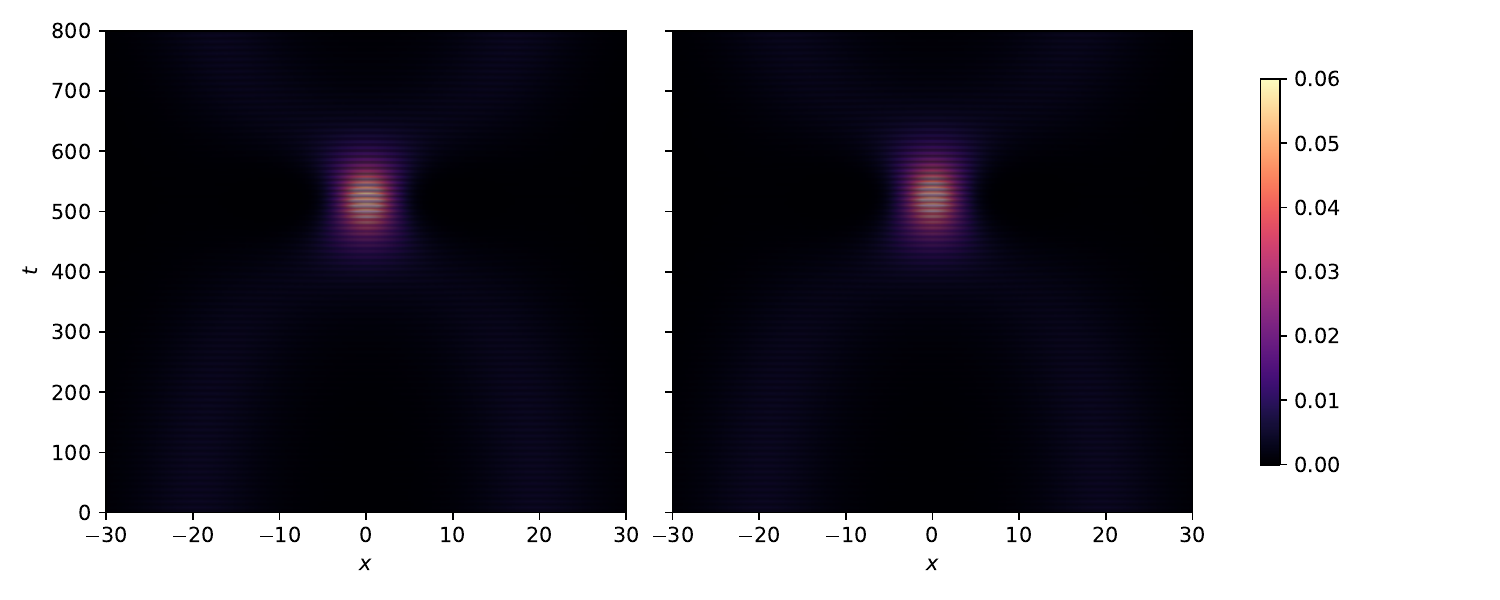}\label{3d_phi4inv_01_01_015}
   \includegraphics[{angle=0,width=8cm,height=4.5cm}]{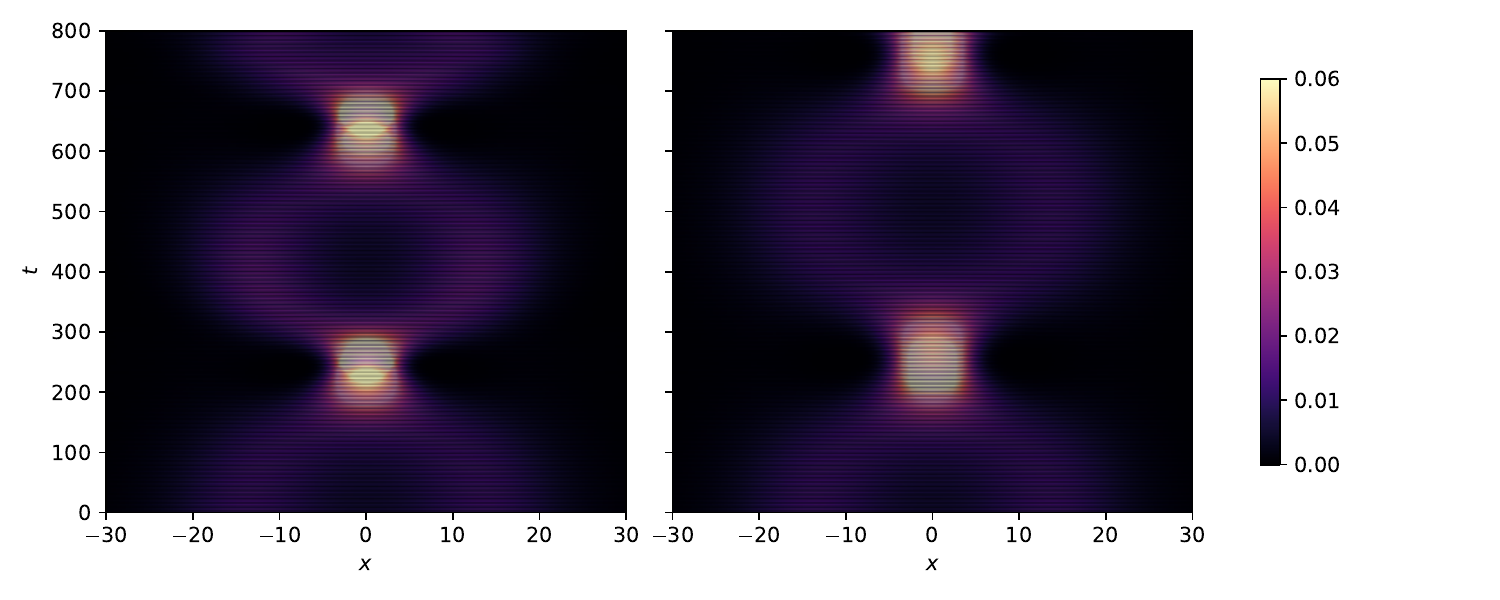}\label{3d_phi4inv_05_01_02}
    \vspace{-0.5cm}
  \caption{Results for the inverse $\phi^4$-potential Eq. (\ref{pi4}). Conventions as in Fig. \ref{fig_1Ca}. %Upper: $\eta=0.10$ with $\lambda_1=0.10$ and $\lambda_2=-0.20$ (left) and $\eta=0.10$ with $\lambda_1=0.15$ and $\lambda_2=-0.20$ (right). Lower: $\eta=0.10$ with $\lambda_1=0.10$ and $\lambda_2=-0.15$ (left) and $\eta=0.50$ with $\lambda_1=0.10$ and $\lambda_2=-0.20$ (right). We plot $|\partial^2 \phi + \phi|$ versus $x$ and $t$ for the inverse $\phi^4$ model.
  }
  \label{fig_2C}
\end{center}
\end{figure*}

This conclusion is corroborated by the results in Figs. \ref{fig_2B} and \ref{fig_2C}. The first one shows the field at the center of mass for different $\lambda_2$ and $\eta$, now with $\lambda_1$ fixed. In addition, the second one presents the numerical evolution of the entire generalized oscillon in comparison to the renormalized analytical solution.

%%%%%%%%%%%%%%%%%%%%%%%%%%%%%%%%%%%%%%%%%%%%%%%%%%%%%%%%%%%%%%%%%%%%%%%%%%% 
%%%%%%%%%%%%%%%%%%%%%%%%%%%%%%%%%%%%%%%%%%%%%%%%%%%%%%%%%%%%%%%%%%%%%%%%%%

Finally, we reconsider the double-well $\phi ^{4}$-potential Eq. (\ref{pdwp}). It gives $a_{3}=3/2$ and $a_{4}=-1$, from which we calculate %\beta$ and $\alpha$ as in Eq. (\ref{badwp}), i.e.
$\beta =(9-4\eta)/2$ and $\alpha =(5+4\eta)/16$. The renormalized oscillon reads%
\begin{equation}
\phi _{R}\left( x,t\right)=2\sqrt{\frac{1}{9-4\eta }}\Psi -\frac{2}{%
9-4\eta }\Psi ^{2}+\frac{6}{9-4\eta }\left\vert \Psi \right\vert
^{2}+\frac{5+4\eta }{2(9-4\eta)} \sqrt{\frac{1}{9-4\eta }}\Psi ^{3}+\text{c.c.,}
\end{equation}%
with $\eta<9/4$.%, see Eq. (\ref{edwp}).

\begin{figure*}[!ht]
\begin{center}
  \centering
    \includegraphics[{angle=0,width=8cm,height=4cm}]{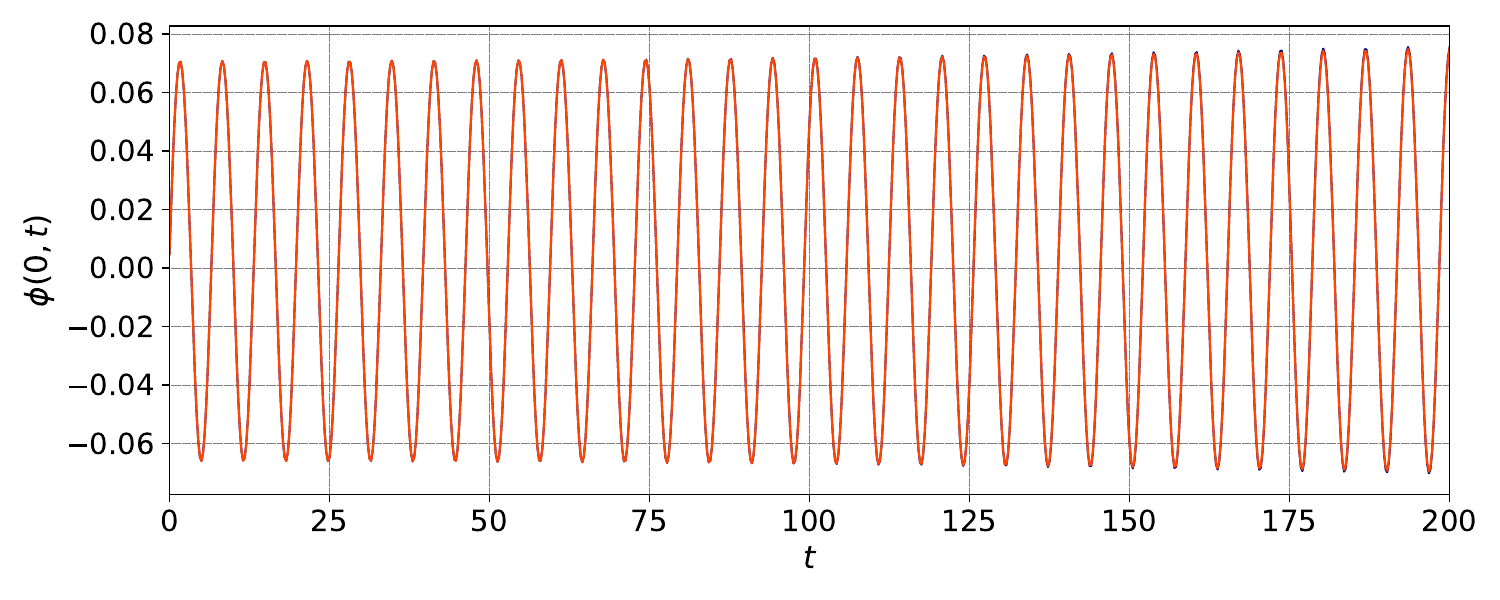}\label{phi4_01_01_005}
    \includegraphics[{angle=0,width=8cm,height=4cm}]{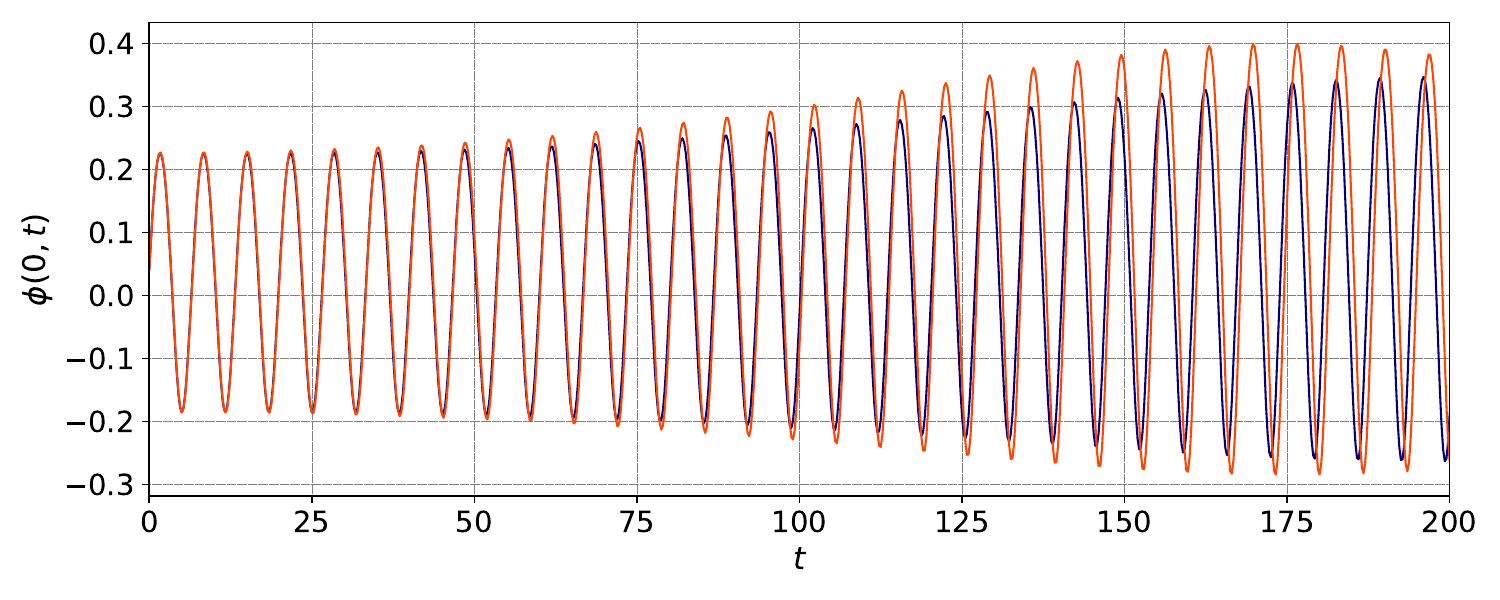}\label{phi4_01_02_005}
    \includegraphics[{angle=0,width=8cm,height=4cm}]{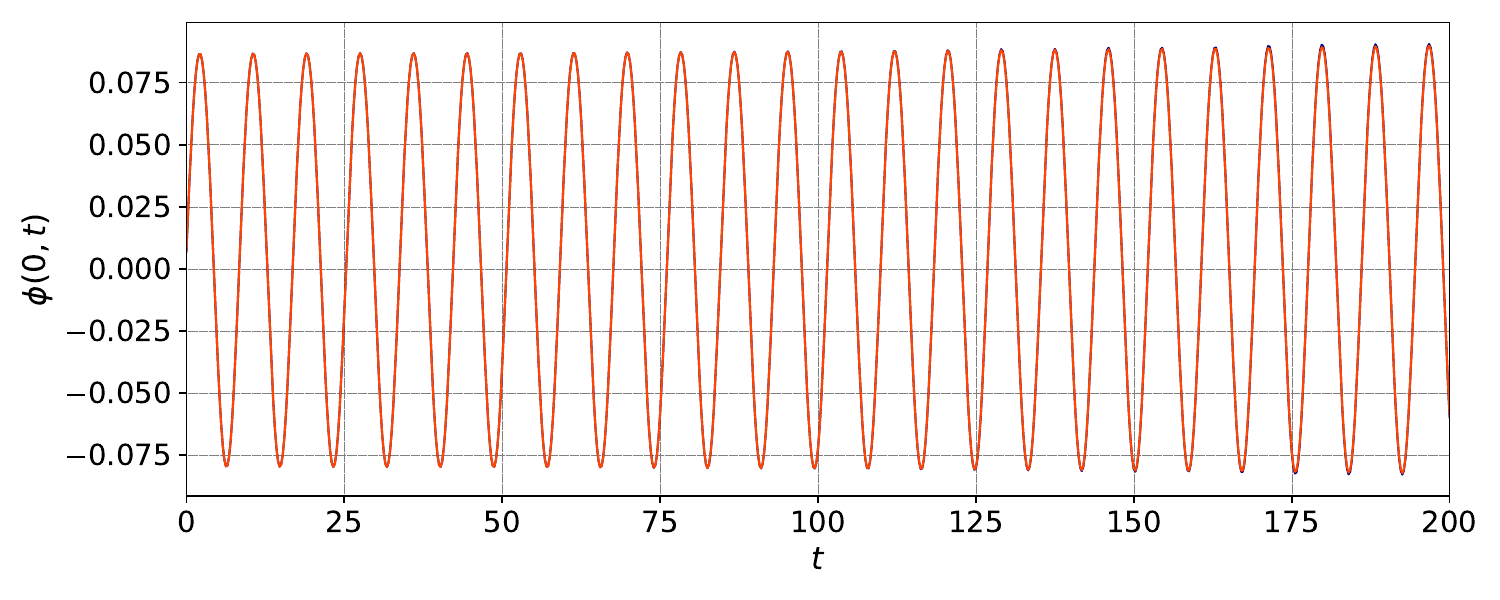}\label{phi4_08_01_005}
    \includegraphics[{angle=0,width=8cm,height=4cm}]{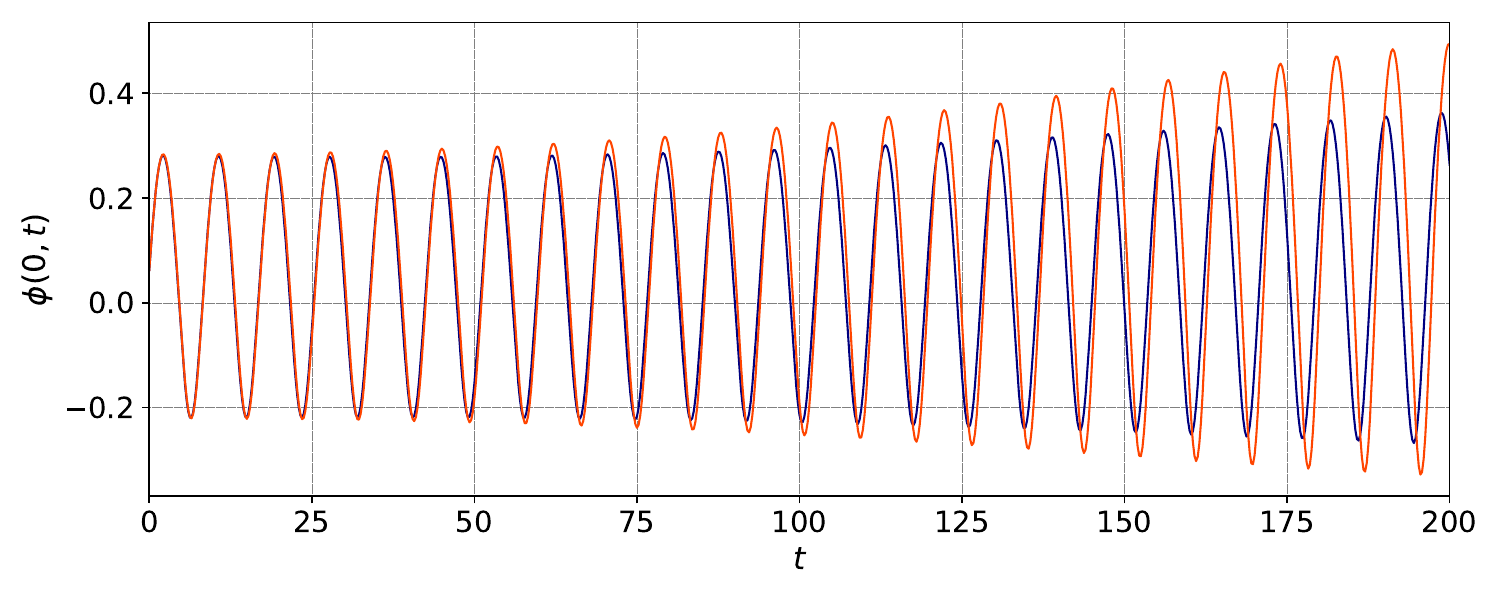}\label{phi4_08_02_005}
    \vspace{-0.5cm}
  \caption{Results for the double-well $\phi^4$-potential Eq. (\ref{pdwp}). Conventions as in Fig. \ref{fig_1B}. %Upper: $\eta=0.10$ with $\lambda_1=0.10$ (left) and $\lambda_1=0.20$ (right). Lower: $\eta=0.80$ with $\lambda_1=0.10$ (left) and $\lambda_1=0.20$ (right). For all figures, we fixed $\lambda_2=-0.05$.
  }
  \label{fig_5B}
\end{center}
\end{figure*}

\begin{figure*}[!ht]
\begin{center}
 \centering
   \includegraphics[{angle=0,width=8cm,height=4cm}]{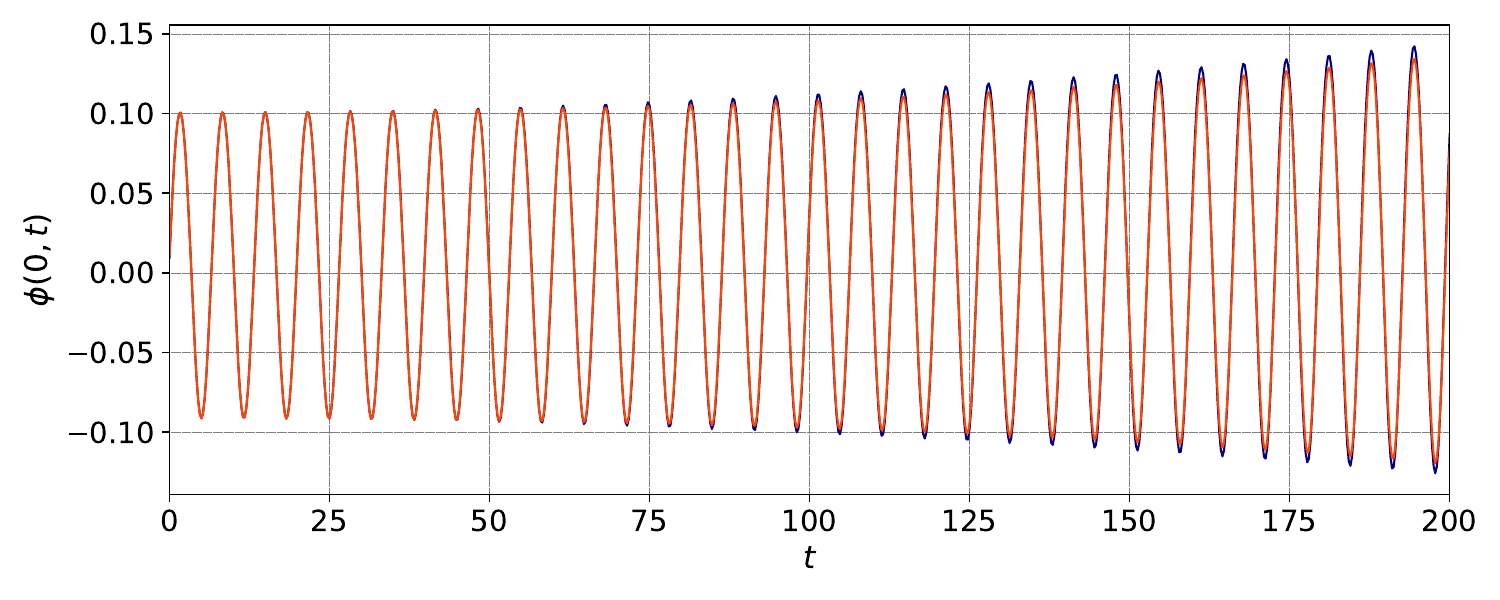}\label{phi4_01_015_008}
    \includegraphics[{angle=0,width=8cm,height=4cm}]{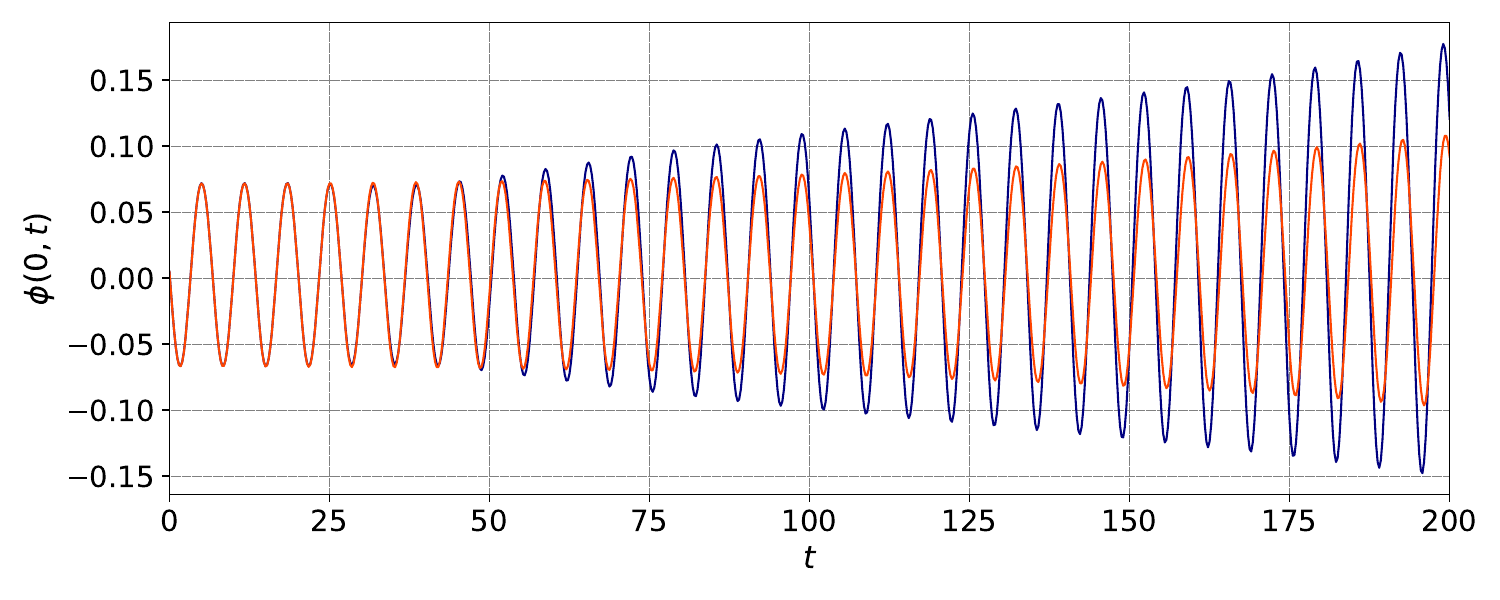}\label{phi4_01_015_02}
    \includegraphics[{angle=0,width=8cm,height=4cm}]{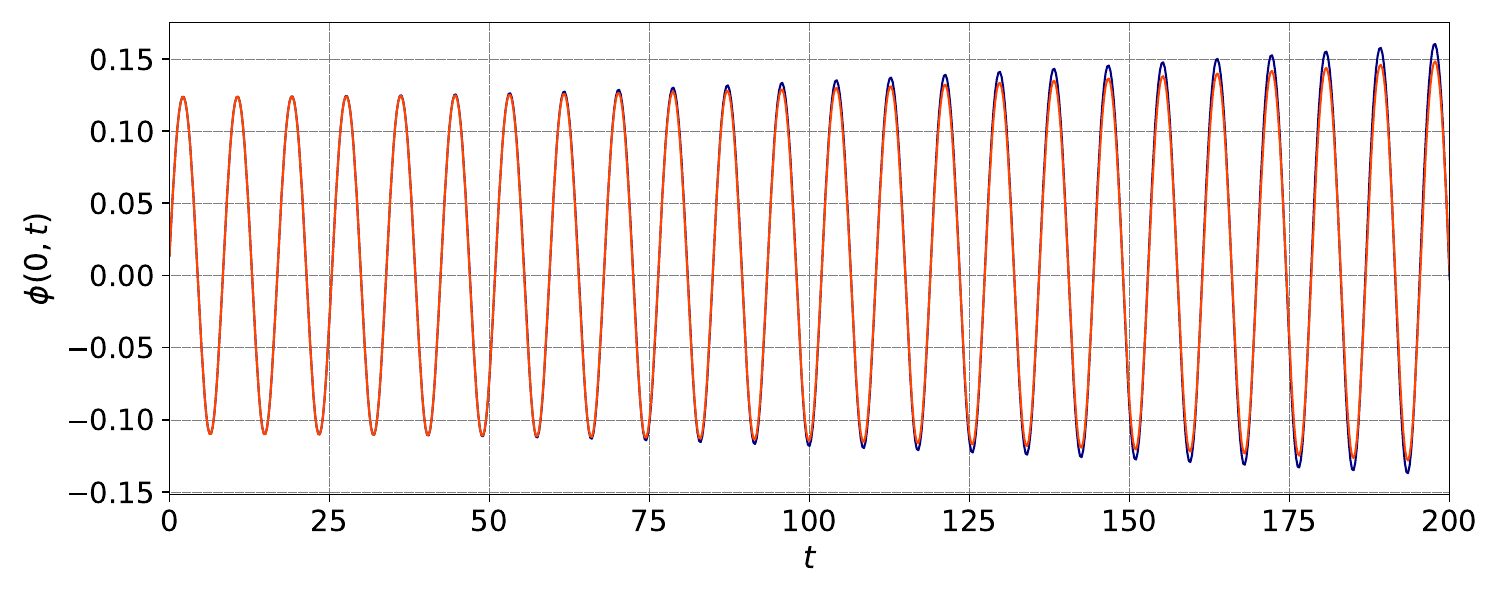}\label{phi4_08_015_008}
    \includegraphics[{angle=0,width=8cm,height=4cm}]{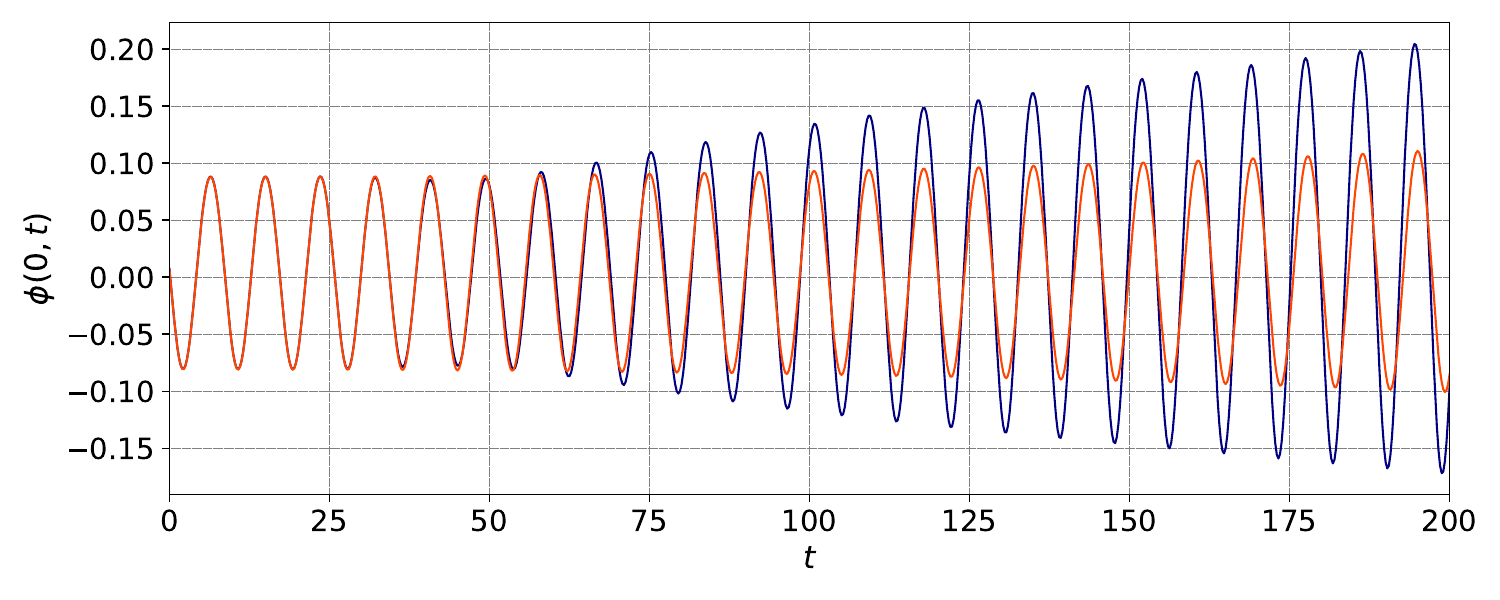}\label{phi4_08_015_02}
    \vspace{-0.5cm}
  \caption{Results for the double-well $\phi^4$-potential Eq. (\ref{pdwp}). Conventions as in Fig. \ref{fig_2Ba}. %Upper: $\eta=0.10$ with $\lambda_2=-0.08$ (left) and $\lambda_2=-0.20$ (right). Lower: $\eta=0.80$ with $\lambda_2=-0.08$ (left) and $\lambda_2=-0.20$ (right). For all figures, we fixed $\lambda_1=0.15$.
  }
  \label{fig_6B}
\end{center}
\end{figure*}

\begin{figure*}[!ht]
\begin{center}
 \centering
   \includegraphics[{angle=0,width=8cm,height=4.5cm}]{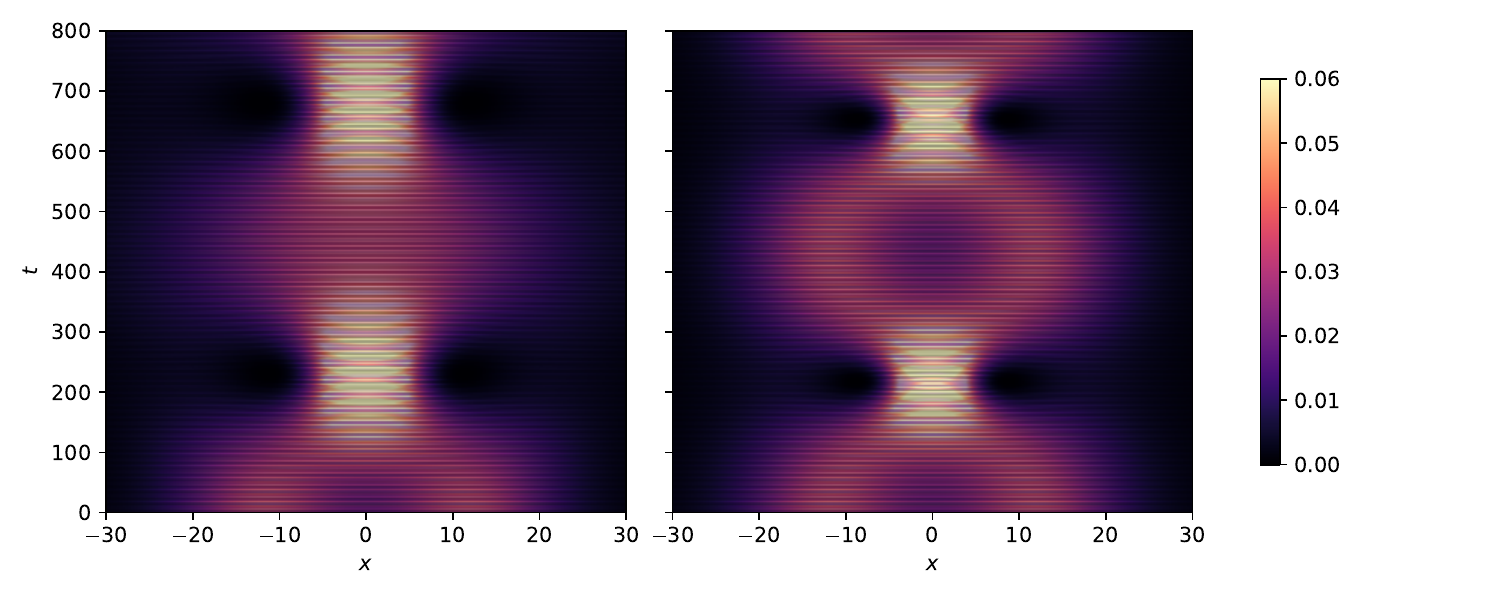}\label{3d_phi4_01_01_02}
   \includegraphics[{angle=0,width=8cm,height=4.5cm}]{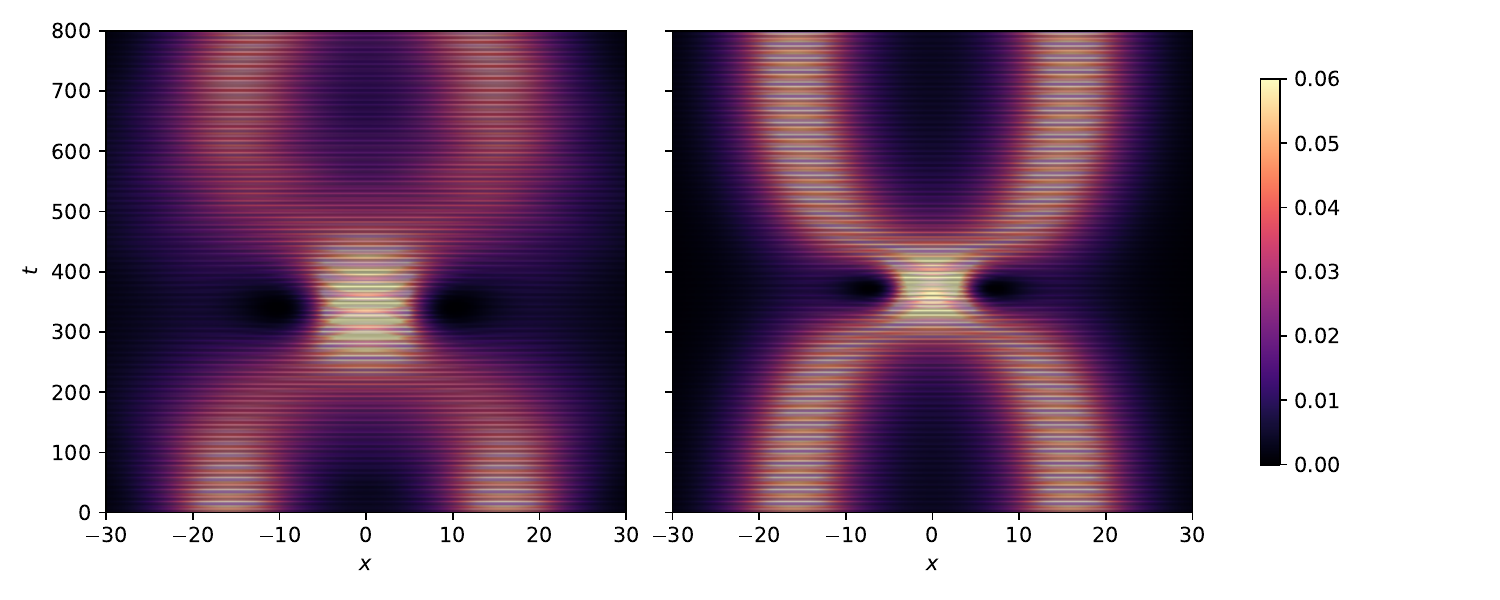}\label{3d_phi4_01_015_02}
   \includegraphics[{angle=0,width=8cm,height=4.5cm}]{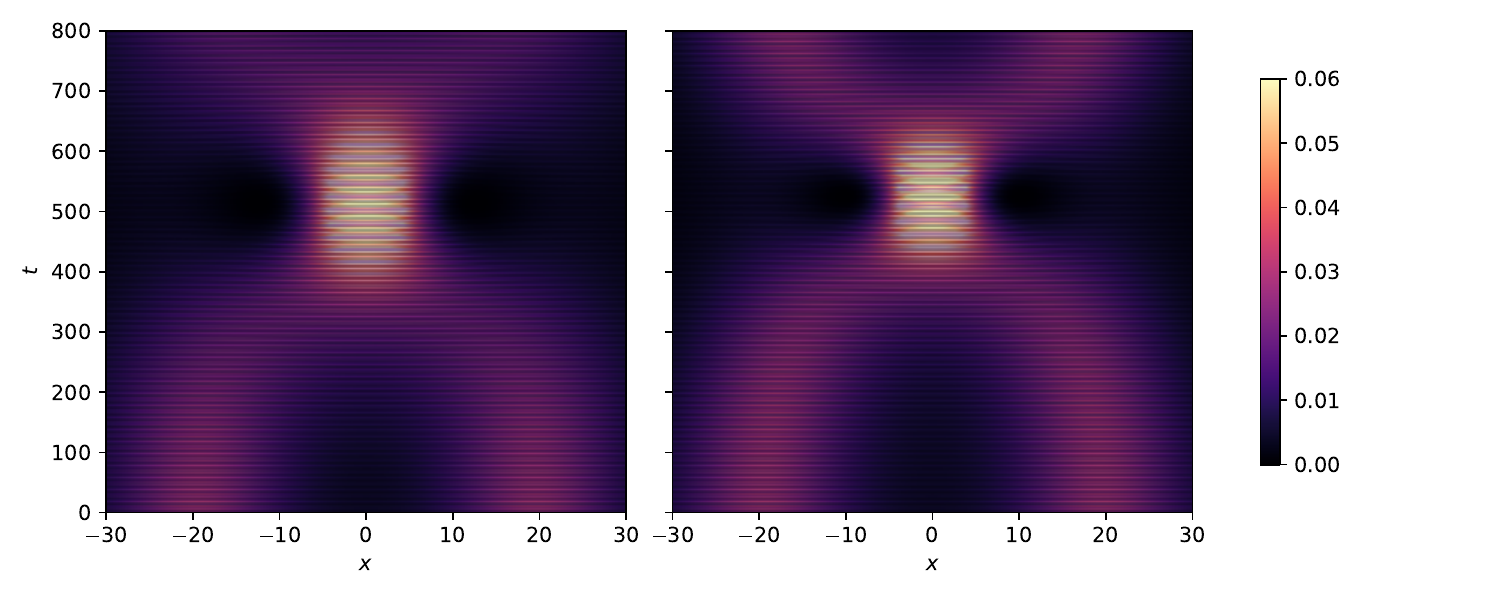}\label{3d_phi4_01_01_015}
   \includegraphics[{angle=0,width=8cm,height=4.5cm}]{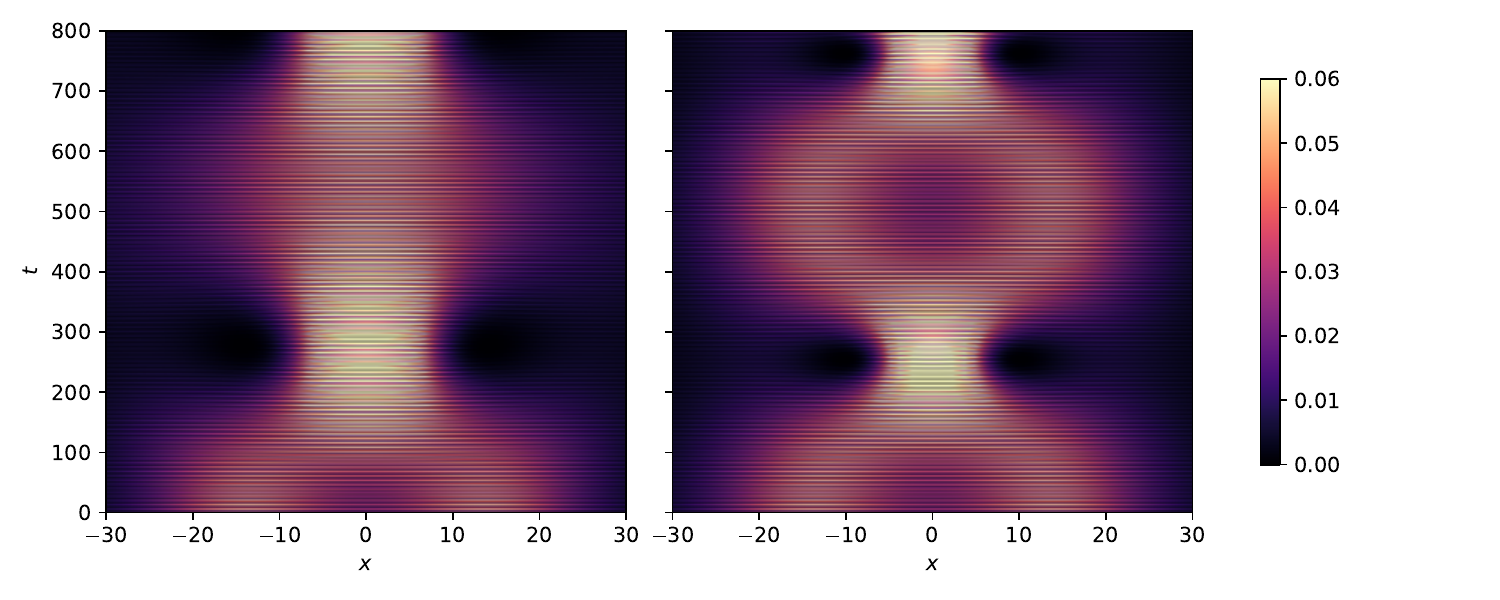}\label{3d_phi4_05_01_02}
    \vspace{-0.5cm}
  \caption{Results for the double-well $\phi^4$-potential Eq. (\ref{pdwp}). Conventions as in Fig. \ref{fig_1Ca}. %Upper: $\eta=0.10$ with $\lambda_1=0.10$ and $\lambda_2=-0.20$ (left) and $\eta=0.10$ with $\lambda_1=0.15$ and $\lambda_2=-0.20$ (right). Lower: $\eta=0.10$ with $\lambda_1=0.10$ and $\lambda_2=-0.15$ (left) and $\eta=0.50$ with $\lambda_1=0.10$ and $\lambda_2=-0.20$ (right). We plot $|\partial^2 \phi + \phi|$ versus $x$ and $t$ for the double-well $\phi^4$ model.
  }
  \label{fig_3C}
\end{center}
\end{figure*}

Figures \ref{fig_5B} and \ref{fig_6B} depict the results for $\phi(x=0,t)$ with $\lambda_1$ and $\lambda_2$ fixed, respectively. while the entire solutions appear in Fig. \ref{fig_3C}. As before, the renormalized profile based on the two $Q$-ball solution mimics the modulated structure almost precisely.

%%%%%%%%%%%%%%%%%%%%%%%%%%%%%%%%%%%%%%%%%%%%%%%%%%%%%%%%%%%%%%%%%%%%%%%%%%% 
%%%%%%%%%%%%%%%%%%%%%%%%%%%%%%%%%%%%%%%%%%%%%%%%%%%%%%%%%%%%%%%%%%%%%%%%%%

%%%%%%%%%%%%%%%%%%%%%%%%%%%%%%%%%%%%%%%%%%%%%%%%%
\section{Additional developments} \label{secIIb}
%%%%%%%%%%%%%%%%%%%%%%%%%%%%%%%%%%%%%%%%%%%%%%%%%

%%%%%%%%%%%%%%%%%%%%%%%%%%%%%%%%%%%%%%%%%%%%%%%%%
\subsection{Canonical $\phi^2$-potential: same universality class} \label{secIIb}
%%%%%%%%%%%%%%%%%%%%%%%%%%%%%%%%%%%%%%%%%%%%%%%%%

The relation between oscillons and $Q$-balls only holds for $\beta>0$, see Eq. (\ref{b1}). In the standard $f=1$ case,
%(i.e. with $B_1=B_2=0$),
this requirement is fulfilled when at least one of the coefficients $a_3$ and $a_4$ does not equal zero. However, in our nonstandard scenario, $\beta > 0$ can be satisfied even when both $a_3$ and $a_4$ vanish simultaneously. In this case, Eq. (\ref{b1}) promptly reduces to
\begin{equation}
B_{1}^2<2B_{2} \text{.}\label{nb}
\end{equation}%
Once Eq. (\ref{nb}) is satisfied, the %relation between
oscillons/$Q$-balls relation is preserved. %In other words, such a relation may follow from the construction based on the third order perturbation series even when $a_3=a_4=0$.
This is a novel effect related to the nonstandard kinematics. So, it does not apply to the usual $f=1$ model.%(in this case, one needs to proceed to a fifth order perturbation series to reestablish the oscillons/$Q$-ball connection).

To illustrate this possibility, we focus on the simplest quadratic potential.
%on an exotic version of the $\phi^6$-potential.
Here, both the cubic and the quartic terms are absent. Concretely, we choose%
\begin{equation}
V\left( \phi \right) =\frac{\phi ^{2}}{2}%-\frac{\phi ^{6}}{6}
\label{ep6}\text{,}
\end{equation}%
which naturally gives $a_3=a_4=0$.%vanise last term does not contribute to the analysis up to the third order in the perturbation series.

%%%%%%%%%%%%%%%%%%%%%%%%%%%%%%%%%%%%%%%%%%%%%%%%%%%%%%%%%%%%%%%%%%%%%%%%%%% 
%%%%%%%%%%%%%%%%%%%%%%%%%%%%%%%%%%%%%%%%%%%%%%%%%%%%%%%%%%%%%%%%%%%%%%%%%%

In regard to noncanonical kinematics, we choose the generalizing function as
\begin{equation}
f\left( \phi \right) = \left( 1+\eta \right)\left(  1 +\eta \phi ^{2} \right)\label{f2}\text{,}
\end{equation}%
which can be obtained from the general form
\begin{equation}
f\left( \phi \right) =\frac{1+\eta }{\left(1+\eta \phi ^{2}\right)^\delta}\label{gf}\text{}
\end{equation}
once we set $\delta=-1$. Note that Eq. (\ref{gf}) reduces to Eq. (\ref{fphi}) for $\delta=1$. So, Eqs. (\ref{fphi}) and (\ref{f2}) may be understood to belong to the same class of generalizations. In the extreme regime $\eta \rightarrow \infty$, Eq. (\ref{f2}) behaves as $f\left( \phi \right) \approx \eta^2 \phi^2$. This particular form was recently used to study kink crystals in the context of a two-field model with a $\phi^4$-potential, see Ref. \cite{hnew1}. The same idea was also applied to the sine-Gordon potential, see \cite{hnew2}.

Equation (\ref{f2}) leads to $b_{0}=1+\eta $, $b_{1}=0$, and $b_{2}=2\eta \left( 1+\eta \right)$.
%(with $b_{3}=b_{4}=0$).
So, we get $B_{1}=0$ and $B_{2}=2\eta$, from which Eq. (\ref{nb}) provides
\begin{equation}
0<\eta \text{.}\label{nb0}
\end{equation}%
It indicates that the oscillon/$Q$-ball connection holds for any $\eta$.
%up to the third order in the perturbation series.
That is, the potential Eq. (\ref{ep6}) does not restrict
%impose any bound on
the strength of the novel effects. %values of $\eta$. However, whether we increase the order in the perturbation series, a novel bound emerges. We explore this possibility later below.

In addition, %view of $a_3=a_4=0$,
both $\beta _{u}$ and $\alpha _{u}$ vanish identically, see Eqs. (\ref{bet}) and (\ref{alf}). As a consequence, one simply gets $\beta =2\eta$ and $\alpha =-\eta/4$.
%\begin{equation}
%\beta =2\eta\text{ \ \ and \ \ }\alpha =-\frac{\eta }{4}\text{,}\label{bet_t}
%\end{equation}%
%where we have implemented $B_1=0$ and $B_2=2\eta$.

The renormalized oscillon Eq. (\ref{de1x}) then reduces to%
\begin{equation}
\phi _{R}\left( x,t \right) =\sqrt{\frac{1}{\eta }}\Psi -\frac{1 }{4}
 \sqrt{\frac{1}{\eta }}\Psi ^{3}+\text{c.c.,}  \label{de1xep6}
\end{equation}
%where we have used Eq. (\ref{ncfx_y1}).
%\begin{equation}
%\phi _{R}\left( x,t \right) =2\sqrt{\frac{1}{\eta}}\frac{\lambda \cos\left(\frac{\omega }{\sqrt{1+\eta }}t\right)}{\cosh \left( \frac{\lambda }{\sqrt{1+\eta }}x \right) } -\frac{1}{2}\sqrt{\frac{1}{\eta}}\frac{\lambda^3 \cos\left(\frac{3\omega }{\sqrt{1+\eta }}t\right)}{\cosh^3 \left( \frac{\lambda }{\sqrt{1+\eta }}x  \right) }\text{,}  \label{de1xep6}
%\end{equation}
where $\Psi$ is the single $Q$-ball solution Eq. (\ref{cosh1}), with $T=t/\sqrt{1+\eta}$ and $X=x/\sqrt{1+\eta}$.

\begin{figure*}[!ht]
\begin{center}
  \centering
    \includegraphics[{angle=0,width=8cm,height=4cm}]{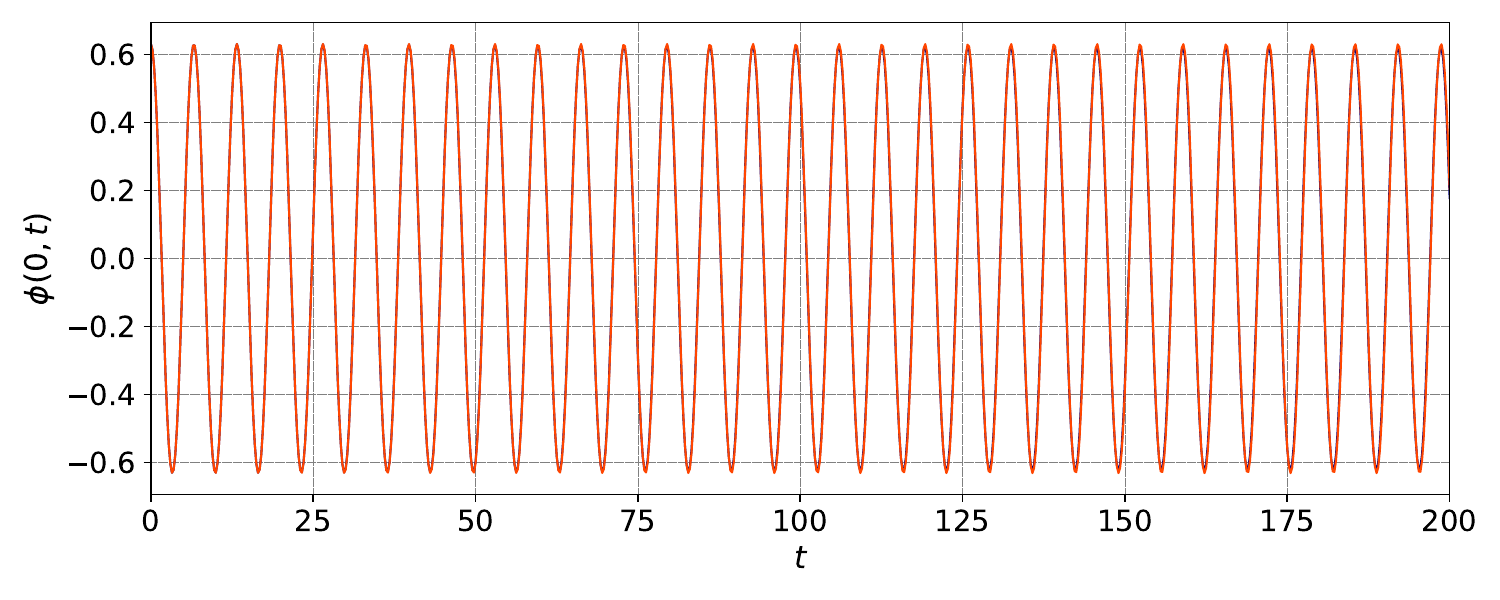}\label{phi6_01_01}
    \includegraphics[{angle=0,width=8cm,height=4cm}]{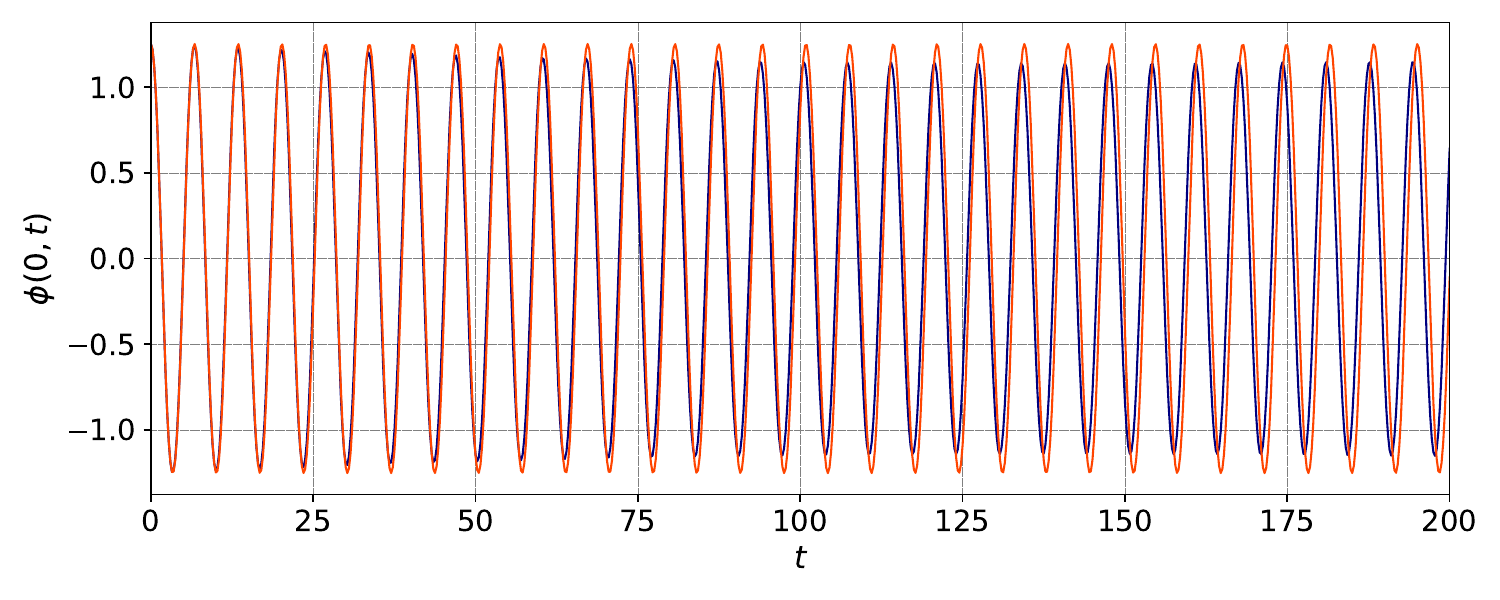}\label{phi6_02_02}
    \includegraphics[{angle=0,width=8cm,height=4cm}]{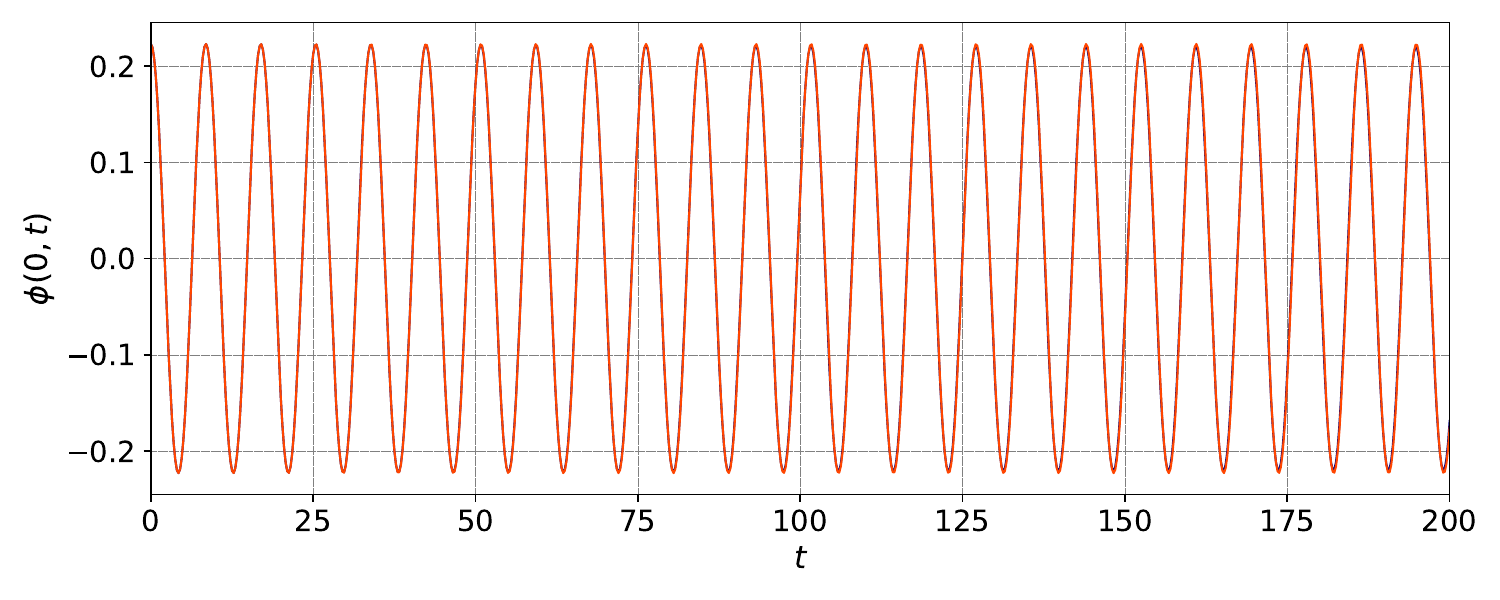}\label{phi6_08_01}
    \includegraphics[{angle=0,width=8cm,height=4cm}]{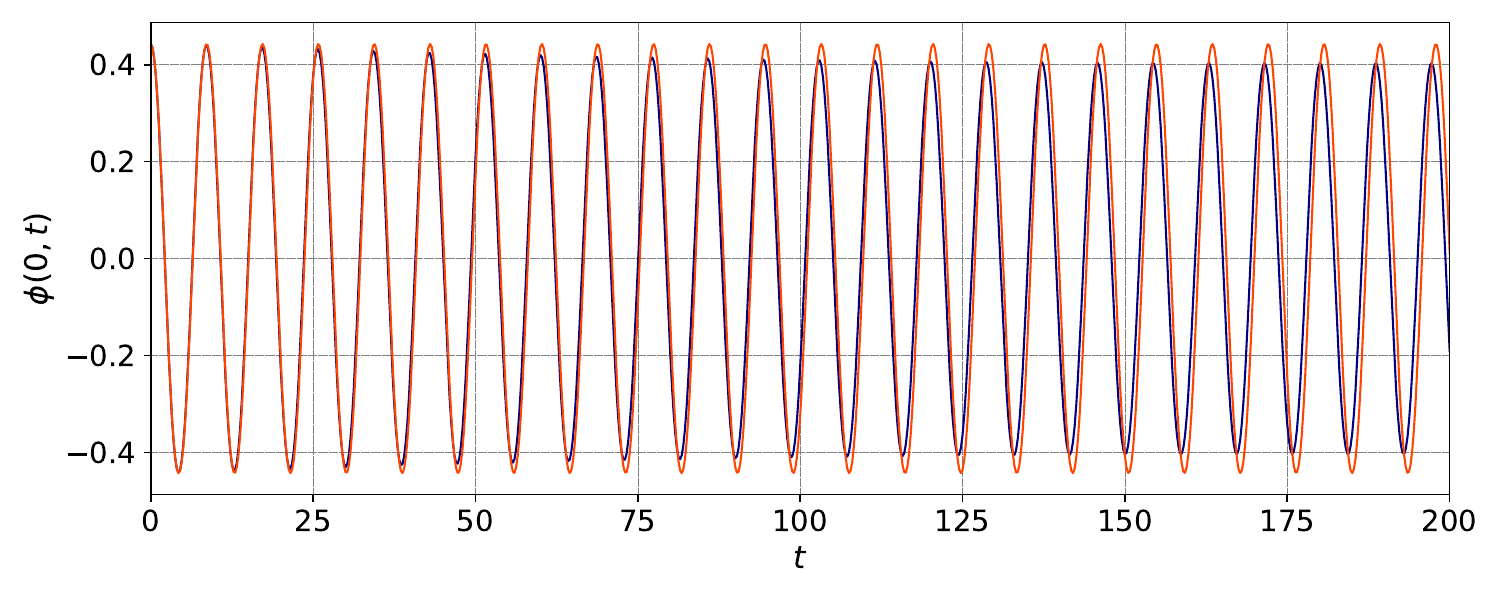}\label{phi6_08_02}
    \vspace{-0.5cm}
  \caption{Results for the $\phi^2$-potential Eq. (\ref{ep6}). Conventions as in Fig. \ref{fig_1A}.%The numerical oscillon (black line) is compared to the renormalized analytical one (red line). Upper: $\eta=0.10$, with $\lambda=0.10$ (left) and $\lambda=0.20$ (right). Lower: $\eta=0.80$, with $\lambda=0.10$ (left) and $\lambda=0.20$ (right).
  }
  \label{fig_phi6A}
\end{center}
\end{figure*}

We use Eq. (\ref{de1xep6}) as the initial state, and study the evolution of a generalized oscillon numerically. The results for different $\lambda$ and $\eta$ appear in Fig. \ref{fig_phi6A}. They reveal that well-behaved oscillations occur even for $a_3=a_4=0$. As we have pointed out, this is a novel effect due to the nonstandard kinematics.%once similar oscillations do not appear in the usual $f=1$ context.

%%%%%%%%%%%%%%%%%%%%%%%%%%%%%%%%%%%%%%%%%%%%%%%%%%%%%%%%%%%%%%%%%%%%%%%%%%% 
%%%%%%%%%%%%%%%%%%%%%%%%%%%%%%%%%%%%%%%%%%%%%%%%%%%%%%%%%%%%%%%%%%%%%%%%%%

Additionally, we compare the numerical solution with the analytical one. As in the previous cases, the mapping is more accurate in the small-amplitude regime. %On the other hand,
As the initial amplitude increases, the renormalized solution %Eq. (\ref{de1xep6})
does not mimic the numerical profile so precisely. This behavior resembles the general one already identified in this manuscript.

In summary, noncanonical kinematics is sufficient to yield well-behaved oscillons in connection to the simplest quadratic potential. %in connection supports induces the emergence of a well-behaved oscillon even in connection to the simplest phi^2-model. the existence of oscillons inherent to the canonical $\phi^2$-model is guaranteed by the noncanonical kinematics.
Moreover, their evolution can be approximated via the analytical profile in the same way as in the previous cases.

As the initial amplitude gets higher, the numerical evolution gives rise to a modulated structure. As before, this fact suggests that we must implement the two $Q$-ball solution. Such an approximation is only possible because the generalized $\phi^2$-oscillon belongs to the same universality class, i.e. it emerges from the RG Eq. (\ref{eom1}).%Naturally, this possibility does not occur in the usual scenario.

\begin{figure*}[!ht]
\begin{center}
  \centering
    \includegraphics[{angle=0,width=8cm,height=4cm}]{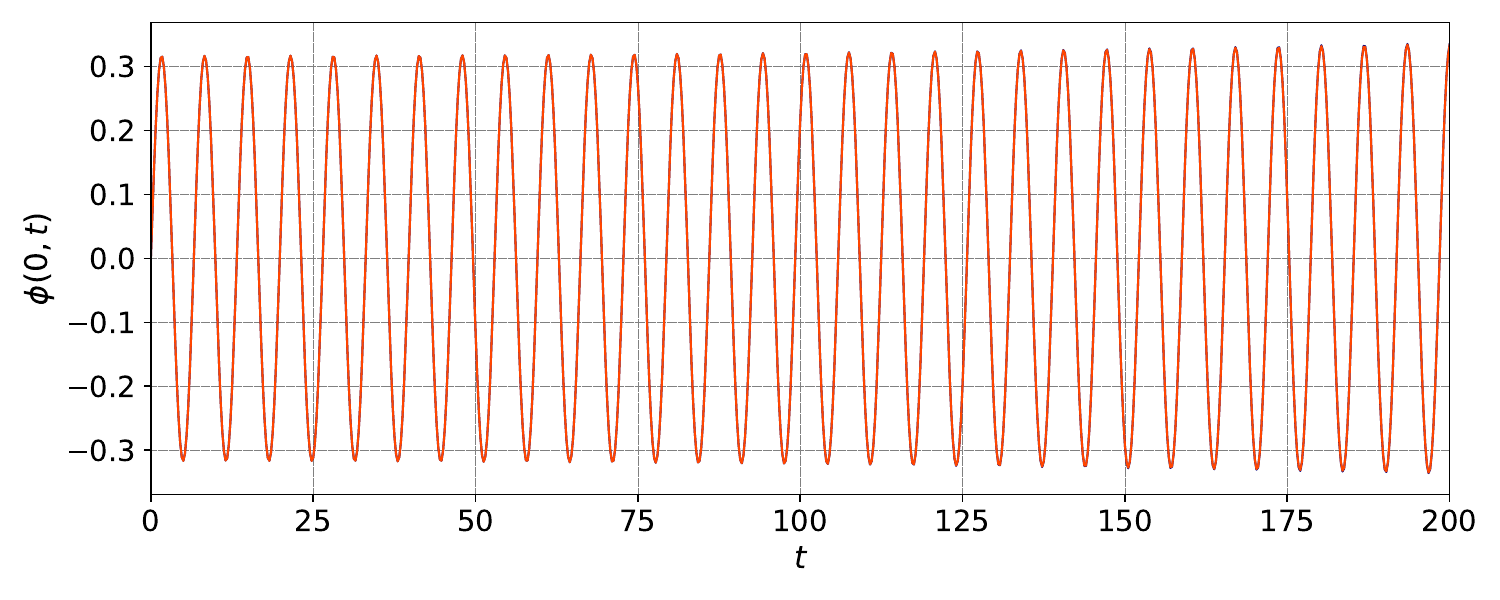}\label{phi6_01_01_005}
    \includegraphics[{angle=0,width=8cm,height=4cm}]{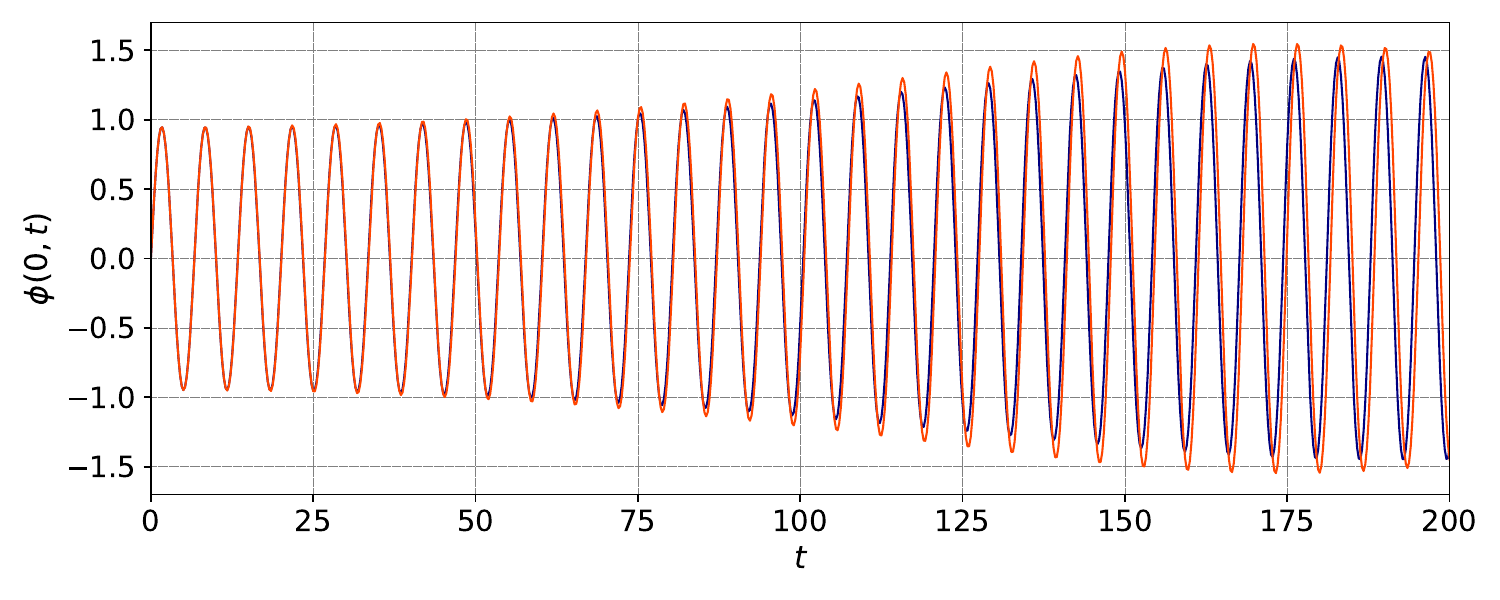}\label{phi6_01_02_005}
    \includegraphics[{angle=0,width=8cm,height=4cm}]{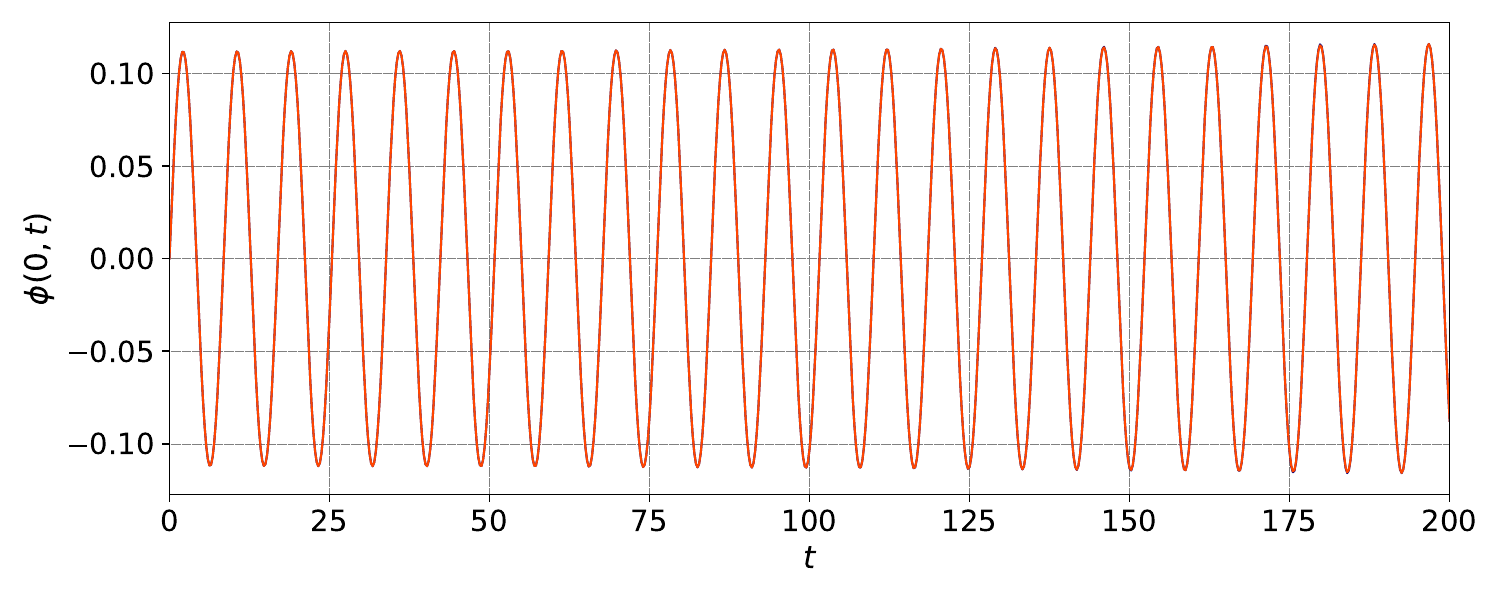}\label{phi6_08_01_005}
    \includegraphics[{angle=0,width=8cm,height=4cm}]{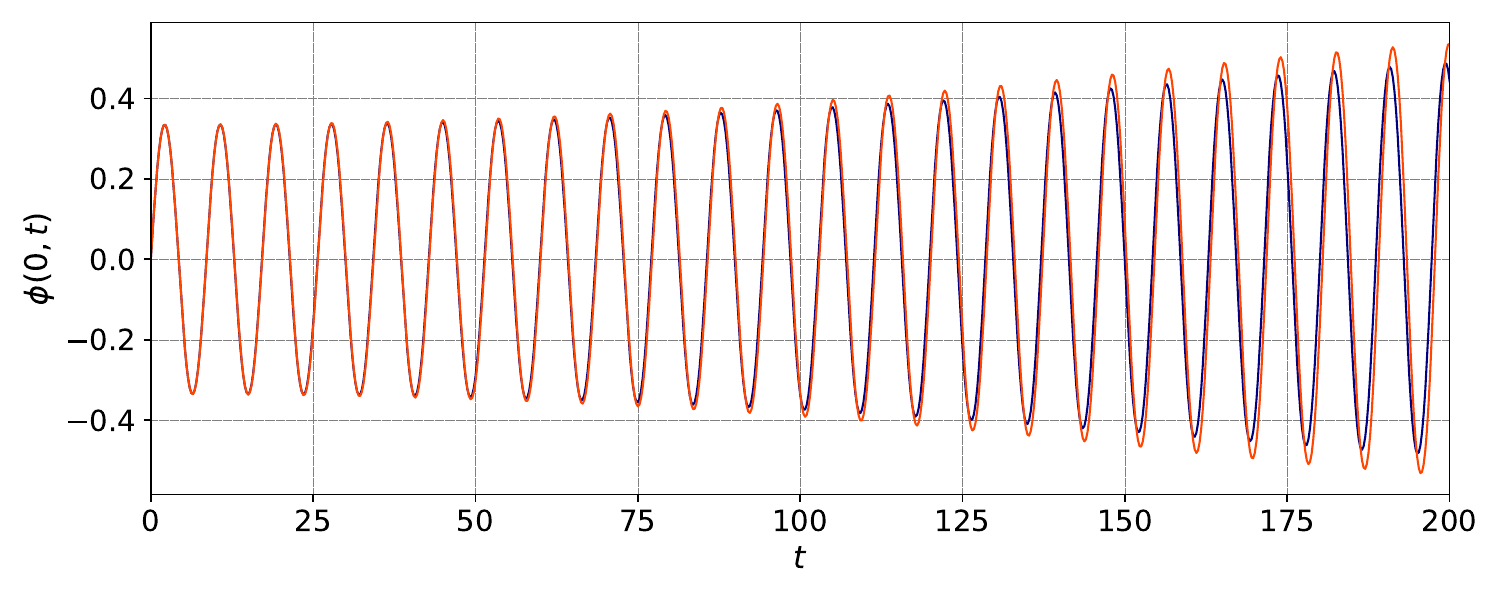}\label{phi6_08_02_005}
    \vspace{-0.5cm}
  \caption{Results for the $\phi^2$-potential Eq. (\ref{ep6}). Conventions as in Fig. \ref{fig_1B}. %The numerical oscillon (black line) is compared to the renormalized analytical one (red line). Upper: $\eta=0.10$ with $\lambda_1=0.10$ (left) and $\lambda_1=0.20$ (right). Lower: $\eta=0.80$ with $\lambda_1=0.10$ (left) and $\lambda_1=0.20$ (right). For all figures, we fixed $\lambda_2=-0.05$.
  }
  \label{fig_phi6Ba}
\end{center}
\end{figure*}

\begin{figure*}[!ht]
\begin{center}
  \centering
    \includegraphics[{angle=0,width=8cm,height=4cm}]{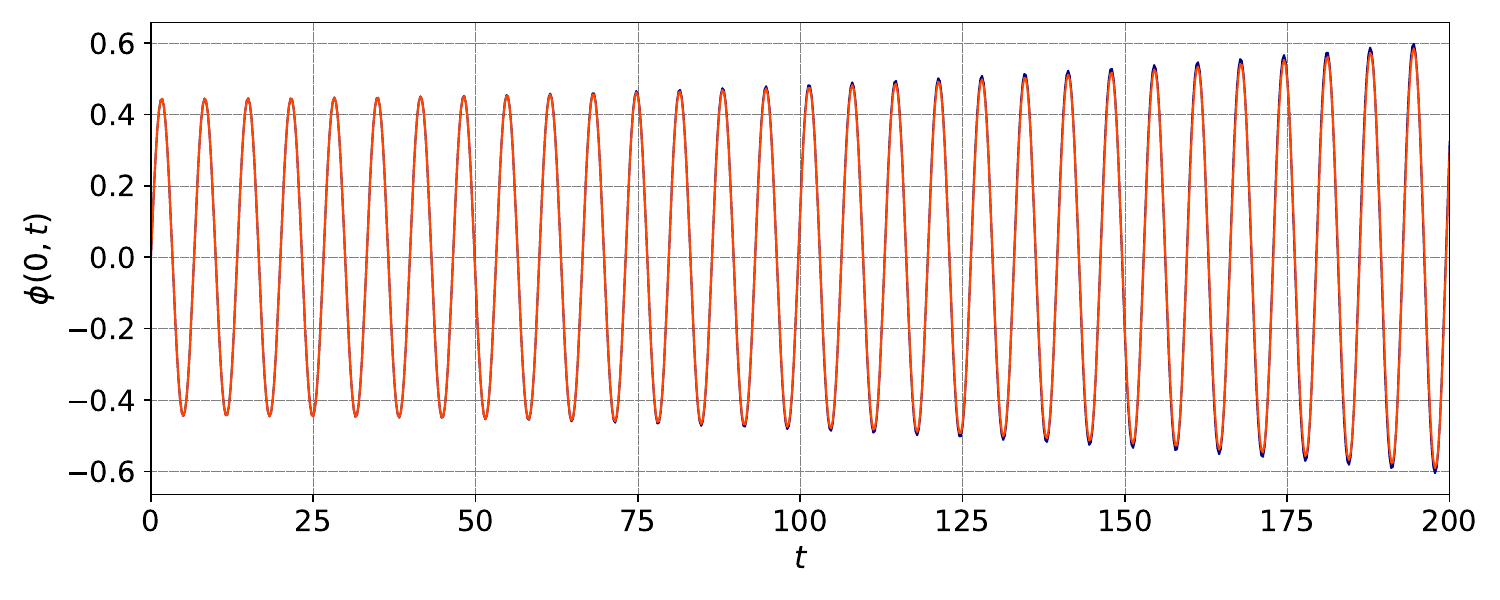}\label{phi6_05_015_008}
    \includegraphics[{angle=0,width=8cm,height=4cm}]{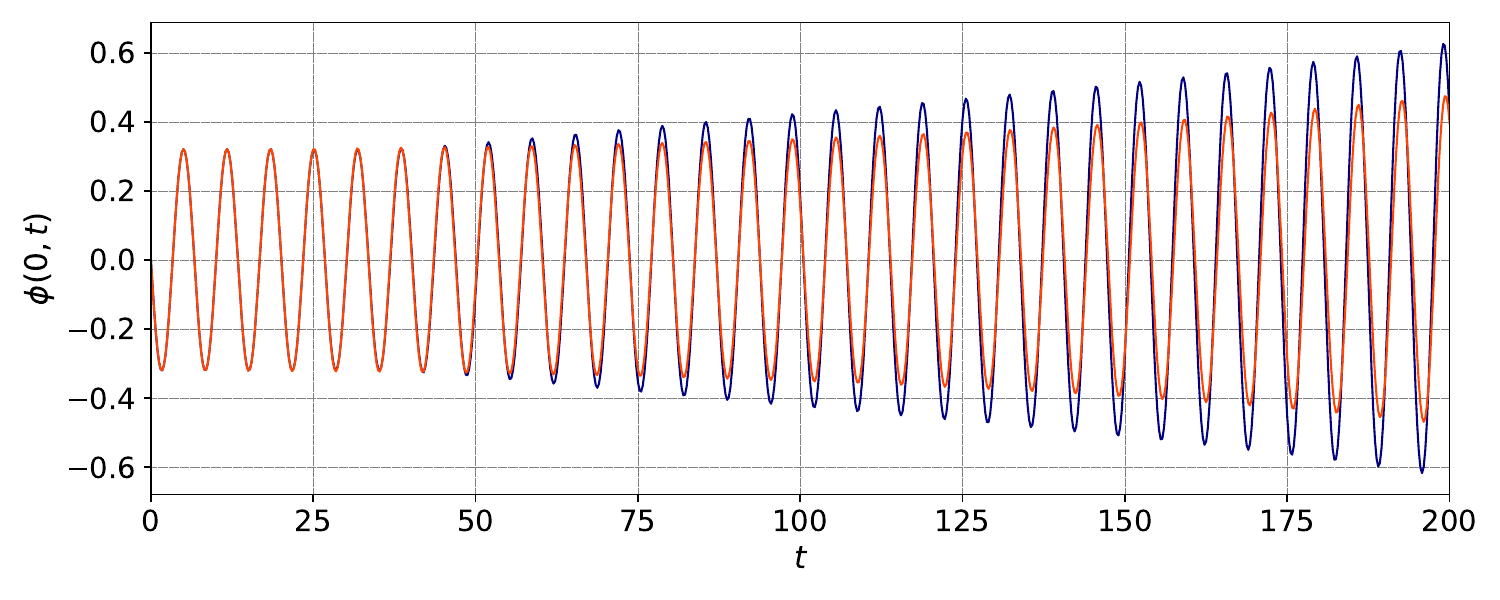}\label{phi6_05_015_02}
    \includegraphics[{angle=0,width=8cm,height=4cm}]{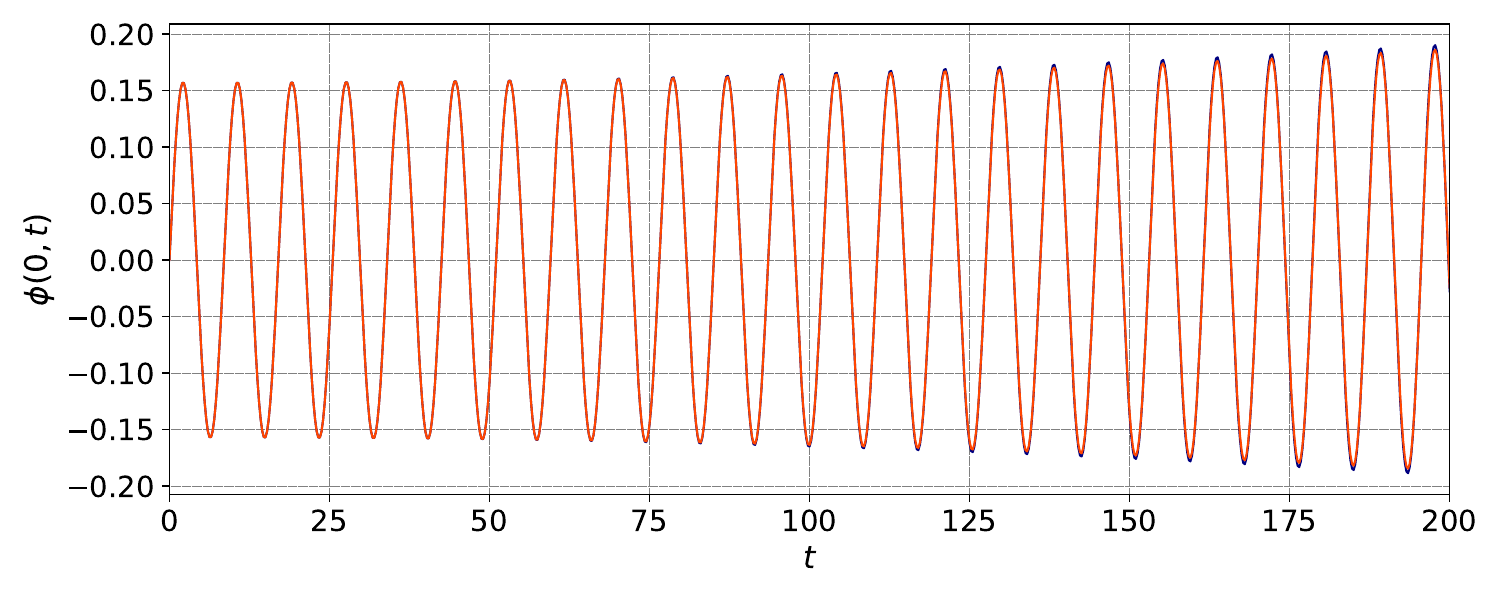}\label{phi6_08_015_008}
    \includegraphics[{angle=0,width=8cm,height=4cm}]{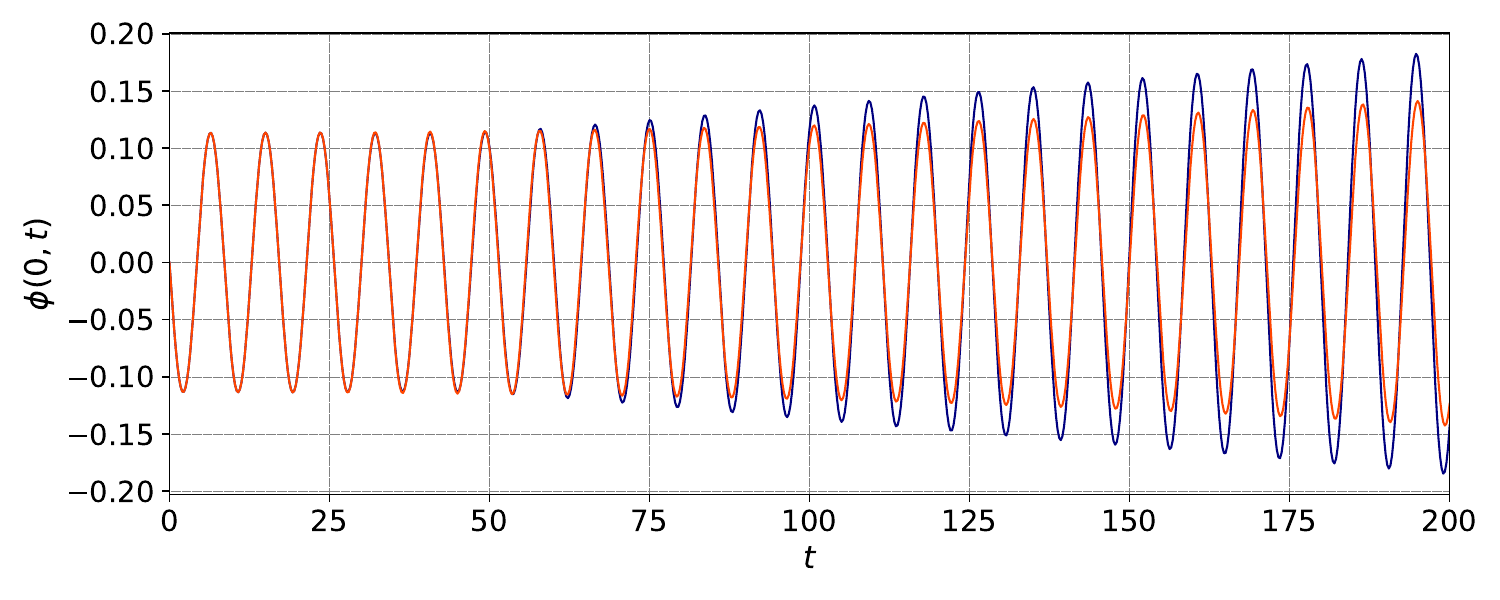}\label{phi6_08_015_02}
    \vspace{-0.5cm}
  \caption{Results for the $\phi^2$-potential Eq. (\ref{ep6}). Conventions as in Fig. \ref{fig_2Ba}. %Results obtained from the $\phi^2$-potential. The numerical oscillon (black line) is compared to the renormalized analytical one (red line). Upper: $\eta=0.10$ with $\lambda_2=-0.08$ (left) and $\lambda_2=-0.20$ (right). Lower: $\eta=0.80$ with $\lambda_2=-0.08$ (left) and $\lambda_2=-0.20$ (right). For all figures, we fixed $\lambda_1=0.15$.
  }
  \label{fig_phi6C}
\end{center}
\end{figure*}

\begin{figure*}[!ht]
\begin{center}
 \centering
   \includegraphics[{angle=0,width=8cm,height=4.5cm}]{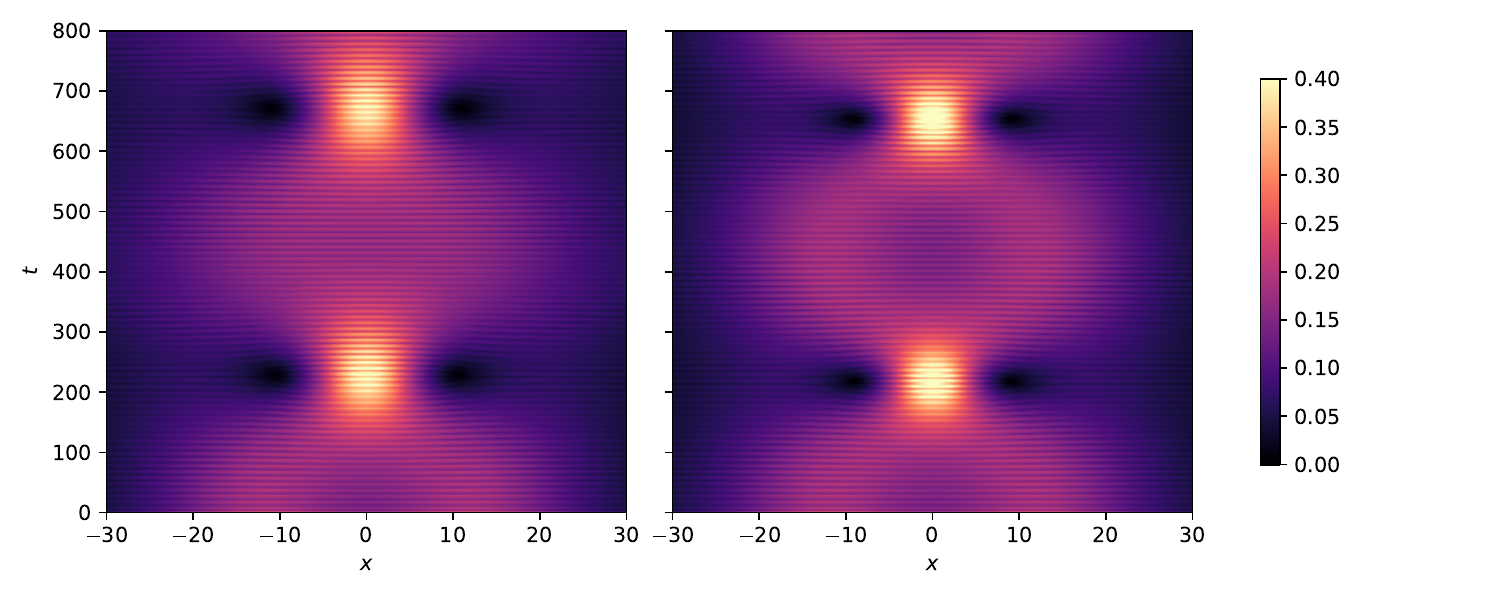}\label{3d_phi6_01_01_02}
   \includegraphics[{angle=0,width=8cm,height=4.5cm}]{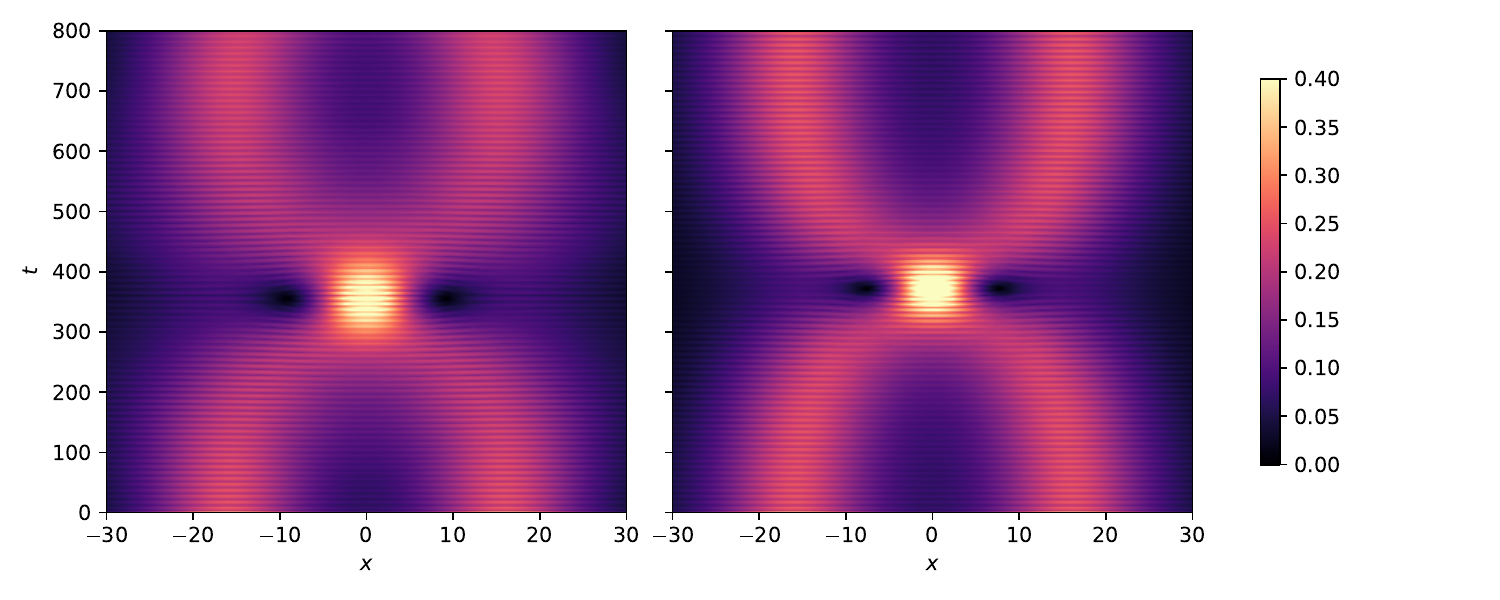}\label{3d_phi6_01_015_02}
   \includegraphics[{angle=0,width=8cm,height=4.5cm}]{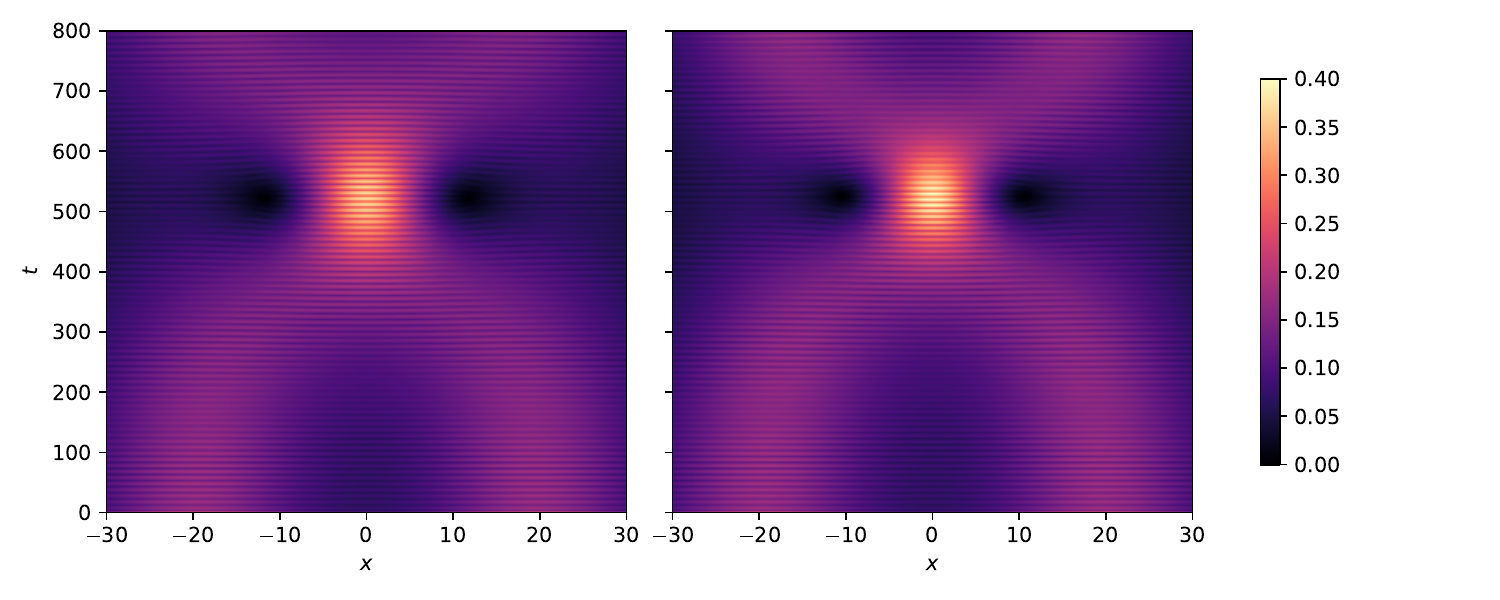}\label{3d_phi6_01_01_015}
   \includegraphics[{angle=0,width=8cm,height=4.5cm}]{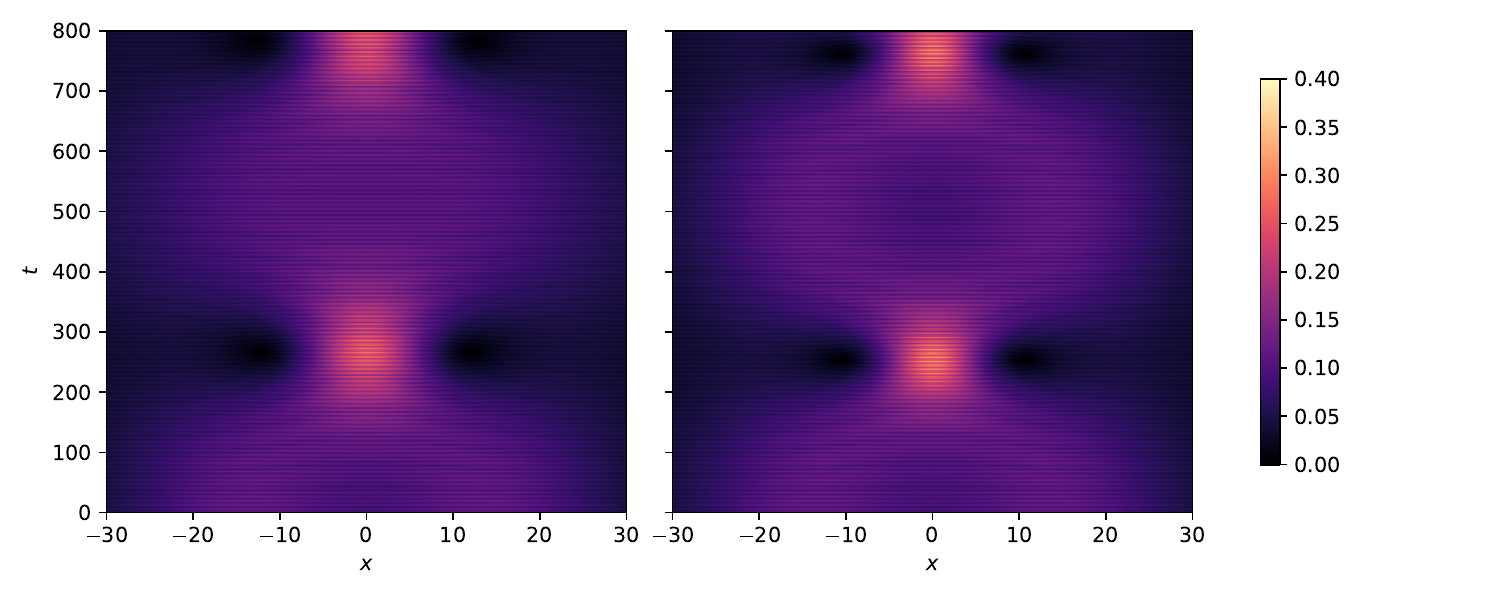}\label{3d_phi6_05_01_02}
    \vspace{-0.5cm}
  \caption{Results for the $\phi^2$-potential Eq. (\ref{ep6}). Conventions as in Fig. \ref{fig_1Ca}.%Comparasion bewtween numerically found modulated oscillon (left) and the renormalized solution obtained from the two Q-balls solution (right). Upper: $\eta=0.10$ with $\lambda_1=0.10$ and $\lambda_2=-0.20$ (left) and $\eta=0.10$ with $\lambda_1=0.15$ and $\lambda_2=-0.20$ (right). Lower: $\eta=0.10$ with $\lambda_1=0.10$ and $\lambda_2=-0.15$ (left) and $\eta=0.5$ with $\lambda_1=0.1$ and $\lambda_2=-0.20$ (right). We plot $|\partial^2 \phi + \phi|$ versus $x$ and $t$ for the $\phi^2$ model.
  }
  \label{fig_phi6D}
\end{center}
\end{figure*}

Therefore, instead of Eq. (\ref{cosh1}), we now use the two $Q$-ball profile Eq. (\ref{tqbs}) to seed the renormalized oscillon Eq. (\ref{de1xep6}). We solve the numerical problem for fixed $\lambda_2$ and $\lambda_1$, separately. The results for $\phi(x=0,t)$ appear in Figs. \ref{fig_phi6Ba} and \ref{fig_phi6C}, respectively, while the entire solutions are shown in Fig. \ref{fig_phi6D}. They reveal that the analytical profile based on the two $Q$-ball solution reproduces the evolution of a generalized $\phi^2$-oscillon with great precision.

%%%%%%%%%%%%%%%%%%%%%%%%%%%%%%%%%%%%%%%%%%%%%%%%%%%%%%%%%%%%%%%%%%%%%%%%%%% 
%%%%%%%%%%%%%%%%%%%%%%%%%%%%%%%%%%%%%%%%%%%%%%%%%%%%%%%%%%%%%%%%%%%%%%%%%%

%%%%%%%%%%%%%%%%%%%%%%%%%%%%%%%%%%%%%%%%%%%%%%%%%
\subsection{Exotic $\phi^6$-potential: different universality class} \label{secIIb}
%%%%%%%%%%%%%%%%%%%%%%%%%%%%%%%%%%%%%%%%%%%%%%%%%

We now briefly discuss the case with $\beta=0$. In this scenario, Eq. (\ref{ncfx_y1}) mutes, and the third order approximation does not apply anymore. To circumvent this issue, it is necessary to implement a higher order approximation. So, in what follows, we approximate the scalar field up to the fifth order in $\varepsilon$.

In the standard $f=1$ model, $\beta=0$ is fully obtained once we set $a_3=a_4=0$. However, in our generalized context, this choice must be followed by the additional condition $B_{1}^{2}=2B_{2}$. So, in order to satisfy this last one, we choose $B_1=B_2=0$, for the sake of simplicity.

In view of $a_3=a_4=0$, Eq. (\ref{gg4}) reduces to%
\begin{equation}
V\left( \phi \right) = \frac{\phi ^{2}}{2}-a_{5}\frac{\phi ^{5}}{5}%
-a_{6}\frac{\phi ^{6}}{6}+\text{... ,}\label{neweq1}
\end{equation}%
while $b_1=b_2=0$ leads to a generalized function as%
\begin{equation}
f\left( \phi \right) = b_{0}-b_{3}\frac{\phi ^{3}}{3}-b_{4}\frac{\phi
^{4}}{4}-b_{5}\frac{\phi ^{5}}{5}-b_{6}\frac{\phi ^{6}}{6}+\text{... ,}\label{neweq2}
\end{equation}%
see Eq. (\ref{gg5}).

As we have argued before, we approximate $\phi$ up to the fifth order in $\varepsilon$, i.e.%
\begin{equation}
\phi = \varepsilon \phi _{1}+\varepsilon ^{4}\phi _{4}+\varepsilon ^{5}\phi _{5}+\text{... ,}\label{neweq3}
\end{equation}%
where we have chosen $\phi _{2}=\phi _{3}=0$, for simplicity (to be justified below).

%%%%%%%%%%%%%%%%%%%%%%%%%%%%%%%%%%%%%%%%%%%%%%%%%%%%%%%%%%%%%%%%%%%%%%%%%%% 
%%%%%%%%%%%%%%%%%%%%%%%%%%%%%%%%%%%%%%%%%%%%%%%%%%%%%%%%%%%%%%%%%%%%%%%%%%

We implement Eqs. (\ref{neweq1}), (\ref{neweq2}) and (\ref{neweq3}) in Eq. (\ref{gceom0x1}). Then, up to the fifth order in $\varepsilon$, we get%
\begin{equation}
\frac{d^{2}\phi _{1}}{d\vartheta ^{2}}+\phi _{1}=0\text{,}
\label{ep1x0_y00a}
\end{equation}%
\begin{equation}
\frac{d^{2}\phi _{4}}{d\vartheta ^{2}}+\phi _{4}-\frac{B_{3}}{3}\phi
_{1}^{3}\frac{d^{2}\phi _{1}}{d\vartheta ^{2}}-\frac{B_{3}}{2}\phi
_{1}^{2}\left( \frac{d\phi _{1}}{d\vartheta }\right) ^{2}=a_{5}\phi _{1}^{4}%
\text{,}  \label{ep2x0_y00a}
\end{equation}%
\begin{equation}
\frac{d^{2}\phi _{5}}{d\vartheta ^{2}}+\phi _{5}-\frac{B_{4}}{4}\phi
_{1}^{4}\frac{d^{2}\phi _{1}}{d\vartheta ^{2}}-\frac{B_{4}}{2}\phi
_{1}^{3}\left( \frac{d\phi _{1}}{d\vartheta }\right) ^{2}=a_{6}\phi _{1}^{5}%
\text{,}  \label{ep3x0_y00a}
\end{equation}%
where we have used $\vartheta =\theta / \sqrt{b_{0}}$ again. We have also defined%
\begin{equation}
B_{3}=\frac{b_{3}}{b_{0}}\text{ \ \ and \ \ }B_{4}=\frac{b_{4}}{b_{0}}\text{.%
}
\end{equation}%
Alternatively, whether we consider both $\phi _{2}$ and $\phi _{3}$ in Eq. (\ref{neweq3}) explicitly, we get that each one of these fields satisfies the same Eq. (\ref{ep1x0_y00a}). Therefore, $\phi _{2}=\phi _{3}=0$ can be chosen as a genuine solution, as we have argued previously.

%%%%%%%%%%%%%%%%%%%%%%%%%%%%%%%%%%%%%%%%%%%%%%%%%%%%%%%%%%%%%%%%%%%%%%%%%%% 
%%%%%%%%%%%%%%%%%%%%%%%%%%%%%%%%%%%%%%%%%%%%%%%%%%%%%%%%%%%%%%%%%%%%%%%%%%

Equations (\ref{ep1x0_y00a}), (\ref{ep2x0_y00a}) and (\ref{ep3x0_y00a}) give rise to the solutions%
\begin{equation}
\phi _{1}\left( \vartheta \right) =A_{0}e^{i\vartheta }+\text{c.c.,}  \label{sp1x_y00aa}
\end{equation}
\begin{equation}
\phi _{4}\left( \vartheta \right) =-\frac{A_{0}^{4}}{15}\left( a_{5}-\frac{5B_{3}}{6}%
\right)e^{4i\vartheta }-\frac{4A_{0}^{2}}{3}\left( a_{5}-\frac{B_{3}}{3%
}\right) \left\vert A_{0}\right\vert ^{2}e^{2i\vartheta }+3\left(
a_{5}-\frac{B_{3}}{6}\right) \left\vert A_{0}\right\vert ^{4}+\text{ c.c.,}\label{xx1}
\end{equation}%
%or%
%\begin{equation}
%\phi _{4}\left( \vartheta \right) =\alpha _{4}A_{0}^{4}e^{4i\vartheta
%}+\beta _{4}\left\vert A_{0}\right\vert ^{2}A_{0}^{2}e^{2i\vartheta }+\gamma
%_{4}\left\vert A_{0}\right\vert ^{4}+\alpha _{4}\overline{A}%
%_{0}^{4}e^{-4i\vartheta }+\beta _{4}\left\vert A_{0}\right\vert ^{2}%
%\overline{A}_{0}^{2}e^{-2i\vartheta }\text{,}
%\end{equation}%
%where we have defined%
%\begin{equation}
%\alpha _{4}=-\frac{1}{15}\left( a_{5}-\frac{5}{6}B_{3}\right) \text{ \ \ and
%\ \ }\beta _{4}=-\frac{4}{3}\left( a_{5}-\frac{B_{3}}{3}\right) \text{,}
%\end{equation}%
%\begin{equation}
%\gamma _{4}=6\left( a_{5}-\frac{B_{3}}{6}\right) \text{.}
%\end{equation}%
%Here, we have assumed that the two integration constants vanish.
\begin{equation}
\phi _{5}\left( \vartheta \right) =-\frac{A_{0}^{5}}{24}\left( a_{6}-\frac{3B_{4}}{4}%
\right) e^{5i\vartheta }-\frac{A_{0}^{3}}{8}\left( 5a_{6}-\frac{7B_{4}}{4}%
\right) \left\vert A_{0}\right\vert ^{2}e^{3i\vartheta
}+2A_{0}\left( 5a_{6}-\frac{3B_{4}}{4}\right) \left\vert A_{0}\right\vert
^{4}\mathcal{S}e^{i\vartheta } +\text{ c.c.,}\label{xx2}
\end{equation}%
where $\mathcal{S}$\ represents the secular term. Note that Eqs. (\ref{xx1}) and (\ref{xx2}) recover Eqs. (48) and (49) of Ref. \cite{prd} for $f=1$.

Then, we write the bare solution Eq. (\ref{neweq3}) as%
\begin{eqnarray}
\phi _{B}\left( \vartheta \right) &=&\varepsilon A_{0}e^{i\vartheta
}-\frac{\varepsilon ^{4}A_{0}^{4}}{15}\left( a_{5}-\frac{5B_{3}}{6}\right)
e^{4i\vartheta }-\frac{4\varepsilon ^{4}A_{0}^{2}}{3}\left( a_{5}-\frac{B_{3}%
}{3}\right) \left\vert A_{0}\right\vert ^{2}e^{2i\vartheta
}  \notag \\
&&
+3\varepsilon ^{4}\left( a_{5}-\frac{B_{3}}{6}\right) \left\vert
A_{0}\right\vert ^{4}-\frac{\varepsilon ^{5}
A_{0}^{5}}{24}\left( a_{6}-\frac{3B_{4}}{4}\right)e^{5i\vartheta }-\frac{\varepsilon ^{5}A_{0}^{3}}{8}\left( 5a_{6}-\frac{7B_{4}}{4}\right) \left\vert A_{0}\right\vert ^{2}e^{3i\vartheta
}  \notag \\
&&+2\varepsilon ^{5}A_{0}\left( 5a_{6}-\frac{3B_{4}}{4}\right) \left\vert
A_{0}\right\vert ^{4}\mathcal{S}e^{i\vartheta }+\text{c.c..}\label{nbs}
\end{eqnarray}

%%%%%%%%%%%%%%%%%%%%%%%%%%%%%%%%%%%%%%%%%%%%%%%%%%%%%%%%%%%%%%%%%%%%%%%%%%% 
%%%%%%%%%%%%%%%%%%%%%%%%%%%%%%%%%%%%%%%%%%%%%%%%%%%%%%%%%%%%%%%%%%%%%%%%%%

In order to remove the secular term, we propose the connection%between $A_0$ and $A$ as%
\begin{equation}
A_{0}=A\left( 1-2\varepsilon ^{4}\left\vert A\right\vert ^{4}\left( 5a_{6}-%
\frac{3B_{4}}{4}\right) \mathcal{S}_{0}+\mathcal{O}\left( \varepsilon
^{5}\right) \right) \text{,}
\end{equation}%
from which we rewrite Eq. (\ref{nbs}) in the dressed form%
\begin{eqnarray}
\phi _{D}\left( \vartheta \right) &=&\varepsilon Ae^{i\vartheta
}-\frac{\varepsilon ^{4}A^{4}}{15}\left( a_{5}-\frac{5B_{3}}{6}\right)
e^{4i\vartheta }-\frac{4\varepsilon ^{4}A^{2}}{3}\left( a_{5}-\frac{B_{3}}{3}%
\right) \left\vert A\right\vert ^{2}e^{2i\vartheta }  \notag
\\
&&+3\varepsilon
^{4}\left( a_{5}-\frac{B_{3}}{6}\right) \left\vert A\right\vert ^{4}
-\frac{\varepsilon ^{5}A^{5}}{24}\left( a_{6}-\frac{3B_{4}}{4}\right)
e^{5i\vartheta }-\frac{\varepsilon ^{5}A^{3}}{8}\left( 5a_{6}-\frac{7B_{4}}{4}\right) \left\vert A\right\vert ^{2}e^{3i\vartheta }  \notag \\
&&+2\varepsilon ^{5}A\left( 5a_{6}-\frac{3B_{4}}{4}\right) \left\vert
A\right\vert ^{4}\left( \mathcal{S}-\mathcal{S}_{0}\right)e^{i\vartheta }+\text{c.c..}\label{dssuc}
\end{eqnarray}
Here, as before, $A$ and $\mathcal{S}_{0}$ are functions of $\vartheta_{0}$ and $\overline{\vartheta}_{0}$.

Again, $\phi _{D}\left( \vartheta \right)$ is required to be independent of the renormalized scales. So, we impose
\begin{equation}
\frac{\partial \phi _{D}}{\partial \vartheta _{0}}=\frac{\partial \phi _{D}}{\partial \overline{\vartheta }_{0}}=0\text{,}
\end{equation}%
which leads to%RG equations of the form
\begin{equation}
\frac{\partial A}{\partial \vartheta _{0}}=2\varepsilon ^{4}\left( 5a_{6}-%
\frac{3B_{4}}{4}\right) \left\vert A\right\vert ^{4}A\frac{\partial \mathcal{%
S}_{0}}{\partial \vartheta _{0}}+\mathcal{O}\left( \varepsilon ^{5}\right)\text{ \ \ and \ \ }\frac{\partial A}{\partial \overline{\vartheta }_{0}}=2\varepsilon
^{4}\left( 5a_{6}-\frac{3B_{4}}{4}\right) \left\vert A\right\vert ^{4}A%
\frac{\partial \mathcal{S}_{0}}{\partial \overline{\vartheta }_{0}}+\mathcal{O}\left( \varepsilon ^{5}\right)\text{.}
\end{equation}%
Note that these RG equations differ from Eq. (\ref{rge1}) not only by the order in $\varepsilon$, but also by the power of $\left\vert A\right\vert$. This fact suggests the emergence of a novel universality class, as we confirm below.

We adopt the minimal subtraction scheme. Following the RGPE algorithm, we also assume%
\begin{equation}
2i\frac{\partial A}{\partial \vartheta _{0}}+\frac{\partial ^{2}A}{\partial
\vartheta _{0}^{2}}-\frac{\partial ^{2}A}{\partial \overline{\vartheta }%
_{0}^{2}}=2\varepsilon ^{4}\left( 5a_{6}-\frac{3B_{4}}{4}\right) \left\vert
A\right\vert ^{4}A\left( 2i\frac{\partial \mathcal{S}_{0}}{\partial
\vartheta _{0}}+\frac{\partial ^{2}\mathcal{S}_{0}}{\partial \vartheta
_{0}^{2}}-\frac{\partial ^{2}\mathcal{S}_{0}}{\partial \overline{\vartheta }%
_{0}^{2}}\right) \text{,}
\end{equation}%
which can be reduced to%
\begin{equation}
2i\frac{\partial A}{\partial \vartheta _{0}}+\frac{\partial ^{2}A}{\partial
\vartheta _{0}^{2}}-\frac{\partial ^{2}A}{\partial \overline{\vartheta }%
_{0}^{2}}=2\varepsilon ^{4}\left( 5a_{6}-\frac{3B_{4}}{4}\right) \left\vert
A\right\vert ^{4}A\text{,}  \label{rgex_ya}
\end{equation}%
where we have used Eq. (\ref{fe_0x_y1}).

In the present case, we define the underlying complex field $\Psi$ as%
\begin{equation}
\Psi \left( \vartheta \right) =\frac{\varepsilon }{\gamma }Ae^{i\vartheta
}\text{,}  \label{ncfx_ya}
\end{equation}%
where we have chosen%
\begin{equation}
\gamma =\sqrt[4]{\frac{3}{2\left( 5a_{6}-\frac{3B_{4}}{4}\right) }}\text{.}\label{dg}
\end{equation}

This definition allows us to rewrite RG Eq. (\ref{rgex_ya}) in the form%
\begin{equation}
\frac{\partial ^{2}\Psi }{\partial \vartheta ^{2}}+\Psi =3\left\vert \Psi
\right\vert ^{4}\Psi \text{,}  \label{eoma}
\end{equation}%
which can be understood to be the EoM that comes from Lagrange density%
\begin{equation}
\mathcal{L}=\left\vert \partial _{\mu }\Psi \right\vert ^{2}-\left\vert \Psi
\right\vert ^{2}+\left\vert \Psi \right\vert ^{6}\label{novelm}\text{.}
\end{equation}%
As before, the coordinates are $T=b_{0}^{-1/2}t$ and $X=b_{0}^{-1/2}x$. Note that, instead of the quartic term, the self-interaction now contains a sixtic one. In this case, stationary $Q$-balls of the form%
\begin{equation}
\Psi =\frac{\sqrt{\lambda }e^{i\omega T}}{\sqrt{\cosh \left( 2\lambda
X\right) }}\text{}  \label{cosh2}
\end{equation}%
emerge as periodic solutions to Eq. (\ref{eoma}).

%Here, $\omega =\sqrt{1-\lambda ^{2}}$ stands for the $Q$-ball's frequency.

In terms of the complex field (\ref{ncfx_ya}), $\phi _{D}\left( \vartheta \right)$ as given by Eq. (\ref{dssuc}) can be written as%
\begin{eqnarray}
\phi _{R}\left( \vartheta \right) &=&\gamma \Psi -\frac{\gamma ^{4}}{15}%
\left( a_{5}-\frac{5B_{3}}{6}\right) \Psi ^{4}-\frac{4\gamma ^{4}}{3}\left(
a_{5}-\frac{B_{3}}{3}\right) \Psi ^{2}\left\vert \Psi \right\vert ^{2}+3\gamma ^{4}\left( a_{5}-\frac{B_{3}}{6}\right) \left\vert \Psi
\right\vert ^{4} 
\notag \\
&&-\frac{\gamma ^{5}}{24}\left( a_{6}-\frac{3B_{4}}{4}\right)
\Psi ^{5}-\frac{\gamma ^{5}}{8}\left( 5a_{6}-\frac{7B_{4}}{4}\right) \Psi
^{3}\left\vert \Psi \right\vert ^{2}+\text{c.c.,}\label{rssuc}
\end{eqnarray}%
which stands for the renormalized oscillon.%Here, we have also implemented the minimal subtraction scheme.

Equation (\ref{rssuc}) indicates that the relation between generalized oscillons and underlying $Q$-balls holds up also to the fifth order in the approximation expansion. The nonstandard result is said to belong to a distinct universality class due to its RG Eq. (\ref{rgex_ya}), which differs from the previous Eq. (\ref{rgex_y1}).

Our generalized oscillon keeps the canonical structure unaltered, see Eq. (56) of Ref. \cite{prd}. Meanwhile, the underlying field still exhibits standard kinematics. So, up to the fifth order (and except for the rescaled coordinates), the same $Q$-ball profile figures as the basic ingredient to both the canonical and generalized oscillons.

%So, as in the previous case, we identify the novel effects in the rescaled parameters.

%This means that the basic correspondence applies well also to those solutions whose amplitudes are not necessarily so small.

%As in the previous cases, Eq. (\ref{rssuc}) means that oscillons in a model with generalized kinematics are related to $Q$-balls in an underlying theory of a complex scalar sector. It is now clear that such a relation holds not only up to the third order in the approximation series expansion, but also up to the fifth order. This indicates that the basic correspondence applies well also to those solutions whose amplitudes are not necessarily so small.

%Moreover, the underlying complex field is still controlled by standard kinematics. So, up to the fifth order in the approximation series, the same $Q$-ball figures as the basic ingredient to both the canonical and generalized oscillons.

%%%%%%%%%%%%%%%%%%%%%%%%%%%%%%%%%%%%%%%%%%%%%%%%%%%%%%%%%%%%%%%%%%%%%%%%%%% 
%%%% ------ sec B ------
\begin{figure*}[!ht]
\begin{center}
  \centering
    \includegraphics[{angle=0,width=8cm,height=4cm}]{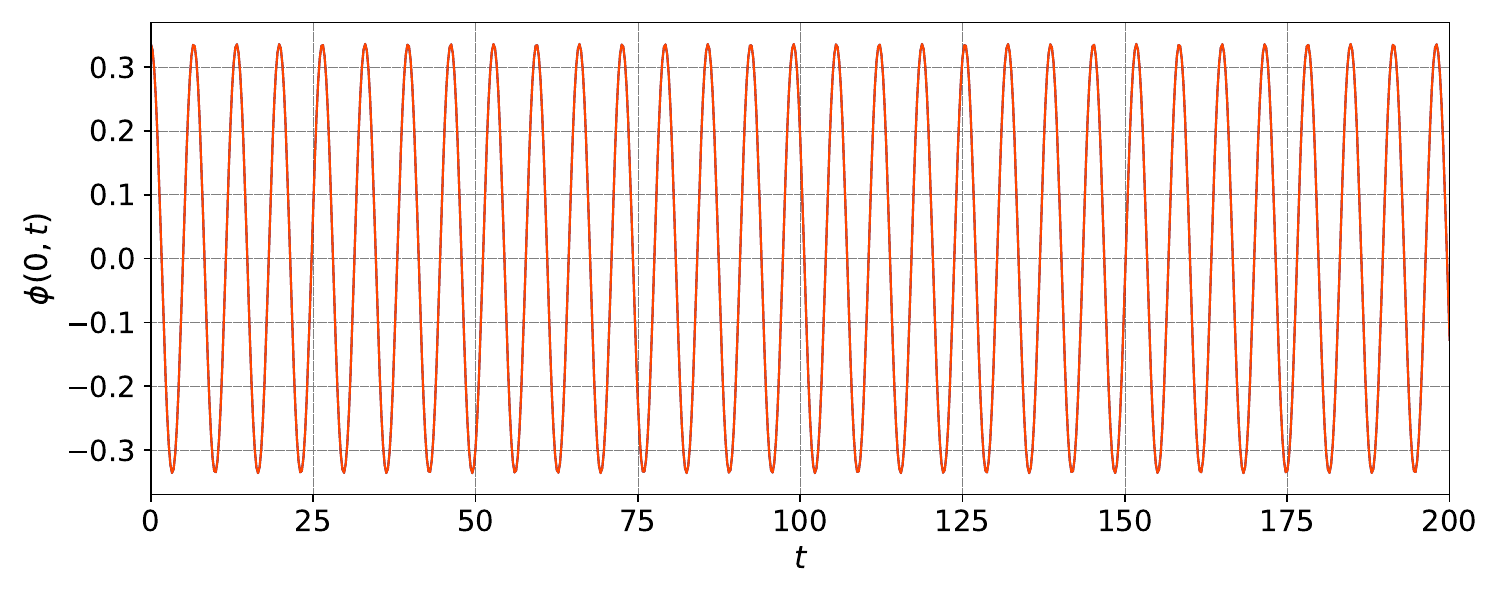}\label{B_phi6_01_005}
    \includegraphics[{angle=0,width=8cm,height=4cm}]{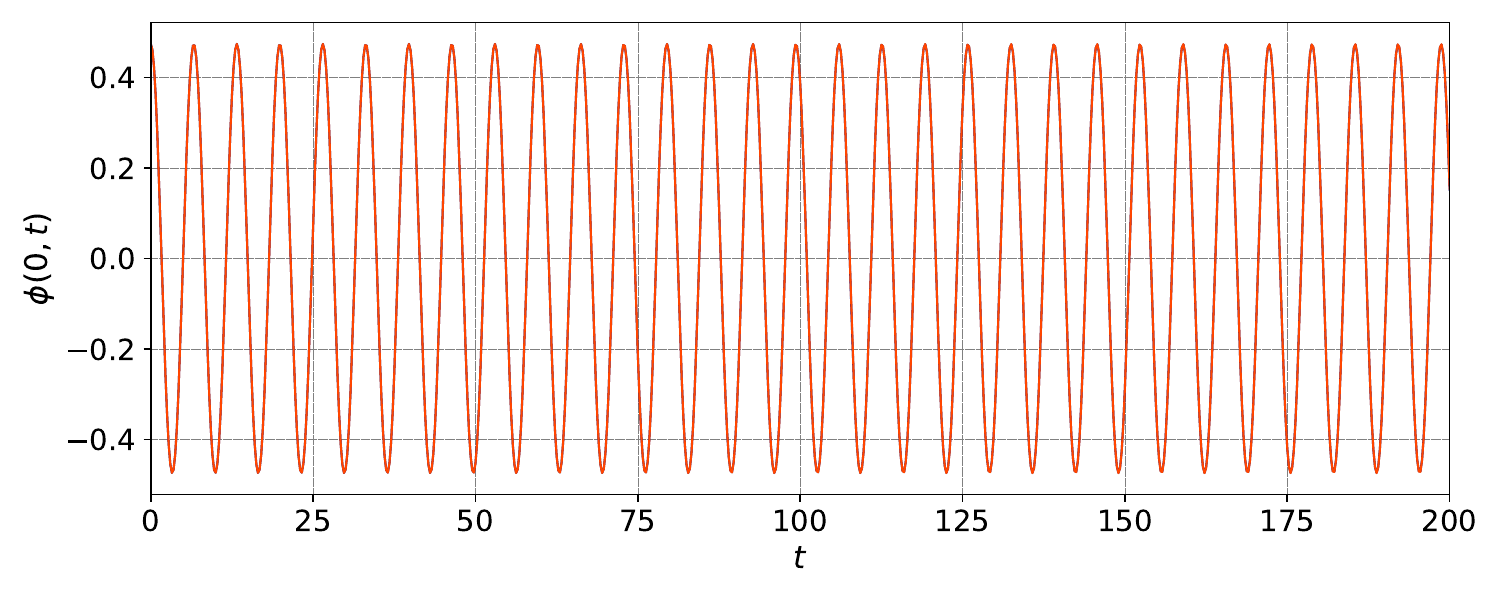}\label{B_phi6_01_01}
    \includegraphics[{angle=0,width=8cm,height=4cm}]{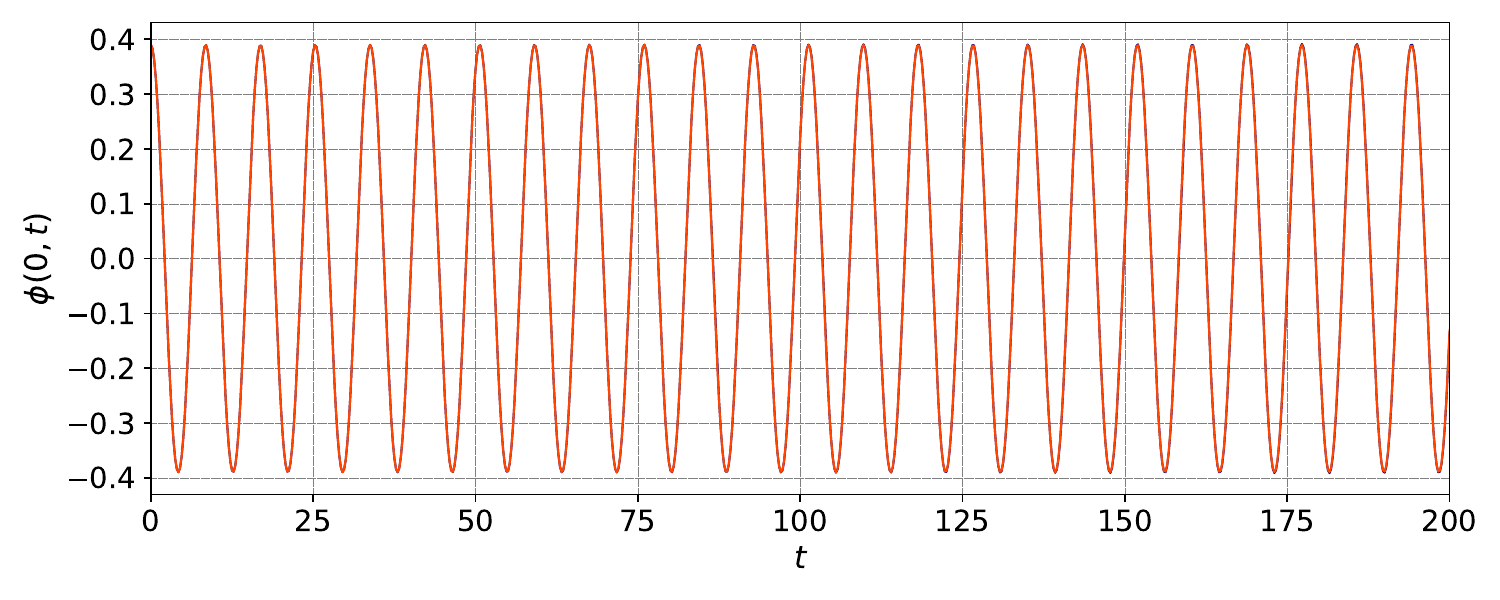}\label{B_phi6_08_005}
    \includegraphics[{angle=0,width=8cm,height=4cm}]{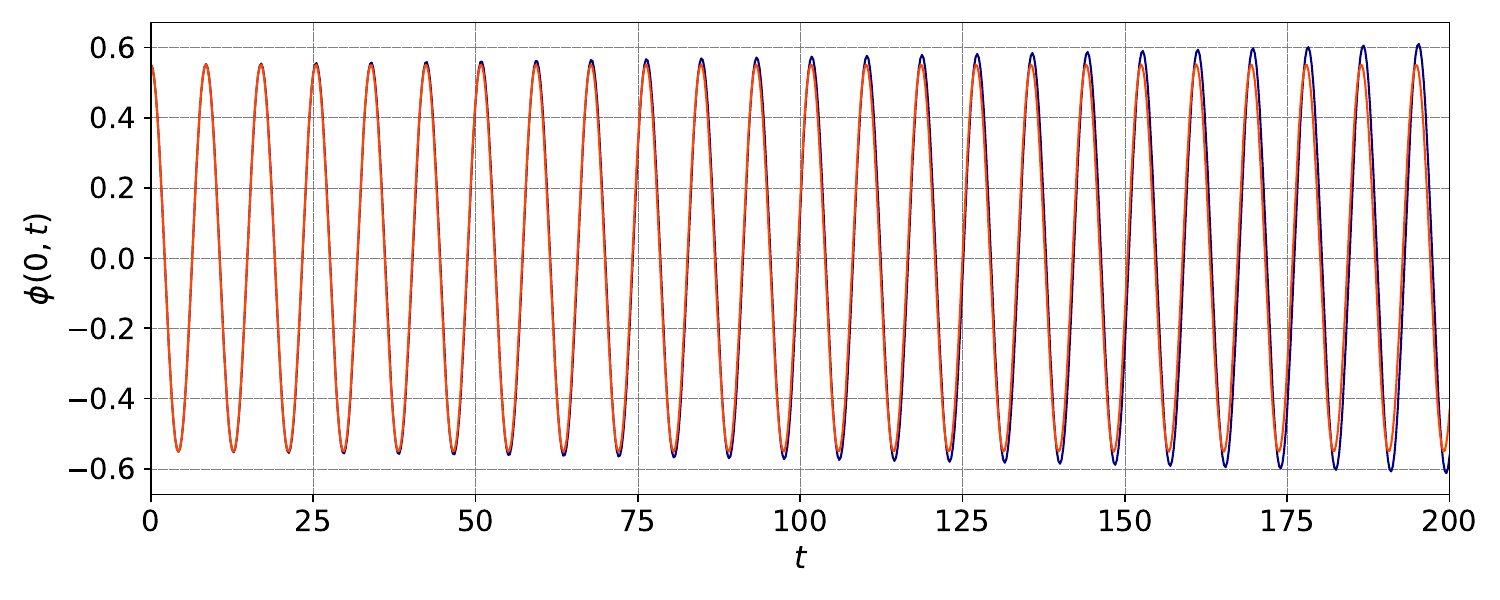}\label{B_phi6_08_01}
    \vspace{-0.5cm}
  \caption{Results for the exotic $\phi^6$-potential Eq. (\ref{ep6aa}). The numerical oscillon (black line) is compared to the renormalized analytical one (red line). Here, we have chosen $\eta=0.10$ (upper line) and $\eta=0.80$ (lower line), with $\lambda=0.05$ (left column) and $\lambda=0.10$ (right column).}
  \label{fig_phi6B}
\end{center}
\end{figure*}
%%%%%%%%%%%%%%%%%%%%%%%%%%%%%%%%%%%%%%%%%%%%%%%%%%%%%%%%%%%%%%%%%%%%%%%%%%

In Eq. (\ref{ncfx_ya}), $\gamma$ is assumed to be real. Therefore, the renormalized solution (\ref{rssuc}) only exists if%
\begin{equation}
5a_{6}>\frac{3B_{4}}{4} \label{ggub}
\end{equation}%
is satisfied, see Eq. (\ref{dg}). That is, the potential again restricts the strength of the effects caused by nontrivial kinematics.

\textit{Effective example}. We now study a particular case. First, we choose%
\begin{equation}
f\left( \phi \right) =\frac{1+\eta }{1+\eta \phi ^{4}}\text{,}
\end{equation}
where $\eta \ge0$ again controls the strength of the novel effects.

When $\phi$ is sufficiently small, we can approximate%
\begin{equation}
f\left( \phi \right) \approx \left( 1+\eta \right) -\eta \left( 1+\eta
\right) \phi ^{4}\text{,}
\end{equation}%
which leads to $b_{0}=1+\eta $, $b_{3}=0$ and $b_{4}=4\eta \left(
1+\eta \right) $. Then, we get $B_{3}=0$ and $B_{4}=4\eta$.

Now, we specify the potential. For the sake of illustration, we choose the exotic $\phi^6$ one, i.e.
\begin{equation}
V\left( \phi \right) =\frac{\phi ^{2}}{2}-\frac{\phi ^{6}}{6}
\label{ep6aa}\text{,}
\end{equation}%
which leads to $a_{5}=0$ and $a_{6}=1$.

In view of these choices, the renormalized oscillon (\ref{rssuc}) assumes the explicit form
\begin{eqnarray}
\phi _{R}\left( x,t\right) &=&2\gamma \sqrt{\lambda} \frac{\cos \left( \frac{%
\omega }{\sqrt{1+\eta }}t\right)}{\sqrt{\cosh \left( 
\frac{2\lambda }{\sqrt{1+\eta }}x\right)}}-\frac{\gamma ^{5}\lambda^{5/2}\left( 5-7\eta
\right)}{4} \frac{ \cos \left( \frac{3\omega }{\sqrt{1+\eta }}t\right)}{\cosh^{5/2} \left( \frac{2\lambda }{\sqrt{1+\eta }}%
x\right) } 
\notag \\
&&-\frac{\gamma ^{5} \lambda^{5/2} \left( 1-3\eta \right)}{12} \frac{ \cos \left( 
\frac{5\omega }{\sqrt{1+\eta }}t\right) }{\cosh^{5/2}
\left( \frac{2\lambda }{\sqrt{1+\eta }}x\right) } \label{ggns} \text{,}
\end{eqnarray}%
which only holds for $\eta < 5/3$, see Eq. (\ref{ggub}). Here, $\gamma$ effectively reads%
\begin{equation}
\gamma =\sqrt[4]{\frac{3}{2\left( 5-3\eta \right) }}\text{,}
\end{equation}
as obtained from Eq. (\ref{dg}).

We assume Eq. (\ref{ggns}) as the initial state, and solve the evolution numerically. The comparison between analytical and numerical profiles appears in Fig. \ref{fig_phi6B}. As in the previous cases, the correspondence applies well for small and moderate amplitudes.

As the initial amplitude gets higher, the correspondence loses accuracy. In this regime, however, the numerical solution does not develop a modulated behavior. Such an aspect is also observed in the standard $f=1$ scenario. It means that also our generalized oscillon has one fundamental frequency only.

In comparison to the previous cases, the absence of modulations is a qualitative property inherent to the different universality class. In this sense, it is closely related to the novel RG Eq. (\ref{rgex_ya}). %(or, equivalently, Eq. (\ref{eoma})).
Furthermore, once the underlying model (\ref{novelm}) can not be approximated by the complex sine-Gordon theory (\ref{lcsg}), there is no reason to use the two $Q$-ball solution as the basis of our generalized analytical oscillon.

%%%%%%%%%%%%%%%%%%%%%%%%%%%%%%%%%%%%%%%%%%%%%%%%%
\section{Summary and perspectives} \label{secIII}
%%%%%%%%%%%%%%%%%%%%%%%%%%%%%%%%%%%%%%%%%%%%%%%%%

We have considered a $(1+1)$-dimensional model with a single real scalar field. Its kinetic term was enlarged to include nonstandard dynamics. In such a context, we have applied the RGPE algorithm to study the connection between generalized oscillons with small-amplitude and an underlying complex model supporting $Q$-ball solutions.

We have assumed that the real field self-interacts according to the general potential Eq. (\ref{gg4}). It presents a global minimum at $\phi=0$ around which the oscillon evolves. Based on the same idea, we have proposed the nonstandard kinematics as in Eq. (\ref{gg5}). It has the advantage to allow us to recover the canonical results promptly.

We have approximated the real sector up to the third order in a book-keeping parameter. Then, we have used the RGPE prescription to obtain the renormalized oscillon Eq. (\ref{de1x}). The generalized result has emerged as a function of an underlying complex field $\Psi$ whose particular model admits the single $Q$-ball solution Eq. (\ref{cosh1}).

In this sense, we have demonstrated that the close relation between oscillons and $Q$-balls remains preserved even in the presence of generalized kinematics. This conclusion is not obvious a priori, especially due to the high nonlinearity introduced by a nonstandard kinetic term.

Surprisingly, the generalized oscillon is mathematically similar to its canonical counter-partner. %exhibits the same mathematical structure as its canonical counter-partner.
The novelty lies on its parameters, which are now rescaled, see Eqs. (\ref{bet}) and (\ref{alf}). So, we have concluded that nontrivial kinematics induces the rescaling of the original constants, with the new ones now carrying the effects of the generalization.

%Moreover, due to such a rescaling,
The generalized dynamics also affects the existence of renormalized oscillons themselves. Now, the condition that supports these profiles contains those coefficients that come from the generalization itself, see Eq. (\ref{b1}). As a consequence, the potential restricts the strength of the effects due to the enlarged kinematics.

To corroborate our results, we have applied them to study some effective cases. First, we have chosen a well-established generalization, see Eq. (\ref{fphi}). Then, we have explored three different potentials. In all cases, our renormalized oscillon has reproduced the evolution of a generalized numerical one with great accuracy in the limits of small and moderate amplitudes.

As the initial amplitude increases, the numerical oscillon develops a modulated behavior, and the correspondence fails. To recover it, we have used an adequate two $Q$-ball solution to seed our renormalized profile. As a consequence, the analytical oscillon has mimicked the numerical modulated structure accurately.

In the sequence, we have focused on the simplest $\phi^2$-potential, see Eq. (\ref{ep6}). This case is of particular interest because it does not support renormalized oscillons in the canonical $f=1$ context. Here, however, we have showed that the generalized kinematics allows it to admit well-behaved renormalized solutions, and that they reproduce the noncanonical numerical ones with great accuracy. Surprisingly, we have found that the $\phi^2$-potential does not restrict the strength of the novel effects. The existence of renormalized oscillons in connection to the $\phi^2$-model %up to the third order in the approximation expansion
represents one of the main outcomes of nonstandard kinematics.

We have also considered the case in which the original third order approach mutes. To circumvent this issue, we have enlarged our analysis to include all contributions up to the fifth order in the approximation, from which we have obtained a renormalized oscillon (\ref{rssuc}) in terms of a complex field now controlled by a sixtic potential, see Eq. (\ref{novelm}). Furthermore, we have found that a new restriction on the nonusual effects appears.

We have used this fifth order solution to examine the exotic $\phi^6$-model, from which we have observed that the analytical profile mimics those numerical oscillons %with small and moderated amplitudes
accurately. In this case, no modulated structure appears as the initial amplitude increases. It means that our generalized oscillon has one fundamental frequency only. In this sense, there is no reason to use a two $Q$-ball profile to seed it. This is due to the different universality class to which such nonstandard oscillons belong.

Interesting issues to be investigated include the question on whether the %relation between
oscillon/$Q$-ball relation still exists in a theory with two mutually interacting scalar fields. In the simplest case, the interaction appears as a crossed term in the potential that defines the model. The application of the RGPE algorithm is expected to be much more intricate. However, preliminary calculations suggest that the $Q$-ball related to one sector may now affect the renormalized oscillon inherent to the other one. This may lead to novel aspects to be studied, especially in regard to the modulated behavior. We are currently working on this topic, and positive results will be presented in a future manuscript.

%%%%%%%%%%%%%%%%%%%%%%%%%%%%%%%%%%%%%%%%%%%%%%

\section*{Acknowledgements}

We thank Fundação de Amparo à Pesquisa e ao Desenvolvimento Científico e Tecnológico do Maranhão (FAPEMA), Conselho Nacional de Desenvolvimento Científico e Tecnológico (CNPq), and Coordenação de Aperfeiçoamento de Pessoal de Nível Superior (CAPES) - Finance Code 001 (Brazilian agencies) for partial financial support. F. C. S. acknowledges the support from the grants PQ-C-12556/25 Produtividade Estaduais CNPq/FAPEMA.

%%%%%%%%%%%%%%%%%%%%%%%%%%%%%%%%%%%%%%%%%%%%%%

%%%%%%%%%%%%%%%%%%%%%%%%%%%%%%%%%%%%%%%%%%%%%%

\end{document}